\documentclass[11pt]{article}
\usepackage[margin=1in]{geometry} 
\usepackage{amssymb,amsfonts,amsmath,bbm,mathrsfs}
\usepackage{stmaryrd}

\usepackage[colorlinks,
             linkcolor=black!50!red,
             citecolor=blue,
             pdftitle={The formal path integral and quantum mechanics},
             pdfauthor={Theo Johnson-Freyd},
             pdfsubject={quantum mechanics},
             pdfkeywords={path integral, quantum mechanics},
             pdfproducer={pdfLaTeX},
             pdfpagemode=None,
             bookmarksopen=true
             bookmarksnumbered=true]{hyperref}

\usepackage[amsmath,thmmarks,hyperref]{ntheorem}
\usepackage[poorman]{cleveref}

\theoremstyle{change}

\newtheorem{defn}[equation]{Definition}
\newtheorem{thm}[equation]{Theorem}
\newtheorem{prop}[equation]{Proposition}
\newtheorem{lemma}[equation]{Lemma}
\newtheorem{cor}[equation]{Corollary}

\theoremstyle{nonumberplain}

\newtheorem{mainthm}{Main Theorem}

\theorembodyfont{\upshape}

\theoremsymbol{\ensuremath{\Box}}
\newtheorem{proof}{Proof}

\qedsymbol{\ensuremath{\Box}}

\creflabelformat{equation}{#2(#1)#3} 

\crefname{equation}{equation}{equations}
\crefname{eg}{example}{examples}
\crefname{defn}{definition}{definitions}
\crefname{prop}{proposition}{propositions}
\crefname{thm}{Theorem}{Theorems}
\crefname{lemma}{lemma}{lemmas}
\crefname{cor}{corollary}{corollaries}
\crefname{section}{Section}{Sections}
\crefname{subsection}{Section}{Sections}

\crefformat{equation}{#2equation~(#1)#3} 
\crefformat{eg}{#2example~#1#3} 
\crefformat{defn}{#2definition~#1#3} 
\crefformat{prop}{#2proposition~#1#3} 
\crefformat{thm}{#2Theorem~#1#3} 
\crefformat{lemma}{#2lemma~#1#3} 
\crefformat{cor}{#2corollary~#1#3} 
\crefformat{section}{#2Section~#1#3} 
\crefformat{subsection}{#2Section~#1#3} 

\Crefformat{equation}{#2Equation~(#1)#3} 
\Crefformat{eg}{#2Example~#1#3} 
\Crefformat{defn}{#2Definition~#1#3} 
\Crefformat{prop}{#2Proposition~#1#3} 
\Crefformat{thm}{#2Theorem~#1#3} 
\Crefformat{lemma}{#2Lemma~#1#3} 
\Crefformat{cor}{#2Corollary~#1#3} 
\Crefformat{section}{#2Section~#1#3} 
\Crefformat{subsection}{#2Section~#1#3}

\numberwithin{equation}{subsection}

\newcommand{\Section}[1]{\section{#1} \setcounter{equation}{0}}

\newcommand{\qedhere}{\hfill\ensuremath{\Box}}

\usepackage{tikz}
\usetikzlibrary{arrows,decorations.pathreplacing,decorations.markings,shapes.geometric,through,fit,shapes.symbols}

\tikzset{
    dot/.style={circle,draw,fill,inner sep=1pt},
    Sdot/.style={circle,draw,fill,inner sep=1pt},
    twice/.style={double, double distance=1pt},
}

\newcommand\Aa{{\mathcal A}}
\newcommand\Bb{{\mathcal B}}
\newcommand\Cc{{\mathcal C}}
\newcommand\Dd{{\mathcal D}}
\newcommand\Ee{{\mathcal E}}
\newcommand\Ff{{\mathcal F}}
\newcommand\Nn{{\mathcal N}}
\newcommand\Oo{{\mathcal O}}
\newcommand\Uu{{\mathcal U}}
\newcommand\Vv{{\mathcal V}}
\newcommand\Ww{{\mathcal W}}

\newcommand\RR{{\mathbb R}}

\newcommand\T{{\rm T}}
\renewcommand{\d}{{\rm d}}

\newcommand\st{{\textrm{ s.t.\ }}}

\DeclareMathOperator{\Aut}{Aut}
\DeclareMathOperator{\cp}{cr}
\DeclareMathOperator{\ev}{ev}
\DeclareMathOperator{\extend}{ext}
\DeclareMathOperator{\flow}{flow}
\DeclareMathOperator{\formal}{formal}
\DeclareMathOperator{\id}{id}
\DeclareMathOperator{\Mat}{Mat}
\DeclareMathOperator{\res}{res}
\DeclareMathOperator{\tr}{tr}
\DeclareMathOperator{\dVol}{dVol}

\newcommand\isom{\overset{\sim}{\to}}
\newcommand\mono{\hookrightarrow}
\newcommand\epi{\twoheadrightarrow}

\newcommand{\define}[1]{{\bf #1}}

\title{The formal path integral and quantum mechanics}
\author{Theo Johnson-Freyd}

\begin{document}
\maketitle

\begin{abstract}
  Given an arbitrary Lagrangian function on $\RR^d$ and a choice of classical path, one can try to define Feynman's path integral supported near the classical path as a formal power series parameterized by ``Feynman diagrams,'' although these diagrams may diverge.  We compute this expansion and show that it is (formally, if there are ultraviolet divergences) invariant under volume-preserving changes of coordinates.  We prove that if the ultraviolet divergences cancel at each order, then our formal path integral satisfies a ``Fubini theorem'' expressing the standard composition law for the time evolution operator in quantum mechanics.  Moreover, we show that when the Lagrangian is inhomogeneous-quadratic in velocity such that its homogeneous-quadratic part is given by a matrix with constant determinant, then the divergences cancel at each order.  Thus, by ``cutting and pasting'' and choosing volume-compatible local coordinates, our construction defines a Feynman-diagrammatic ``formal path integral'' for the nonrelativistic quantum mechanics of a charged particle moving in a Riemannian manifold with an external electromagnetic field.
\end{abstract}

\Section{Introduction}

The primary goal of this paper is to clarify the definition and construction of the formal path integral as it applies to quantum mechanics on possibly-curved spaces.  We will prove that in the most important cases, the formal path integral is well-defined and satisfies a Fubini-style composition law.  In a companion piece \cite{meshort}, we address the formal-path-integral approach to quantum mechanics on $\RR^d$, and prove that the output of the formal path integral satisfies Shr\"odinger's equation with the correct initial value; the arguments there translate {\em mutatis mutandis} to the generality in this paper.

Feynman introduced the path integral in his thesis \cite{Feynman2005} (first published as \cite{Feynman1948} in 1948) as a new formalism for quantum mechanics.  In 1949, based on his path integral and his powerful physical intuition, Feynman introduced his famous diagrams as a tool for studying quantum electrodynamics \cite{Feynman1949b}.  In the subsequent years, path integrals and Feynman diagrams became universal in the study of quantum field theories; for a detailed history, see \cite{Kaiser2005}.  These applications are usually ``formal,'' in the sense that they return formal power series in the physical variables: analytic definitions of path integrals remain elusive in most cases.  Among physically important quantum field theories, only quantum mechanics (a one-dimensional quantum field theory) exists analytically (see e.g.\ \cite{Takhtajan2008}).  But the diagrammatic methods have not been rigorously checked against the analytic theory.

For us, the formal path integral is a machine that inputs a classical physical system --- a smooth manifold $\Nn$, called the \define{configuration space}, and a smooth function $L:\RR\times \T\Nn \to \RR$, called the \define{Lagrangian}, which is required to be convex along fibers, along with an extra bit of data: a smooth volume-form on $\Nn$ --- and a nonfocal classical trajectory $\gamma: [t_0,t_1] \to \Nn$.  The output of the formal path integral is essentially a formal power series in a formal variable $\hbar$ (``Planck's constant'').  We write ``essentially'' for two reasons.  Less importantly, the output is actually of the form:
\[ U_\gamma = \pm (i\hbar)^{-(\dim \Nn)/2} \exp\left( \sum_{n=-1}^\infty c_n(i\hbar)^n\right)\]
The $c_n$ are real coefficients that depend on $\gamma$, and are given by finite sums of finite-dimensional integrals.

More importantly, for $n\geq 1$, the coefficients $c_n$ are integrals of products of distributions (``generalized functions,'' like Dirac's $\delta$-function), and can diverge: the $c_n$ are formal polynomials in the infinite quantity $\delta(0)$.  These ``ultraviolet'' divergences are not usually considered in the physics literature on quantum mechanics, where ultraviolet divergence are normally thought of as a feature of higher-dimensional quantum field theories.
(An important case in which divergences have been considered is when the corresponding classical mechanics involves singular potentials.  Certain examples have been studied thoroughly within the framework of renormalization \cite{MT1994,DKlong}.)
  In \cref{divergenteg} we give a natural example with no classical singularities in which there are quantum divergences.  But one of the main theorems of this paper (\cref{divfreethm}) is that if $L$ is inhomogeneous quadratic on fibers and the chosen volume form is the one arising from the homogeneous-quadratic part of $L$ (a Riemannian metric), then the divergences cancel at each order.  In physical jargon, this situation corresponds to the nonrelativistic quantum mechanics of a single charged particle confined to a manifold, moving through an external electromagnetic field.
  
  We remark that even when there are no ultraviolet divergences, in general the power series given by the formal path integral has zero radius of convergence, an issue we will not discuss further.

\subsection{Detailed outline of the paper}

By a \define{path} we will always mean a piecewise-smooth parameterized path in $\Nn$; i.e.\ a continuous function $\varphi: [t_0,t_1] \to \Nn$ such that there exists a subdivision $t_0 = \tau_0 < \tau_1 < \dots < \tau_n = t_1$ of $[t_0,t_1]$ so that all the restrictions $\varphi|_{[\tau_j,\tau_{j+1}]}$ for $j=0,\dots,n-1$ are smooth.  When $\varphi$ is a path, we write its canonical lift to $\T\Nn$ as $(\dot \varphi, \varphi)$; this lift is not continuous, but rather has discontinuities like Heaviside's step function $\Theta(x)$, which we interpret, along with its derivatives, in the sense of distributions.  A \define{Lagrangian} $L: \RR \times \T\Nn \to \RR$ associates to each path $\varphi: [t_0,t_1] \to \Nn$ an \define{action}, given by:
\begin{equation} \label{actionfirst}
  \Aa(\varphi) = \int_{\tau = t_0}^{t_1} L\bigl(\tau,\dot \varphi(\tau), \varphi(\tau)\bigr)\,\d\tau
\end{equation}

Let $t_0 < t_1$ be real numbers and $q_0,q_1 \in \Nn$.  Temporarily, consider $\hbar$ as a non-zero real number (for most of this paper, $\hbar$ is a formal parameter).  Feynman's proposal for the path integral is to consider the following (ill-defined) infinite-dimensional integral:
\begin{equation} \label{FeynmanIntegral}
  U(t_0,q_0,t_1,q_1) = \int_{\substack{\text{paths } \varphi:[t_0,t_1]\to \Nn \text{ with} \\ \varphi(t_0) = q_0\text{ and } \varphi(t_1) = q_1}}\limits \exp\left( \frac i \hbar \Aa(\varphi)\right)\,\d \varphi
\end{equation}
The ``measure'' $\d\varphi$ is supposed to be given by the infinite product $\d \varphi = \prod_{t_0 < \tau < t_1} \d \varphi(\tau)$, where $\d\varphi(\tau) = \dVol$ is a copy of the volume form on $\Nn$.  Feynman asserts that if such an integral can be defined, then $U$ will be a fundamental solution to the Schr\"odinger equation corresponding to the Lagrangian $L$: i.e.\ $U$ will be the kernel of the ``time evolution'' operator, a unitary operator on the Hilbert space ${\rm L}^2(\Nn, \dVol)$.  His justifications in \cite{Feynman1948,FeynmanHibbs1965} hold only to a ``physical'' level of rigor, and break down when the Lagrangian is not inhomogeneous quadratic along fibers with flat quadratic part, and even in this special case the arguments break down when the Lagrangian grows too quickly in the position coordinates. The problem is well-illustrated by a particle moving in one dimension under a quartic potential.  The reader is invited to check for herself that in this situation, for any $t>0$ there are infinitely many classical trajectories of duration $t$ connecting a chosen pair of points, most of which involve the particle flying very far away very quickly.  These ``high energy'' classical solutions invalidate Feynman's estimations.

When $\hbar \to 0$, a comparison with finite-dimensional oscillating integrals suggests that the integral in \cref{FeynmanIntegral} should be supported near critical points of the action function $\Aa$.  These critical points are precisely the classical trajectories for the corresponding physical system.  In this paper, we treat $\hbar$ as a formal variable, and take the suggestion as a definition.  Let $\gamma: [t_0,t_1] \to \Nn$ be a classical trajectory with $\gamma(t_a) = q_a$ for $a=0,1$.  Our formal path integral $U_\gamma$ will correspond physically to an integral as in \cref{FeynmanIntegral}, where the domain of integration includes only those paths that are ``infinitely close'' to $\gamma$.  The size of ``infinitely close'' is controlled by $\hbar$.

Let $A$ be a smooth real-valued function on a finite-dimensional manifold with a unique critical point, which is nondegenerate in the Morse-theoretic sense.  Then the asymptotic expansion as $\hbar\to 0$ of the oscillating integral $\int \exp\bigl(-(i\hbar)^{-1}A\bigr)$ is well-understood.  In particular, the coefficients depend only on the $\infty$-jet of $A$ at the critical point.  After choosing a coordinate systems, the coefficients are straightforward to compute: each coefficient is described succinctly by a finite sum of ``Feynman diagrams,'' which were generalized by Penrose \cite{Penrose1971} to describe tensor contractions in arbitrary vector spaces. (A famous aesthetic split developed in the theoretical physics community over the interpretation of Feynman diagrams \cite{Kaiser2005}.  Feynman thought of his diagrams as pictures of fundamental interactions of basic particles.  On the other hand, Dyson \cite{Dyson1949a,Dyson1949b}, who deserves most of the credit for codifying and popularizing the use of Feynman diagrams in quantum electrodynamics, believed that the diagrams were devoid of physical meaning, representing only a useful nemonic for the complicated integrals in Schwinger's field theory.  We are firmly in Dyson's camp: the diagrams provide a powerful notation, which we use throughout this paper, but do not represent particle interactions.)  We review this material in \cref{finitedim}.

In the rest of \cref{coordfulldefn}, we translate this asymptotic expansion to the infinite-dimensional integral of \cref{FeynmanIntegral} and compute the necessary components; we then take the translated expansion as the definition of the formal path integral.  The translation applies only when the classical trajectory (corresponding to the critical point in the finite-dimensional case) is nondegenerate for the Morse theory given by $\Aa$.  We recall the proof that any classical trajectory $\gamma$ is nondegenerate if and only if it is \define{nonfocal}: it can be extended to a family of nondegenerate classical paths that vary smoothly with the boundary conditions $\gamma(t_0) = q_0$, $\gamma(t_1) = q_1$.  The coefficients are now given by finite sums of finite-dimensional (possibly divergent) integrals, and we can interpret the formal path integral $U_\gamma$ as a function $U_\gamma(t_0,q_0,t_1,q_1)$, and each coefficient is smooth (if it converges).  The Feynman diagrams provide a convenient notation for computations, hiding the details of the finite-dimensional integrals for which they code and highlighting the general structure.  We will address these divergences in \cref{divsection}.  The best situation is when for each order in $\hbar$, the divergences contributing to that order cancel.  When this happens, we say that the path integral \define{has no ultraviolet divergences}.

{\em A priori}, the definition of the formal path integral depends on a choice of coordinates on $\Nn$, and only applies to classical paths that are contained entirely within the same coordinate patch.  Let $\Oo$ be an open subset of $\Nn$ and $q : \Oo \to \RR^{\dim \Nn}$ a system of coordinates on $\Oo$.  We say that the coordinates $q^i$ are \define{compatible with the volume} if $\dVol = \d q^1\cdots \d q^{\dim \Nn}$.  By a theorem of Moser \cite{Moser1965}, compatible coordinate systems exists for any volume form.  Let $\gamma$ be a nonfocal classical path in $\Oo$ and $q,\tilde q: \Oo \to \RR^{\dim \Nn}$ two coordinate systems that are both compatible with the volume form $\dVol$, and suppose that the image $q(\Oo)$ is \define{star-shaped}: for each $x\in q(\Oo)$ and $s\in [0,1]$, $sx \in q(\Oo)$.  In \cref{coordfree}, we will prove that the formal path integrals $U_\gamma$ and $\tilde U_\gamma$, computed in the coordinate systems $q,\tilde q$, agree.  Thus, we have \cref{coordfreethm}: for a classical nonfocal path that can be contained within some star-shaped coordinate patch, the corresponding path integral depends only on the choice of volume form.

In a slightly different setup, the problem of whether the path integral is invariant under changes of coordinates has also been addressed by Kleinert and Chervyakov \cite{KC1,KC2,KC3,KC4,KC5,KC6}.  Under changes of coordinates, the dependence of the path integral shows up in the form of divergent quantities.  Motivated by higher-dimensional field theories, Kleinert and Chervyakov adopt a ``dimensional renormalization'' scheme from the beginning to handle these divergences.  They consider a more restricted case of examples than we do: they consider only situations of the form ``kinetic energy minus potential energy.''  Moreover, their methods are equivalent to forcing the potential energy to be infinitesimal --- indeed, any approach that begins by setting $\hbar = 1$ must then find some other parameter to use in the perturbation series, and the standard choice is the strength of the potential energy function.

We handle general classical trajectories by a ``cut-and-paste'' method.  In \cref{fubini}, we prove the following ``Fubini'' theorem for formal path integrals, provided that there are no ultraviolet divergences (or rather, provided that the integral represented by the Feynman diagram \begin{tikzpicture}[baseline=(X)]
    \path node[dot] (O) {} ++(0pt,1pt) coordinate (X);
    \draw (O) .. controls +(7pt,7pt) and +(0pt,10pt) .. (O);
    \draw (O) -- ++(-4pt,7pt);
\end{tikzpicture}\!\! converges, as then our {\em ad hoc} choice in \cref{pathintdefn} for the value of a certain ``determinant'' satisfies the correct differential equation).
Let $q: \Oo\to \RR^{\dim \Nn}$ be a coordinate patch that is compatible with the volume form, and let $\gamma: [t_0,t_1] \to \Oo$ be classical and nonfocal.  If $t \in (t_0,t_1)$, then the restrictions $\gamma_0 = \gamma|_{[t_0,t]}$ and $\gamma_1 = \gamma|_{[t,t_1]}$ are classical.  Supposing that $\gamma_0,\gamma_1$ are nonfocal, we can define three path integrals $U_\gamma(t_0,q_0,t_1,q_1)$, $U_{\gamma_0}(t_0,q_0,t,q)$, and $U_{\gamma_1}(t,q,t_1,q_1)$.  We will prove that the product $U_{\gamma_0}(t_0,q_0,t,q)\,U_{\gamma_1}(t,q,t_1,q_1)$, thought of as a function of $q$, has a nondegenerate critical point at $q = \gamma(t)$, and moreover that:
\begin{equation}\label{fubinieqn}
  U_\gamma(t_0,q_0,t_1,q_1) = \int^{\formal} U_{\gamma_0}(t_0,q_0,t,q)\,U_{\gamma_1}(t,q,t_1,q_1)\,\d q
\end{equation}
The integral ranges over a small neighborhood of $\gamma(t)$ and is interpreted formally in the sense of \cref{formalintdefn}.  \Cref{fubinieqn} is a ``composition law'' for the formal path integral, and constitutes our \cref{fubinithm}.  By a standard argument (e.g.\ \cite{Milnor1963}), there are only finitely many $t \in (t_0,t_1)$ for which the restrictions $\gamma_0,\gamma_1$ fail to be nondegenerate, provided the Lagrangian $L$ is convex along the fibers of $\RR\times \T\Nn \to \RR \times \Nn$.  Thus, to define the path integral for a general nondegenerate classical path, we can cut it into small pieces, each contained within some coordinate patch, compute each path integral, and integrate the answer.  \Cref{coordfreethm,fubinithm} guarantee that the resulting power series is independent of the choice of cuts and coordinates.

The results of \crefrange{coordfulldefn}{fubini} largely ignore the issue of ultraviolet divergences.  We begin \cref{divsection} with a seemingly natural example to illustrate that these divergences really are an issue.  In fact, the ultraviolet divergences arise because of the dependence of the formal path integral on the choice of volume form.  When the Lagrangian is (inhomogeneous) quadratic along the tangent fibers, its homogeneous-quadratic part is a Riemannian metric, and thus determines a canonical choice of measure.  If we use this measure to define the formal path integral, there are no ultraviolet divergences, a fact we prove in \cref{divfreethm}.  In particular, \cref{coordfreethm,fubinithm} hold for these Lagrangians.  As any such Lagrangian is of the form ``kinetic plus magnetic plus potential,'' where the ``kinetic'' term is determined by a Riemannian metric, these systems describe the motion of ``nonrelativistic charged particles in external electromagnetic fields.''  All together, we have:
\begin{mainthm}
  Let $(\Nn,a)$ be a Riemannian manifold with a chosen differential one-form $b$ and a chosen function $c$, and let $L(v,q) = \frac12 a(q)\cdot v^2 + b(q)\cdot v + c(q)$ be the corresponding Lagrangian on $\Nn$ and $\Aa(\gamma) = \int L(\dot\gamma,\gamma)$ its action.  Then for each classical nonfocal path $\gamma:[t_0,t_1] \to \Nn$, Feynman's heuristic diagrammatic expansion yields a well-defined formal power series $U_\gamma$ in $\hbar$, representing the ``near $\gamma$'' contribution to the path integral $\int \exp\bigl( \frac i \hbar \Aa(\gamma)\bigr)\d\gamma$ with the measure $\d\gamma = \prod_{t_0 < \tau < t_1} \sqrt{\det a(\gamma(\tau))}$.  This ``formal path integral'' is independent of any further choices, and satisfies the composition law of \cref{fubinieqn}.
\end{mainthm}

Finally, our results from \cite{meshort} hold in the generality of this paper.  For a given pair of points $q_0,q_1$ and given times $t_0<t_1$, there can by infinitely many nondegenerate classical paths $\gamma$ with $\gamma(t_a) = q_a$.  But the arguments in \cite{meshort} show that the sum $\sum_\gamma U_\gamma$ converges in the sense of distributions.  Moreover, if we are in the quadratic case of the Main Theorem, then for each path $\gamma$, the methods of \cite{meshort} show that $U_\gamma$ satisfies the corresponding Schr\"odinger equation, and as $(t_1-t_0)\to 0$, the sum $\sum_\gamma U_\gamma(t_0,q_0,t_1,q_1)$ approaches Dirac's $\delta$ distribution.

The results in this paper apply {\em a priori} only when the Lagrangian $L : \RR \times \T\Nn \to \RR$ is convex along the fibers of the projection $\RR\times\T\Nn \to \RR\times\Nn$; in fact, we insist that the matrix $\frac{\partial^2 L}{\partial v^2}(t,v,q)$ be everywhere positive definite.  Although we have tried to use this condition sparingly, let us briefly highlight the places where it seems necessary.  A sign ambiguity in the definition of the formal path integral can presumably be fixed by using Maslov's index rather than Morse's; alternately, one can simply understand the formal path integral as double-valued.  Also, it is no longer automatically the case that the Legendre transform determines an injection $\RR \times \T\Nn \to \RR \times \T^*\Nn$, preventing \cref{classicalglue} from holding as stated; however, for \cref{fubinithm} we need only part of \cref{classicalglue}, and this part requires only that the matrix $\frac{\partial^2 L}{\partial v^2}$ be everywhere invertible.  Finally, in the indefinite case, extra assumptions on the classical mechanics determined by $L$ are also necessary in order to get from \cref{fubinithm} to a definition of the formal path integral on an arbitrary manifold: there might not be enough medium-length nonfocal paths for our argument to go through \cite{Helfer1994}.  In all cases, the convexity condition on $L$ simply assures that the classical mechanics is suitably well-behaved, and we believe that results similar to ours apply even in well-behaved indefinite examples.

We leave open the following three questions, each of which deserves its own paper:
\begin{enumerate}
  \item When there are ultraviolet divergences in the formal path integral, what do they measure?
  \item When the Lagrangian is not inhomogeneous quadratic, is there a choice of measure in which the formal path integral converges?  Does this require a ``measure'' on path-space that is more general than ``$\d\gamma = \prod_{t_0 < \tau < t_1}\d\gamma(\tau)$,'' say by depending explicitly on the velocities of paths?
  \item When the Lagrangian is not inhomogeneous quadratic, does the diagrammatic formal path integral still yield a solution to Schr\"odinger's equation?
\end{enumerate}

\subsection{Acknowledgments}

It is with great pleasure that I thank my adviser, Nicolai Reshetikhin, who suggested this project and has provided support, advice, and ideas throughout.  While working on this project, I frequently described parts of it on the website \url{mathoverflow.net} and received valuable feedback; in particular, Greg Kuperberg and Chris Schommer-Pries each provided important suggestions online.  I would also like to thank Kiril Datchev, George Thompson, and Ivan Ventura for discussions.  Finally, I would like to thank the anonymous referee for the very useful feedback.  This work was supported by the NSF grant DMS-0901431.  I am grateful to Aarhus University for their hospitality.

\Section{The coordinate-full definition of the formal path integral}  \label{coordfulldefn}

In this section, we define the formal path integral for Lagrangians on open subsets of $\RR^d$, where the volume form is taken to be the usual one.
Although this definition is essentially well-known, it seems not to be carefully written down in full generality.
Indeed, this section ought to be given as an exercise in an advanced quantum mechanics textbook.  The closest we can find is Exercise 10.11 from \cite{Hori2003}, in which only Lagrangians on $\RR^d$ of the form ``kinetic energy minus potential energy'' are considered: ``Formulate Feynman diagram perturbation theory for quantum mechanics by following steps similar to those for the zero-dimensional QFT.''  The authors do not suggest any of our \cref{coordfreethm,fubinithm,divfreethm} nor the results from \cite{meshort}, and we suspect they did not try to prove these theorems themselves.

We begin in \cref{finitedim} by recalling the description of the asymptotics of finite-dimensional oscillating integrals (compare \cite{EZ2007,Polyak2005,KolyaLong}): as $\hbar\to 0$, an integral of the form $\int_{\RR^N} \exp\bigl( \frac i \hbar A(x)\bigr)\d x$ is supported near the critical points $c$ of $A$, and depends only on the Taylor expansion of $A$ at $c$, and the coefficients of the asymptotic integral can be succinctly described in terms of sums of ``Feynman diagrams''.  In the remainder of the section we translate the asymptotic expansion to the infinite-dimensional case $\int_{\text{paths}} \exp\bigl( \frac i \hbar \Aa(\gamma)\bigr)\d\gamma$, where $\Aa$ is the action from \cref{actionfirst}.
In \cref{actiondersub} we describe the derivatives of the action $\Aa$ and recall the following well-known fact: the first derivative $\Aa^{(1)}$ can be understood as a nonlinear second-order differential operator on paths in $\RR^d$; setting $\Aa^{(1)}(\gamma) = 0$ gives the Euler-Lagrange equations for the Lagrangian $L$, and the solutions are the classical trajectories.

Only a few difficulties present themselves in making this translation.  In the finite-dimensional case, the usual asymptotic expansion of the integral $\int \exp\bigl( \frac i \hbar A(x)\bigr)\d x$ requires the inverse to the Hessian $A^{(2)}(c)$ at each critical point $c$; thus $A$ is required to be a Morse function.  In the infinite-dimensional case, the notion of Hessian is still well-defined, but a nonsingular operator need not have an inverse.  In \cref{GreenFnSub} we will interpret the Hessian $\Aa^{(2)}(\gamma)$ of the action $\Aa$ at a classical path $\gamma$ as a second-order linear differential operator $\Dd_\gamma$ on paths in $\RR^d$.  We call a classical path $\gamma: [t_0,t_1] \to \RR^d$ \define{nondegenerate} if $\Dd_\gamma$ has no zero modes $\xi$ with $\xi(t_0) = 0 = \xi(t_1)$.  We will prove that $\gamma$ is nondegenerate if and only if it is \define{nonfocal} --- extends to a family of paths --- and that in this case $\Dd_\gamma$ has a \define{Green's function}, i.e.\ a function $G_\gamma: [t_0,t_1]^{\times 2} \to (\RR^d)^{\otimes 2}$ that vanishes along the boundary of the square and so that $\Dd_\gamma[G_\gamma]$ is the product of the Kronecker-$\delta$ element of $(\RR^d)^{\otimes 2}$ with the diagonal Dirac-$\delta$ distribution on the square $[t_0,t_1]^{\times 2}$.  Simultaneously, we will explicitly compute the Green's function in terms of the classical trajectory $\gamma$.

The finite-dimensional integral requires computing $\sqrt{\lvert\det A^{(2)}(c)\rvert}$ and, when $A^{(2)}(c)$ is not positive-definite, picking the correct sign.  In the infinite-dimensional case, one common approach is to use $\zeta$-function regularization to compute the analogous quantities \cite{Elizalde1994}, but we will not do so here.  Rather, in \cref{defUsection} we will declare a value for ``$\sqrt{\lvert\det \Aa^{(2)}(c)\rvert}$'' more-or-less {\em ad hoc}.  When the Lagrangian $L$ is convex along fibers of $\T\RR^d$, the usual arguments (c.f.\ \cite{Milnor1963}) guarantee that each nondegenerate classical trajectory has a well-defined Morse index, which we use to pick the sign.  The justification for these choices comes in \cref{fubinithm}.  We sum up all the results of the section in \cref{pathintdefn}.

Henceforth, we adopt Einstein's summation convention: $p_iq^i = q^ip_i = \sum_{i=1}^d q^ip_i$ (or $\sum_{i=1}^N$ in \cref{finitedim}).

\subsection{Finite-dimensional oscillating integrals} \label{finitedim}

Let $A: \RR^N \to \RR$ be a smooth function with finitely many critical points, and consider the integral $\int_{\RR^N} \exp(\frac i \hbar A(x))\,\d x$.  Under mild assumptions on the growth rate of the gradient $A^{(1)}$, the integral converges conditionally for non-zero $\hbar$.  As $\hbar$ goes to $0$, for most values of $x$ the integral oscillates rapidly, so only small neighborhoods of the critical points will contribute to the asymptotics of the integral.  Thus to compute the asymptotics of the integral, we take $\hbar$ to be a formal variable, and expand $A$ in Taylor series around each critical point $c$ of $A$:
\begin{align}
  \int_{\RR^N} \exp\left(\frac i \hbar A(x)\right)\d x & \approx \sum_{\textrm{critical points } c} \int_{\text{small nbhd of }c}  \exp\left(\frac i \hbar A(x)\right)\d x \\
  & \approx    \sum_{\textrm{critical points } c} \int_{\T_c\RR^N} \exp\left( \frac i \hbar \sum_{n=0}^\infty \frac1{n!}\left.A^{(n)}_{i_1\cdots i_n}\right|_c x^{i_1}\cdots x^{i_n} \right) \d x \label{line1}
\end{align}
In \cref{line1}, we have switched coordinates $x \mapsto c+x$.
We write the degree-three and higher terms in the exponential in terms of Taylor series:
\begin{align}
  \int & \approx \sum_c e^{\frac i \hbar A(c)} \int_{\RR^N} \sum_{m=0}^\infty \frac1{m!} \left(  \frac i \hbar \sum_{n=3}^\infty \frac1{n!}\left.A^{(n)}_{i_1\cdots i_n}\right|_c x^{i_1}\cdots x^{i_n} \right)^m \exp\left( \frac i \hbar \frac12 \left. A^{(2)}_{ij}\right|_c x^ix^j\right)\d x \label{line2}
\end{align}

We now recall the following elementary fact, provable by, for example, diagonalization and integration by parts.  Let $a_{ij}$ be a positive-definite symmetric bilinear form on $\RR^N$.  Then it is invertible, and for any symmetric tensor $b_{i_1\cdots i_n}$:
\begin{multline}
  \int_{\RR^N} \frac1{n!} b_{i_1\cdots i_n}x^{i_1}\cdots x^{i_n} \exp \left( -\frac12 a_{ij}x^ix^j\right)\d x = \\ = \begin{cases}
    0, & n\textrm{ odd} \\
    \displaystyle \sqrt{\det(2\pi a^{-1})}\frac{1}{2^kk!}  \,b_{i_1\cdots i_n} (a^{-1})^{i_1i_2}\cdots (a^{-1})^{i_{n-1}i_n},  & n = 2k
  \end{cases} \label{gaussianintegral}
\end{multline}

A \define{pairing} of $\{1,\dots,n\}$ is a partition of the set into $n/2$ blocks of size $2$; if $n$ is odd there are no pairings, and if $n=2k$ is even there are $n!/(2^kk!)$ pairings.  We can sort first the elements of each block and then the blocks by first entry; then a pairing is a bijection $P: \{1,\dots,n\} \to \{1,\dots,n\}$ such that $P(2j-1) < P(2j)$ if $j \leq n/2$ and $P$ is increasing on $\{1,3,\dots,n-1\}$.   We can generalize \cref{gaussianintegral} to the case of an arbitrary tensor $b_{i_1\dots i_n}$:
\begin{multline}  \label{firstpairingintegral}
  \int_{\RR^N}  b_{i_1\cdots i_n}x^{i_1}\cdots x^{i_n} \exp \left( -\frac12 a_{ij}x^ix^j\right)\d x = \\ = \sqrt{\det(2\pi a^{-1})} \sum_{\textrm{pairings }P} b_{i_1\cdots i_n} (a^{-1})^{i_{P(1)}i_{P(2)}}\cdots(a^{-1})^{i_{P(n-1)}i_{P(n)}}
\end{multline}

A better notation is given by \define{Feynman diagrams}, which we define by example.   We draw the tensor $b :(\RR^N)^{\otimes n} \to \RR$ as a vertex with $n$ upward-pointing edges, and the bivector $a^{-1} \in(\RR^N)^{\otimes 2}$ as an edge:
\begin{align*}
  b_{i_1\cdots i_n} & =
  \begin{tikzpicture}[baseline=(X)]
    \path node[dot] (O) {} ++(0pt,4pt) coordinate (X);
    \draw (O) -- ++(-15pt,15pt) +(0,3pt) node[anchor=base] {$\scriptstyle i_1$};
    \draw (O) -- ++(-6pt,15pt) +(0,3pt) node[anchor=base] {$\scriptstyle i_2$};
    \draw (O) -- ++(15pt,15pt) +(0,3pt) node[anchor=base] {$\scriptstyle i_n$};
    \path (O) ++(4pt,13pt) node {$\scriptstyle \ldots$};
  \end{tikzpicture} &
  (a^{-1})^{ij} & =
  \begin{tikzpicture}[baseline=(X)]
    \path coordinate (A) ++(0pt,4pt) coordinate (X);
    \path (A) +(0,-8pt) node[anchor=base] {$\scriptstyle i$};
    \path (A) ++(20pt,0) coordinate (B) +(0,-8pt) node[anchor=base] {$\scriptstyle j$};
    \draw (A) .. controls +(0,15pt) and +(0,15pt) .. (B);
  \end{tikzpicture} &
  \text{a pairing:} & 
  \begin{tikzpicture}[baseline=(X)]
    \path node[dot] (O) {} ++(0pt,4pt) coordinate (X);
    \path (O) +(-15pt,-5pt) coordinate (a) +(11pt,20pt) coordinate (b);
    \clip (a) rectangle (b);
    \draw (O) .. controls +(-30pt,20pt) and +(30pt,20pt) .. (O);
    \draw (O) .. controls +(-10pt,15pt) and +(10pt,15pt) .. (O);
  \end{tikzpicture} = b_{i_ii_2i_3i_4}(a^{-1})^{i_1i_4}(a^{-1})^{i_2i_3}
\end{align*}
The vertical connections correspond to tensor contractions.

We observe that in \cref{gaussianintegral}, the $n!$ and $2^kk!$ terms count the number of symmetries of the corresponding diagrams.  More generally, each summand in the expanded-out sum from \cref{line2} ---
\begin{multline}
\sum_{m=0}^\infty \frac1{m!} \left(  \frac i \hbar \sum_{n=3}^\infty \frac1{n!}\left.A^{(n)}_{i_1\cdots i_n}\right|_c x^{i_1}\cdots x^{i_n} \right)^m = 1  + \frac i \hbar \left( \frac1{3!} \left.A^{(3)}_{ijk}\right|_c x^ix^jx^k + \frac1{4!} \left.A^{(4)}_{ijkl}\right|_c x^ix^jx^kx^l +  \ldots \right)  \\  + \left(\frac i \hbar\right)^2 \left( \frac1{2!\,3!^2} \left(\left.A^{(3)}_{ijk}\right|_c x^ix^jx^k\right)^2 + \ldots \right) + \ldots
\end{multline}
--- corresponds to some collection of vertices (the $1$ corresponds to the empty collection), the power on $i\hbar$ counts the number of vertices, and the factorial prefactor counts the number of symmetries of each collection.  This suggests the following definition:

\begin{defn} \label{formalintdefn}
  Let $A:\RR^N\to\RR$ be a smooth function and $c\in \RR^N$ a critical point of $A$ such that $A^{(2)}_{ij}(c)$ is nondegenerate.  Let $\eta(c)$ be the number of negative eigenvalues of $A^{(2)}(c)$.  The \define{formal integral of $\exp\bigl( -(i\hbar)^{-1}A(x)\bigr)$ near $c$} is:
  \begin{multline}
    \int_c^{\formal} \exp\bigl( -(i\hbar)^{-1}A(x)\bigr)\, \d x = \\ = (2\pi i \hbar)^{N/2} e^{-(i\hbar)^{-1}A(c)} (-i)^{\eta(c)} \left|\det A^{(2)}(c)\right|^{-1/2} \sum_{\Gamma} \frac{(i\hbar)^{-\chi(\Gamma)} \ev(\Gamma)}{\left|\Aut \Gamma\right|} \label{FeynmanInt0}
  \end{multline}
  The sum ranges over combinatorial graphs $\Gamma$ with every vertex of degree three or more, $\chi(\Gamma) = |V_\Gamma| - |E_\Gamma|$ is the Euler characteristic of $\Gamma$, and $|\Aut \Gamma|$ is the number of symmetries of $\Gamma$.  We evaluate $\ev(\Gamma)$ via the following \define{Feynman rules}, and contract indices along each vertical edge, so that each diagram is a picture of a tensor contraction:
  \begin{align} \label{FeynmanInt1}
   \ev\biggl(  
    \begin{tikzpicture}[baseline=(X)]
      \path node[dot] (O) {} ++(0pt,4pt) coordinate (X);
      \draw (O) -- ++(-15pt,15pt) +(0,3pt) node[anchor=base] {$\scriptstyle i_1$};
      \draw (O) -- ++(-6pt,15pt) +(0,3pt) node[anchor=base] {$\scriptstyle i_2$};
      \draw (O) -- ++(15pt,15pt) +(0,3pt) node[anchor=base] {$\scriptstyle i_n$};
      \path (O) ++(4pt,13pt) node {$\scriptstyle \ldots$};
    \end{tikzpicture} 
    \biggr) &
    = -A^{(n)}_{i_1\cdots i_n}(c),\;\; n\geq 3 &
   \ev\biggl(  
    \begin{tikzpicture}[baseline=(X)]
      \path coordinate (A) ++(0pt,4pt) coordinate (X);
      \path (A) +(0,-8pt) node[anchor=base] {$\scriptstyle i$};
      \path (A) ++(20pt,0) coordinate (B) +(0,-8pt) node[anchor=base] {$\scriptstyle j$};
      \draw (A) .. controls +(0,15pt) and +(0,15pt) .. (B);
    \end{tikzpicture} 
    \biggr) &
    = \left( \left( A^{(2)}(c) \right)^{-1}\right)^{ij}
  \end{align}
  
  If $B: \RR^N \to \RR$ is another smooth function, the formal integral $\int_c^{\formal} (i\hbar)^{-1}B(x)\, \exp\bigl( -(i\hbar)^{-1}A(x)\bigr)\, dx$ is given by the right-hand side of \cref{FeynmanInt0} with one modification: the sum now ranges over diagrams with a unique marked vertex of arbitrary valence, with an added Feynman rule for the marked vertex:
  \begin{align}
   \begin{tikzpicture}[baseline=(X)]
    \path node[shape=circle,inner sep=1pt] (O) {$\star$} ++(0pt,4pt) coordinate (X);
    \draw (O) -- ++(-15pt,15pt) +(0,3pt) node[anchor=base] {$\scriptstyle i_1$};
    \draw (O) -- ++(-6pt,15pt) +(0,3pt) node[anchor=base] {$\scriptstyle i_2$};
    \draw (O) -- ++(15pt,15pt) +(0,3pt) node[anchor=base] {$\scriptstyle i_n$};
    \path (O) ++(4pt,13pt) node {$\scriptstyle \ldots$};
  \end{tikzpicture} &  
  = \left.B^{(n)}_{i_1\cdots i_n}\right|_c,\;\; n \text{ arbitrary}
\end{align}
\end{defn}

We have written $\frac i\hbar A(x) = -(i\hbar)^{-1}A(x)$ to make the integrand look more like \cref{firstpairingintegral}.  One can derive the $(-1)^\eta$ term by diagonalizing and considering the behavior of the integral under $A\mapsto -A$; it picks out a branch of the square root $\sqrt{\det A^{(2)}}$. When $B(c) > 0$, one can incorporate $\log B(x)$ as an $O(\hbar)$ correction to $A(x)$, and then the two definitions agree as power series in $\hbar$.  Indeed, if $A(x) = A_0(x) + O(\hbar)$ is a formal power series, then the critical points of $A$ differ from critical points of $A_0$ by terms of order $\hbar$, as do the corresponding derivatives.

Since at most one vertex (the marked one) has valence less than three, the Euler characteristic of any diagram in $\int^{\formal} (i\hbar)^{-1}\,B\exp\bigl(-(i\hbar)^{-1}A\bigr)$ is at most $1$, and there are finitely many diagrams at each characteristic.  The precise statement that \cref{FeynmanInt0} gives an asymptotic expansion of an oscillating integral is:
\begin{lemma} \label{asymptoticslemma}
  Let $A,B:\RR^N\to\RR$ be smooth functions and $\Cc \subseteq \RR^N$ a compact neighborhood containing precisely one critical point $c$ of $A$, and suppose that the second derivative $A^{(2)}(c)$ is nondegenerate. 
  Let $I$ be the value of the Riemann integral $\int_\Cc (i\hbar)^{-1}B(x)\, \exp\bigl( -(i\hbar)^{-1}A(x)\bigr)\, \d x$, and let $I_M$ be the value of the formal integral $\int_c^{\formal} (i\hbar)^{-1}B(x)\, \exp\bigl( -(i\hbar)^{-1}A(x)\bigr)\, \d x$ with the sum truncated to include only those diagrams $\Gamma$ with $-\chi(\Gamma) \leq M$.  Then $I_M / I = 1 + o(\hbar^M)$.
\end{lemma}
For details, see \cite{EZ2007}.

The reader is invited to check directly, without appealing to \cref{asymptoticslemma}, that the right-hand-side of \cref{FeynmanInt0} does not change under a volume-preserving change-of-coordinates on $\RR^N$.  For comparison, see \cref{coordfreethm}.

\subsection{The classical action and its derivatives}  \label{actiondersub}

In this section we recall the notion of functional derivative and use it to write down the derivatives of the action.

We work on the configuration space $\RR^d$ with tangent bundle $\T\RR^d = \RR^{2d}$;  the standard coordinates on $\T\RR^d$ are $(v^i,q^i)$ for $i=1,\dots,d$.  A  \define{(piecewise-smooth parameterized) path in $\RR^d$} is a continuous map $\gamma: [t_0,t_1] \to \RR^d$ such that there exists a finite division $t_0 = \tau_0 < \tau_1 < \dots < \tau_n=t_1$ with $\gamma|_{[\tau_j,\tau_{j+1}]}$ smooth for each $j=0,\dots,n-1$.  
We think of the space of paths as an infinite-dimensional manifold.  For fixed $t_0<t_1$, the space of paths $\gamma: [t_0,t_1] \to \RR^d$ is an infinite-dimensional vector space, and for fixed $q_0,q_1 \in \RR^d$, the space of paths $\gamma$ with $\gamma(t_a) = q_a$, $a =0,1$ is an affine subspace thereof; this affine subspace is modeled on the vector space $\{\gamma:[t_0,t_1] \to \RR^d \st \gamma(t_0) = 0 = \gamma(t_1) \}$ of \define{piecewise-smooth loops based at $0$}.  Thus we can identify this loop space with the tangent space at any path to the subspace of paths with the same boundary conditions. We abbreviate ``piecewise-smooth loop based at $0$'' by \define{based loop}.

A \define{Lagrangian function} is any smooth function $L: \RR\times\T\RR^d \to \RR$.  For a chosen Lagrangian function $L$, the corresponding  \define{action} $\Aa$ assigns to each path $\gamma: [t_0,t_1] \to \RR^d$ the number $\Aa(\gamma) = \int_{t_0}^{t_1} L(\tau,\dot\gamma(\tau),\gamma(\tau))\d\tau$.

Let $L: \RR\times\T\RR^d \to \RR$ be a Lagrangian on $\RR^d$ and $\Aa: \gamma \mapsto \int_{t_0}^{t_1} L\bigl(\tau,\dot\gamma(\tau),\gamma(\tau)\bigr)\d\tau$ the corresponding action.  If $\xi: [t_0,t_1] \to \RR^d$ is another path, then by the chain rule:
\begin{align} \Aa(\gamma + \xi) = \Aa(\gamma) +  \int_{t_0}^{t_1} \left( \left.\frac{\partial L}{\partial v^i}\right|_{\gamma(\tau)} \dot\xi^i(\tau) + \left. \frac{\partial L}{\partial q^i}\right|_{\gamma(\tau)} \xi^i(\tau) \right)\d\tau + o(\xi) \end{align}
where $o(\xi)$ is a quantity that vanishes faster than linearly under rescaling $\xi^i \mapsto \epsilon\xi^i$.  Thus we define the \define{functional derivative} $\delta\Aa/\delta\gamma = \Aa^{(1)}(\gamma)$ to be the following linear operator:
\begin{align}
  \Aa^{(1)}(\gamma) \cdot \xi = \Aa^{(1)}_i(\gamma) \cdot \xi^i = \int_{t_0}^{t_1} \left( \dot\xi^i(\tau) \left.\frac{\partial L}{\partial v^i}\right|_{\gamma(\tau)}  +  \xi^i(\tau) \left. \frac{\partial L}{\partial q^i}\right|_{\gamma(\tau)} \right)\d\tau
\end{align}
Differentiating repeatedly gives:
\begin{align}
  \Aa^{(n)}_{i_1\dots i_n}(\gamma) \cdot \xi_1^{i_1}\dots \xi_n^{i_n} & = \int_{t_0}^{t_1} \prod_{k=1}^n \left. \left( \dot\xi_k^{i_k}(\tau) \frac{\partial}{\partial v^{i_k}}  +  \xi_k^{i_k}(\tau) \frac{\partial}{\partial q^{i_k}} \right) L \right|_{\gamma(\tau)} \d\tau \label{Anformula}
\end{align}
These will correspond to the vertices in \cref{FeynmanInt1}.
It should be understood that the partial derivatives act only on $L$ and commute with the $\xi$s.  For example:
\begin{align} \label{A2}  \Aa^{(2)}_{ij} \cdot \xi^i\zeta^j  = \int_{t_0}^{t_1} \left( \left.\frac{\partial^2 L}{\partial v^i\partial v^j}\right|_{\gamma} \dot\xi^i\dot\zeta^j + \left. \frac{\partial^2 L}{\partial q^i\partial v^j}\right|_{\gamma} \xi^i\dot\zeta^j +  \left.\frac{\partial^2 L}{\partial v^i\partial q^j}\right|_{\gamma} \dot\xi^i\zeta^j + \left. \frac{\partial^2 L}{\partial q^i\partial q^j}\right|_{\gamma} \xi^i\zeta^j \right)\d\tau \end{align}

\Cref{formalintdefn} suggests that \cref{FeynmanIntegral} should be supported near only those paths $\gamma$ for which, for any based loop $\xi$, $\Aa^{(1)}(\gamma) \cdot \xi$ vanishes.  Such a path is \define{classical}; as is well-known, such paths are precisely the classically-allowed trajectories for the mechanical system with Lagrangian $L$.
By integrating by parts, a path $\gamma$ is classical if and only if it satisfies the \define{Euler-Lagrange equations}:
  \begin{align}
    \label{EL} \left.\frac{\partial L}{\partial q^i}\right|_{(\tau,\dot\gamma(\tau),\gamma(\tau))} = \frac{\d}{\d\tau} \left[ \left. \frac{\partial L}{\partial v^i} \right|_{(\tau,\dot\gamma(\tau),\gamma(\tau))} \right]
  \end{align}
We will always assume that \cref{EL} is a nondegenerate second-order differential equation; equivalently, we assume that the symmetric matrix $\frac{\partial^2 L}{\partial v^i\partial v^j}(\tau,v,q)$ is invertible for every $(\tau,v,q) \in \RR\times \T\RR^d = \RR^{2d+1}$.  For Newtonian systems, \cref{EL} reduces to Newton's law $F = ma$.

Our convention is that near a corner of a piecewise-smooth path $\gamma$, the velocity $\dot\gamma$ is discontinuous like Heaviside's step function $\Theta$, and the acceleration $\frac{\d^2}{\d\tau^2}\gamma(\tau)$ has a discontinuity like Dirac's delta function $\delta(\tau) = \frac{\d}{\d\tau}\Theta(\tau)$.  For a path to be classical, we impose the Euler-Lagrange equations even at these points of discontinuity, understanding the equation in the sense of distributions.  Provided $\frac{\partial^2 L}{\partial v^i\partial v^j}(v,q)$ is invertible, all classical paths are smooth, by a classical ``bootstrapping'' argument: the acceleration enters only once in \cref{EL}, as $\frac{\partial^2 L}{\partial v^i\partial v^j}\frac{\d^2\gamma^j}{\d\tau^2}$, and so can have a discontinuity no worse than the step function, but then $\dot\gamma$ is continuous, hence so is $\frac{\d^2\gamma^j}{\d\tau^2}$, etc.

In \cref{formalintdefn}, we insisted that each critical point be nondegenerate.  Let us say that a classical path $\gamma$ is \define{nondegenerate} if, when restricted to based loops, the operator $\Aa^{(2)}(\gamma)$ defined in \cref{A2} has no zero modes.  We will describe this operator in more detail and compute its inverse in the next section.

\subsection{The Green's function} \label{GreenFnSub}

Let $\gamma: [t_0,t_1] \to \RR^d$ be classical and let $\xi,\zeta: [0,t] \to \RR^d$ be based loops.  By integrating \cref{A2} by parts, we have $\Aa^{(2)}\cdot \xi\zeta = \int_{t_0}^{t_1} \Dd_\gamma[\xi(\tau)]_j\,\zeta^i(\tau)\,d\tau$, where $\Dd_{\gamma}$ is the second-order linear differential operator given by:
\begin{multline}
  \Dd_{\gamma}[\xi]_j(\tau)  = -\left.\frac{\partial^2 L}{\partial v^i\partial v^j}\right|_{\gamma(\tau)} \ddot \xi^i(\tau)  + \left( -\frac{\d}{\d\tau} \left[ \left.\frac{\partial^2 L}{\partial v^i\partial v^j}\right|_{\gamma(\tau)}\right] - \left. \frac{\partial^2 L}{\partial q^i\partial v^j}\right|_{\gamma(\tau)} + \left.\frac{\partial^2 L}{\partial v^i\partial q^j}\right|_{\gamma(\tau)}  \right) \dot \xi^i(\tau) + \phantom{\xi} \\  + \left( -\frac{\d}{\d\tau} \left[ \left. \frac{\partial^2 L}{\partial q^i\partial v^j}\right|_{\gamma(\tau)}\right]  + \left. \frac{\partial^2 L}{\partial q^i\partial q^j}\right|_{\gamma(\tau)} \right) \xi^i(\tau) \label{Ddefined}
\end{multline}
As we mentioned in the paragraph following \cref{EL}, the second derivative of a piecewise-smooth function can have a discontinuity like $\delta(\tau)$, and the integral expression for $\Aa^{(2)}$ should be understood accordingly.
We will show that when $\gamma$ is nondegenerate and $\frac{\partial^2 L}{\partial v^2}$ is invertible, then $\Dd_\gamma$ has an inverse.

\begin{defn} \label{greendefn}
  Let $\Dd$ be a second-order linear differential operator on the space of paths $[t_0,t_1] \to \RR^d$.  A \define{Green's function for $\Dd$} is a matrix-valued function of two variables $g: [t_0,t_1]^2 \to \Mat(\RR^d)= \RR^{d^2}$ such that $\Dd[g(\varsigma,-)]^j_k(\tau) = \delta^j_k \delta(\varsigma,\tau)$ (the product of Dirac's delta function with Kronecker's delta matrix), and $g(\varsigma,t_0) = 0 = g(\varsigma,t_1)$, so that $g(\varsigma,-)$ is a based loop for each $\varsigma \in [t_0,t_1]$.
  
  Let $L$ be a Lagrangian on $\RR^d$ and $\gamma: [t_0,t_1] \to \RR^d$ a classical path.  The \define{Green's function for $\gamma$}, if it exists, is the Green's function $G_\gamma$ for the operator $\Dd_\gamma$ in \cref{Ddefined}.
\end{defn}

Let us justify the word ``the'' in the previous sentence.  Since $\Aa^{(2)} \cdot \xi\zeta = \Aa^{(2)} \cdot \zeta\xi$, we see that if $G_\gamma^{ij}(\varsigma,\tau)$ is a Green's function for $\Dd_\gamma$, then so is $G_\gamma^{ji}(\tau,\varsigma)$.  Now suppose that $G,G'$ are two Green's functions for $\Dd_\gamma$, and consider $\Aa^{(2)} \cdot G(\varsigma,-)G'(-,\varsigma')$.  By integrating by parts, this equals both $G(\varsigma,\varsigma')$ and $G'(\varsigma,\varsigma')$.  Moreover, by uniqueness, we see that $\gamma$ cannot have a Green's function if it is not nondegenerate.  We will prove the converse in \cref{greenprop}.

The best way to solve an inhomogeneous linear differential equation, if the solutions to the corresponding homogeneous equation are known, is to use the method of ``variation of parameters,'' which works for matrix-valued functions just as well as it does for scalars:
\begin{lemma}  \label{Gexists}
  Let $\Dd$ be a second-order linear differential operator on the space of paths $[t_0,t_1] \to \RR^d$ of the form $\Dd[\varphi]^j = \ddot\varphi^j + B^j_i\dot\varphi^i + C^j_i\varphi^i$, where $B,C$ are smooth matrix-valued functions on $[t_0,t_1]$.  Suppose that there are functions $\phi^a: [t_0,t_1] \to \Mat(\RR^d)$, $a = 0,1$, satisfying $\Dd[\phi^a] = 0$ with boundary values $\phi^{a,i}_k(t_b) = \delta^i_k \delta^a_b$ (here the indices $i,j,k$ range from $1,\dots,d$, but $a,b \in \{0,1\}$).  Then the $2d\times 2d$ matrix $ M(\tau) = \begin{pmatrix}
    \phi^{0}(\tau) & \phi^{1}(\tau)  \\ \dot\phi^{0}(\tau)& \dot\phi^{1}(\tau)
  \end{pmatrix}$ is invertible for each $\tau \in [t_0,t_1]$.  Let $\psi_0,\psi_1: [t_0,t_1] \to \Mat(\RR^d)$ comprise the right $d\times 2d$ half of $M^{-1}$.   The function $g: [t_0,t_1]^2 \to \Mat(\RR^n)$ given by
  \begin{equation}  \label{Gexistseqn}
    g^i_k(\varsigma,\tau) =   \Theta(\tau - \varsigma) \,\phi^{0,i}_j(\tau)\,\psi_{0,k}^j(\varsigma)- \Theta(\varsigma-\tau)\, \phi^{1,i}_j(\tau)\,\psi_{1,k}^j(\varsigma)
  \end{equation}
is a Green's function for $\Dd$.  (In \cref{Gexistseqn}, $\Theta$ is Heaviside's step function.)
\end{lemma}

\begin{proof}
We first prove that $M(\tau)$ is invertible for each $\tau$.  A solution $\varphi(\tau)$ to $\Dd[\varphi] = 0$ with $\varphi(t_0) = 0$ is determined by $\dot \varphi(t_0)$, and thus $\Dd$ determines a (constant) matrix $D$ satisfying $D^j_i\dot \varphi^i(t_0) = \varphi^j(t_1)$.  In particular, $D^j_i \dot\phi^{1,i}_k(t_0) = \phi^{1,j}_k(t_1) = \delta^j_k$, and so $\dot\phi^{1,i}_k(t_0)$ has full rank.  Thus $M(t_0) = \begin{pmatrix}
    \delta & 0  \\ \dot\phi^{0}(t_0)& \dot\phi^{1}(t_0)
  \end{pmatrix}$ is invertible.  By Liouville's formula, $ \det M(\tau) =  \exp\bigl( -\int_{t_0}^\tau \tr B\bigr)\det M(t_0) $, and in particular it is never $0$.

The boundary conditions for $g$ are immediate --- the $\psi_a$ satisfy $\psi_0(t_0) = 0 = \psi_1(t_1)$, because by definition $\phi^{0,i}_j(\tau)\psi_{0,k}^j(\tau) + \phi^{1,i}_j(\tau)\psi_{1,k}^j(\tau) = 0$ --- and $g$ is continuous near the diagonal $\varsigma = \tau$ for the same reason.  Finally, one must check the derivatives of $g$, but the only terms in $\Dd[g]$ that survive are $\delta(\tau - \varsigma) \dot\phi^{0,i}_j(\tau)\psi_{0,k}^j(\varsigma) + \delta(\tau - \varsigma) \dot\phi^{1,i}_j(\tau)\psi_{1,k}^j(\varsigma) = \delta(\tau - \varsigma) \delta^i_k$, because $\dot\phi^{0,i}_j(\tau)\psi_{0,k}^j(\tau) + \dot\phi^{1,i}_j(\tau)\psi_{1,k}^j(\tau) = \delta^i_k$.
\end{proof}

To show that solutions to $\Dd_\gamma[\phi^a] = 0$, $\phi^a(t_b) = \delta^a_b$ exist when $\gamma$ is nondegenerate, we use the following fact (we reproduce the proof from \cite{Chris2009}):

\begin{lemma} \label{bundlelemma}
  Let $\pi: \Ee \to \Bb$ be a smooth bundle, where $\Bb$ is a finite-dimensional smooth manifold and $\Ee$ is a possibly-infinite-dimensional smooth manifold.  Let $f: \Ee \to \RR$ be a smooth map.  For each $b\in \Bb$, consider the restriction $f|_{\pi^{-1}(b)}$ of $f$ to the fiber $\pi^{-1}(b) \subseteq \Ee$.  Define $\Cc \subseteq \Ee$ to be the set of $c\in \Ee$ so that $\bigl(f|_{\pi^{-1}(\pi(c))}\bigr)^{(1)}(c) = 0$ --- here $\pi^{-1}(\pi(c))$ is the fiber containing $p$, and $\bigl(f|_{\pi^{-1}(\pi(c))}\bigr)^{(1)}$ is the first derivative of $f$ along the fiber, so that $\bigl(f|_{\pi^{-1}(\pi(c))}\bigr)^{(1)}(c) \in \T^*_c\bigl(\pi^{-1}(\pi(c))\bigr)$.  Assume that $\Cc$ is a manifold of the same dimension as $\Bb$.
  
  Let $c \in \Cc$ be nondegenerate in the sense that the second derivative $\bigl(f|_{\pi^{-1}(\pi(c))}\bigr)^{(2)}(c)$, thought of as a map $\T_c\bigl(\pi^{-1}(\pi(c))\bigr) \to  \T^*_c\bigl(\pi^{-1}(\pi(c))\bigr)$, has zero kernel.  Then $\pi|_\Cc: \Cc \to \Bb$ is a local diffeomorphism near $c\in \Cc$; i.e.\ there are open neighborhoods $c \in \Uu \subseteq \Cc$ and $\pi(c) \in \Oo \subseteq \Bb$ with $\pi|_\Uu: \Uu \isom \Oo$.
\end{lemma}
\begin{proof}
Since the statement is local, to save space we restrict $\Bb$ to an open neighborhood of $\pi(c)$ and choose a trivialization $\Ee = \Ff \times \Bb$, so that we can identify all fibers $\pi^{-1}(b)$ with $\Ff$.  Then for $e\in \Ee$, the sequence $\Ff \to \Ee \to \Bb$ gives a short-exact sequence $\ker \pi = \T_e\Ff \to \T_e\Ee \to \T_{\pi(e)}\Bb$.  The function $f: \Ee \to \RR$ defines a map $f^{(1)} = \d f: \T\Ee \to \T\RR = \RR \times \RR$, and $\Cc = \{ c\in \Ee \st$the restriction of $\d f$ to $\T_c\Ff$ is $0\}$.

Pick $c\in \Cc$.  Then $f^{(2)}$ determines a linear map $h: \T_c\Ee \to \T_c^*\Ff$, which deserves to be called the \define{Hessian}; it transforms as a tensor because $f^{(1)}$ vanishes on $\T_c\Ff$.
The key fact is that $\T_c\Cc \subseteq \ker h$, easily checked by considering the derivative of $f^{(1)}$ along paths in $\Cc$.  But $c$ is nondegenerate if and only if $\T_c\Ff \cap \ker h = 0$.  Thus if $c$ is nondegenerate, then $\d\pi: \T_c\Cc \to \T_{\pi(c)}\Bb$ is an injection.  On the other hand, by assumption the dimensions of $\T_c\Cc$ and $\T_{\pi(c)}\Bb$ agree, so $\d\pi$ is full-rank and $\pi$ is a local diffeomorphism.
\end{proof}

Suppose that $L$ is a Lagrangian on $\RR^d$ such that the matrix $\frac{\partial^2L}{\partial v^i\partial v^j}(\tau,v,q)$ is invertible for every $(\tau,v,q) \in \RR\times\T\RR^d$.  Then a classical path $\gamma: [t_0,t_1] \to \RR^d$ is determined by its initial conditions $\bigl(\dot\gamma(t_0),\gamma(t_0)\bigr) \in \T\RR^d$.  Let $\flow: \RR \times \RR \times \T\RR^d \to \RR \times \RR^d \times \RR \times \RR^d$ be the smooth function satisfying
$$ \flow\bigl( t_0,t_1,\dot\gamma(t_0),\gamma(t_0)\bigr) = \bigl(t_0,\gamma(t_0),t_1,\gamma(t_1)\bigr) $$
for classical paths $\gamma$ --- we write this as having domain $\RR^2 \times \T\RR^d$, but of course really the domain is some open neighborhood in $\RR^2\times\T\RR^d$ containing $\{(t_0,t_1,v,q) \st t_0 = t_1\}$.

\begin{defn} \label{nonfocaldef}
  Let $L: \RR\times \T\RR^d \to \RR$ be a Lagrangian such that $\frac{\partial^2L}{\partial v^2}$ is everywhere invertible.  A classical path $\gamma: [t_0,t_1] \to \RR^d$ is \define{nonfocal} if the function $\flow$ defined above is a local diffeomorphism near $\bigl(t_0,t_1, \dot\gamma(t_0),\gamma(t_0)\bigr) \in \RR^2 \times \T\RR^d$.
\end{defn}
In fact, it suffices that for fixed $t_0,t_1$ the function $\flow_{[t_0,t_1]}: \T\RR^d \to \RR^d \times \RR^d$ be a local diffeomorphism near $\bigl(\dot\gamma(t_0),\gamma(t_0)\bigr)$, as this is clearly an open condition in $t_0,t_1$.

By identifying classical paths with their initial conditions (and domains), we see that a classical path $\gamma$ is nonfocal if and only if it extends to a family of classical paths smoothly parameterized by ``Dirichlet'' boundary conditions.  More precisely, if $\gamma: [t_0,t_1] \to \RR^d$ is classical and nonfocal, then there is an open neighborhood $\Oo$ of $\bigl(t_0,\gamma(t_0),t_1,\gamma(t_1)\bigr) \in \RR \times \RR^d \times \RR \times \RR^d$ and a smooth function
$$ \hat\gamma: \bigl\{ (t_0',q_0,t_1',q_1,\tau) \in \RR^{2d+3} \st (t_0',q_0,t_1',q_1) \in \Oo \text{ and } \tau \in [t_0',t_1']\bigr\} \to \RR^d$$ with the following properties: (i) for each $(t_0',q_0,t_1',q_1) \in \Oo$, the path $\hat\gamma(t_0',q_0,t_1',q_1;-)$ is classical; (ii) for $a = 0,1$, we have $\hat\gamma(t_0',q_0,t_1',q_1;t_a') = q_a$; (iii) $\hat\gamma(t_0,\gamma(t_0),t_1,\gamma(t_1);-) = \gamma$.

Henceforth we will drop the \,$\hat{}$\,s and $'$s, and we will feel free to confuse nonfocal classical paths with their extensions.  Let $\gamma$ be a nonfocal classical path and $\Oo$ the corresponding neighborhood in $\RR^{2d+2}$.  Then the corresponding \define{Hamilton principal function} $S_\gamma: \Oo \to \RR$ is:
\begin{equation} \label{Sdefined}
  S_\gamma(t_0,q_0,t_1,q_1) = \Aa\bigl(\gamma(t_0,q_0,t_1,q_1;-)\bigr) = \int_{t_0}^{t_1} L\bigl(\tau, \dot\gamma(t_0,q_0,t_1,q_1;\tau),\gamma(t_0,q_0,t_1,q_1;\tau)\bigr)\,\d\tau
\end{equation}
Here and throughout by $\dot\gamma(t_0,q_0,t_1,q_1;\tau)$ we mean $\frac{\partial \gamma}{\partial \tau}(t_0,q_0,t_1,q_1;\tau)$.  The following equations are well-known, and can be checked by differentiating under the integral and applying \cref{EL}:
\begin{equation} \label{derivativesofS}
  \frac{\partial S_\gamma}{\partial q_0} = - \frac{\partial L}{\partial v} \biggl|_{(\tau,v,q) = (t_0,\dot\gamma(t_0),\gamma(t_0))} \quad\quad\quad\quad \frac{\partial S_\gamma}{\partial q_1} = \frac{\partial L}{\partial v} \biggl|_{(\tau,v,q) = (t_1,\dot\gamma(t_1),\gamma(t_1))}
\end{equation}

\begin{prop} \label{greenprop}
  Let $L$ be a Lagrangian on $\RR^d$ with $\frac{\partial^2 L}{\partial v^2}$ everywhere invertible, and let $\gamma$ be a classical path.  Then the following are equivalent:
  \begin{enumerate}
    \item $\gamma$ is nondegenerate.
    \item $\gamma$ is nonfocal.
    \item A Green's function $G_\gamma$ exists for $\gamma$.
  \end{enumerate}
  Moreover, $G_\gamma$ is given explicitly by:
  \begin{multline} \label{greeneqn}
    G^{ij}(\varsigma,\tau) = \Theta(\tau - \varsigma)\, \frac{\partial \gamma^i}{\partial q_1^k}(\varsigma) \biggl( \! \Bigl( \frac{\partial^2 (-S_\gamma)}{\partial q_1\partial q_0}\Bigr)^{\!-1}\biggr)^{\!kl} \frac{\partial \gamma^j}{\partial q_0^l}(\tau) + \mbox{} \\ \mbox{} +  \Theta(\varsigma - \tau) \, \frac{\partial \gamma^i}{\partial q_0^k}(\varsigma) \biggl( \! \Bigl( \frac{\partial^2 (-S_\gamma)}{\partial q_0\partial q_1}\Bigr)^{\!-1}\biggr)^{\!kl} \frac{\partial \gamma^j}{\partial q_1^l}(\tau)
  \end{multline}
\end{prop}
In \cref{greeneqn}, the indices on the inverse matrix 
are given by: $\Bigl( \! \bigl( \frac{\partial^2 (-S)}{\partial q_1\partial q_0}\bigr)^{\!-1}\Bigr)^{\!kl} \frac{\partial^2(-S)}{\partial q_0^l \partial q_1^m} = \delta^k_m$.

\begin{proof}
  We argued already (in the paragraph following \cref{greendefn}) that 3 implies 1.
  To show that 1 implies 2, we use \cref{bundlelemma}: we let $\Ee$ be the space of all paths in $\RR^d$ (with arbitrary domain), $\Bb = \RR \times \RR^d \times \RR \times \RR^d$ with the natural projections, and $f = \Aa$.  Then $\Cc$ is the set of classical paths, and it is a naturally a ($2d+2$)-dimensional manifold (in fact, an open subset of $\RR^2 \times \T\RR^d$) by the remarks before and after \cref{nonfocaldef}.
  Finally, to show that 2 implies 3, we observe that if $\gamma$ is nonfocal, then the paths $\phi^{a,i}_k = \frac{\partial \gamma^i}{\partial q_a^k}$ satisfy $\Dd_\gamma[\phi^a] = 0$ and $\phi^{a,i}_k(t_b) = \delta^i_k\delta^a_b$.  To apply \cref{Gexists}, we use the fact that if $A: [t_0,t_1] \to \Mat(\RR^d)$ is a smooth function such that $A(\tau)$ is invertible for every $\tau$, then a Green's function for $A\frac{\d^2}{\d \tau^2} + AB\frac{\d}{\d \tau} + AC$ is given by $G(\varsigma,\tau) = g(\varsigma,\tau) (A(\varsigma))^{-1}$, where $g$ is the Green's function from \cref{Gexists}.
  
  Therefore, taking advantage of the Einstein index notation to permute some factors and adopting the notation of \cref{Gexists}, the Green's function for $\gamma$ is given by:
  \begin{multline} \label{Greeneqntheother}
    G^{ij}(\varsigma,\tau)  = - \left( \biggl( \left.\frac{\partial^2 L}{\partial v^2}\right|_{\gamma(\varsigma)}\biggr)^{\!-1}\right)^{\!ik}\, \psi_{0,k}^l(\varsigma)\,\, \frac{\partial \gamma^j}{\partial q_0^l}(\tau)\,\, \Theta(\tau - \varsigma) + \mbox{}
    \\  \mbox{} + \left( \biggl( \left.\frac{\partial^2 L}{\partial v^2}\right|_{\gamma(\varsigma)}\biggr)^{\!-1}\right)^{\!ik}\, \psi_{1,k}^l(\varsigma)\,\, \frac{\partial \gamma^j}{\partial q_1^l}(\tau)\,\, \Theta(\varsigma - \tau)
  \end{multline}
But $G^{ij}(\varsigma,\tau) = G^{ji}(\tau,\varsigma)$ by the symmetry of $\Aa^{(2)}$, and so $\eta_a^{jl} =\Bigl( \bigl( \frac{\partial^2 L}{\partial v^2}\bigr|_{\gamma}\bigr)^{-1}\Bigr)^{jk}\, \psi_{a,k}^l$ is a solution to $\Dd_{ij}[\eta_a^{jl}] = 0$ and therefore a linear combination of the $\phi^b = \frac{\partial \gamma}{\partial q_b}$.  By checking the boundary conditions, we see that:
\begin{align}
  \left( \biggl( \left.\frac{\partial^2 L}{\partial v^2}\right|_{\gamma(\varsigma)}\biggr)^{\!-1}\right)^{\!ik}\, \psi_{0,k}^l(\varsigma) & = \frac{\partial \gamma^i}{\partial q_1^j}(\varsigma) \left( \biggl( \left.\frac{\partial^2 L}{\partial v^2}\right|_{\gamma(t_1)}\biggr)^{\!-1}\right)^{\!jk}\, \psi_{0,k}^l(t_1)  \notag \\
   & = \frac{\partial \gamma^i}{\partial q_1^j}(\varsigma)\left( \biggl( \left.\frac{\partial^2 L}{\partial v^2}\right|_{\gamma(t_1)}\biggr)^{\!-1}\right)^{\!jk}\left(\left( \frac{\partial \dot\gamma}{\partial q_0}(t_1)\right)^{\!-1}\right)^{\!l}_{\!k}
   \\
   \left( \biggl( \left.\frac{\partial^2 L}{\partial v^2}\right|_{\gamma(\varsigma)}\biggr)^{\!-1}\right)^{\!ik}\, \psi_{1,k}^l(\varsigma) & = \frac{\partial \gamma^i}{\partial q_0^j}(\varsigma) \left( \biggl( \left.\frac{\partial^2 L}{\partial v^2}\right|_{\gamma(t_0)}\biggr)^{\!-1}\right)^{\!jk}\left(\left( \frac{\partial \dot\gamma}{\partial q_1}(t_0)\right)^{\!-1}\right)^{\!l}_{\!k}
\end{align} 
Finally, we should study $\Bigl( \bigl(\frac{\partial^2 L}{\partial v^2}\bigr|_{\gamma(t_1)}\bigr)^{\!-1}\Bigr)^{\!jk}\Bigl(\bigl( \frac{\partial \dot\gamma}{\partial q_0}(t_1)\bigr)^{\!-1}\Bigr)^{\!l}_{\!k}$ and $\Bigl( \bigl(\frac{\partial^2 L}{\partial v^2}\bigr|_{\gamma(t_0)}\bigr)^{\!-1}\Bigr)^{\!jk}\Bigl(\bigl( \frac{\partial \dot\gamma}{\partial q_1}(t_0)\bigr)^{\!-1}\Bigr)^{\!l}_{\!k}$.  The former is the inverse matrix to $\frac{\partial^2 L}{\partial v^j\partial v^k}\bigr|_{\gamma(t_1)} \frac{\partial \dot\gamma^k}{\partial q_0^l}(t_1)$.  But:
\begin{equation}
  \frac{\partial}{\partial q_0^l} \left[ \left.\frac{\partial L}{\partial v^j}\right|_{\gamma(t_1)}\right] = \left.\frac{\partial^2 L}{\partial v^j\partial v^k}\right|_{\gamma(t_1)} \frac{\partial \dot\gamma^k(t_1)}{\partial q_0^l} + \left.\frac{\partial^2 L}{\partial v^j\partial q^k}\right|_{\gamma(t_1)} \frac{\partial\gamma^k(t_1)}{\partial q_0^l} = \left.\frac{\partial^2 L}{\partial v^j\partial v^k}\right|_{\gamma(t_1)} \frac{\partial \dot\gamma^k}{\partial q_0^l}(t_1)
\end{equation}
since $\frac{\partial\gamma(t_1)}{\partial q_0} = \frac{\partial q_1}{\partial q_0} = 0$.  Then \cref{derivativesofS} completes the proof.
\end{proof}

\subsection{The Morse index of a classical path}\label{MorseSection}

Finally, let $\gamma$ be a classical path; then $\Aa^{(2)}(\gamma)$ is a symmetric bilinear pairing on the space of loops based at $0$.  The \define{Morse index} $\eta(\gamma)$ is the maximal dimension of any subspace of the space of based loops for which $\Aa^{(2)}(\gamma)$ is negative definite. To show that $\eta(\gamma) < \infty$, we reproduce the classical argument (c.f.\ \cite{Milnor1963}).

\begin{lemma} \label{morselemma}
  Let $\Vv$ be any vector space and $A: \Vv \otimes \Vv \to \RR$ a symmetric pairing.  Suppose that $A$ is negative-definite on a finite-dimensional subspace $\Vv_- \subseteq \Vv$, and that $\Vv_-$ cannot be extended to any larger subspace on which $A$ is negative-definite.  Then any subspace of $\Vv$ on which $A$ is negative-definite has dimension at most $\dim \Vv_-$.
\end{lemma}
\begin{proof} It suffices to consider the kernel $(\Vv_-)^\perp$ of the map $\Vv \to (\Vv_-)^*$ given by $v \mapsto A(v,-)$.  If $\Ww\subseteq \Vv$ has dimension $> \dim \Vv_- = \dim (\Vv_-)^*$, then it intersects nontrivially with $(\Vv_-)^\perp$ as it cannot inject into $(\Vv_-)^*$, but if $v\in \Ww \cap (\Vv_-)^\perp$ has $A(v,v) < 0$, then $A$ is negative-definite on $\Vv_- \oplus v\RR$.
\end{proof}

\begin{prop}\label{morseprop}
  Pick a Lagrangian $L$ on $\RR^d$ and let $\gamma: [t_0,t_1] \to \RR^d$ be classical.  Suppose that the symmetric matrix $\frac{\partial^2 L}{\partial v^i\partial v^j}\bigl(\tau,\dot\gamma(\tau),\gamma(\tau)\bigr)$ is positive-definite for each $\tau\in [t_0,t_1]$.  Then for sufficiently fine subdivisions $t_0 = \tau_0 \leq \tau_1 \leq \dots \leq \tau_n = t_1$ of the interval $[t_0,t_1]$, the pairing $\Aa^{(2)}(\gamma)$ is positive-definite on the space of paths $\xi$ with $\xi(\tau_k) = 0$ for $k=0,\dots,n$.
\end{prop}
\begin{proof}
We will find $\epsilon>0$ so that the statement holds whenever $\tau_k - \tau_{k-1} < \epsilon$.  Let $\Vv$ be the space of based loops with domain $[t_0,t_1]$.  The space $\Vv_{\vec{\tau}}$ of paths $\xi \in \Vv$ that vanish at each $\tau_k$ splits as a direct sum $\Vv_{\vec{\tau}} = \bigoplus_{k=1}^n \Vv_k$, where $\Vv_k$ is the space of based loops with domain $[\tau_{k-1}, \tau_{k}]$, and the direct summands are mutually orthogonal with respect to the pairing $\Aa^{(2)}(\gamma)$.  Thus it suffices to show that $\Aa^{(2)}(\gamma)\cdot \xi\xi > 0$ whenever $\xi \in \Vv$ has support a subinterval of $[t_0,t_1]$ of length less than $\epsilon$.  Upon integrating $\Aa^{(2)}(\gamma)\cdot \xi\xi$ by parts, the $\dot\xi\xi$ integrands cancel out, so for $C_{ij}(\tau) = \frac{\partial^2 L}{\partial q^i \partial q^j}(\gamma(\tau)) - \frac{\d}{\d\tau}\bigl[ \frac{\partial^2 L}{\partial q^i\partial v^j}(\gamma(\tau))\bigr]$, there is some $t' \in [0,t]$ such that:
\begin{equation}
  \Aa^{(2)}(\gamma)\cdot \xi\xi = \int_{t'}^{t'+\epsilon} \left( \left.\frac{\partial^2 L}{\partial v^i\partial v^j}\right|_{\gamma(\tau)} \, \dot\xi^i(\tau)\,\dot\xi^j(\tau) + C_{ij}(\tau)\,\xi^i(\tau)\,\xi^j(\tau)\right)d\tau
\end{equation}
Let $\lambda_1>0$ be the minimal eigenvalue of $\frac{\partial^2 L}{\partial v^i\partial v^j}\bigl(\tau,\dot\gamma(\tau),\gamma(\tau)\bigr)$ as $\tau$ ranges over $[t_0,t_1]$, and let $\lambda_2 > 0$ be the maximum eigenvalue of the $-\frac12\bigl(C_{ij}(\tau)+C_{ji}(\tau)\bigr)$ for $\tau\in[t_0,t_1]$ (if $C$ is always positive-semidefinite, then \cref{morseprop} is trivial).  Then:
\begin{equation}
  \Aa^{(2)}(\gamma) \cdot \xi\xi \geq \int_{t'}^{t'+\epsilon} \left( \lambda_1 \bigl|\dot\xi(\tau)\bigr|^2 - \lambda_2 \bigl|\xi(\tau)\bigr|^2 \right)d\tau \geq \lambda_1 \int_{t'}^{t'+\epsilon} \bigl|\dot\xi(\tau)\bigr|^2d\tau - \epsilon \lambda_2 \sup |\xi|^2
\end{equation}
But by the Cauchy-Schwarz inequality, $\int_{t'}^{t'+\epsilon} |\dot\xi(\tau)|^2d\tau \geq \frac1\epsilon \bigl(\int_{t'}^{t'+\epsilon} |\dot\xi(\tau)|d\tau\bigr)^2 \geq \frac1\epsilon \bigl(2\sup |\xi|\bigr)^2$.  Thus:
\begin{equation}
  \Aa^{(2)}(\gamma) \cdot \xi\xi \geq \left( \frac{\lambda_1}\epsilon - \epsilon\lambda_2\right) \sup_{\tau \in [t',t'+\epsilon]} \bigl|\xi(\tau)\bigr|^2
\end{equation}
Taking $\epsilon < \sqrt{\lambda_1\lambda_2}$ completes the proof of \cref{morseprop}.
\end{proof}

Thus within the space $\Vv$ of based loops with domain $[t_0,t_1]$ we have found a large subspace $\Vv_{\vec{\tau}}$ on which $\Aa^{(2)}(\gamma)$ is positive-definite.  The restriction of $\gamma$ to each interval $[\tau_{k-1},\tau_k]$ gives a nondegenerate classical path $\gamma_k$.  Let $\Ww\subseteq V$ be the space of based loops $\xi$ that are solutions to $\Dd_\gamma[\xi]=0$ except at the times $\tau_k$, $k=1,\dots,n-1$, where $\Dd_\gamma$ is the second-order differential operator from \cref{greeneqn}.  By \cref{greenprop}, such a path $\xi$ depends only on its values at the times $\tau_k$, so that $\Ww \cong (\RR^d)^{\otimes (n-1)}$.  The vector spaces $\Ww$ and $\Vv_{\vec{\tau}}$ are mutually orthogonal with respect to $\Aa^{(2)}(\gamma)$, and $\Ww \oplus \Vv_{\vec{\tau}} = \Vv$.  Since $\Aa^{(2)}(\gamma)$ is positive-definite on $\Vv$, any subspace of $\Vv \oplus \Ww$ on which it is negative-definite cannot have dimension greater than that of $\Ww$.  In particular:
\begin{cor} \label{MorseCor}
  Let $L$ be a Lagrangian function on $\RR^d$ so that the matrix $\frac{\partial^2 L}{\partial v^i\partial v^j}(\tau,v,q)$ is positive-definite for each $(\tau,v,q) \in \RR\times \T\RR^d$.  Then every classical path has finite Morse index. \qedhere
\end{cor}

\subsection{The diagrammatic definition of \texorpdfstring{$U_\gamma$}{U_\gamma}} \label{defUsection}

We proposed to define the path integral of \cref{FeynmanIntegral} in analogy with \cref{formalintdefn}, and so in \cref{actiondersub} we described the derivatives of the action $\Aa$ and in \cref{GreenFnSub} we described the Green's function for $\Aa^{(2)}$.  We recap those formulas, and declare \define{Feynman rules} for the formal path integral supported near the classical nonfocal path $\gamma: [t_0,t_1] \to \RR^d$ with $\gamma(t_a) = q_a$:
\begin{align} \label{FeynmanRulesA}
   \ev\biggl( \begin{tikzpicture}[baseline=(X)]
    \path node[dot] (O) {} ++(0pt,4pt) coordinate (X);
    \draw (O) -- ++(-15pt,15pt) +(0,3pt) node[anchor=base] {$\scriptstyle \xi_1$};
    \draw (O) -- ++(-6pt,15pt) +(0,3pt) node[anchor=base] {$\scriptstyle \xi_2$};
    \draw (O) -- ++(15pt,15pt) +(0,3pt) node[anchor=base] {$\scriptstyle \xi_n$};
    \path (O) ++(4pt,13pt) node {$\scriptstyle \ldots$};
  \end{tikzpicture}  \biggr)
   & = -\Aa^{(n)} \cdot \xi_1\cdots \xi_n = -\int_{\tau=0}^t \left. \prod_{k=1}^n \left( \dot\xi^{i_k}_k(\tau) \frac{\partial}{\partial v^{i_k}} + \xi^{i_k}_k(\tau) \frac{\partial}{\partial q^{i_k}} \right)L\right|_{(\dot\gamma(\tau),\gamma(\tau))} \d t  \\
  \ev\biggl(  \begin{tikzpicture}[baseline=(X)]
    \path coordinate (A) ++(0pt,4pt) coordinate (X);
    \path (A) +(0,-8pt) node[anchor=base] {$\scriptstyle \varsigma,i$};
    \path (A) ++(20pt,0) coordinate (B) +(0,-8pt) node[anchor=base] {$\scriptstyle \tau,j$};
    \draw (A) .. controls +(0,15pt) and +(0,15pt) .. (B);
  \end{tikzpicture}  \biggr)
  & = G^{ij}(\varsigma,\tau)
  =  
  \Theta(\tau - \varsigma)\, \frac{\partial \gamma^i}{\partial q_1^k}(\varsigma) \biggl( \! \Bigl( \frac{\partial^2 (-S_\gamma)}{\partial q_1\partial q_0}\Bigr)^{\!-1}\biggr)^{\!kl} \frac{\partial \gamma^j}{\partial q_0^l}(\tau) + \mbox{} \notag \\
  && \hspace{-3in} \mbox{} + \Theta(\varsigma - \tau) \, \frac{\partial \gamma^i}{\partial q_0^k}(\varsigma) \biggl( \! \Bigl( \frac{\partial^2 (-S_\gamma)}{\partial q_0\partial q_1}\Bigr)^{\!-1}\biggr)^{\!kl} \frac{\partial \gamma^j}{\partial q_1^l}(\tau) \;\; \label{FeynmanRulesG}
\end{align}

Continuing to copy \cref{FeynmanInt0}, in \cref{MorseSection} we defined the Morse index of a classical path and proved that it is finite provided the matrix $\frac{\partial^2 L}{\partial v^i\partial v^j}$ is everywhere positive-definite.  We have only to define the ``dimension'' of the space of based loops, and the ``determinant'' of $\Aa^{(2)}(\gamma)$.  We make these definitions {\em ad hoc}: $\dim\{\text{based loops in $\RR^d$}\} = -d$ and $\left| \det \Aa^{(2)}\right|^{-1} = \left| \det \frac{\partial^2[-S_\gamma]}{\partial q_0\partial q_1}\right|$.  The justification will come in \cref{coordfreethm,fubinithm,divfreethm}.  All together, we have:

\begin{defn} \label{pathintdefn}
  Let $L$ be a Lagrangian on the configuration space $\RR^d$ such that the matrix $\frac{\partial^2 L}{\partial v^2}(\tau,v,q)$ is positive definite for every $(\tau,v,q) \in \RR\times\T\RR^d$, and let $\gamma: [t_0,t_1] \to \RR^d$ be a nonfocal classical path, extended to a smooth family of classical paths parameterized by the Dirichlet boundary conditions $\gamma(t_0) = q_0$ and $\gamma(t_1) = q_1$.  Then the \define{formal path integral supported near $\gamma$} is:
\begin{equation} \label{Udefined}
    U_\gamma(t_0,q_0,t_1,q_1) = (2\pi i \hbar)^{-d/2} \exp\!\left(\frac i \hbar S_\gamma(t_0,q_0,t_1,q_1)\right) (-i)^{\eta(\gamma)} \sqrt{\left| \det \frac{\partial^2[-S_\gamma]}{\partial q_0\partial q_1}\right|}  \sum_\Gamma \frac{(i\hbar)^{-\chi(\Gamma)}\ev(\Gamma)}{\left|\Aut \Gamma\right|}
\end{equation}
The sum ranges over combinatorial diagrams in which all vertices have valence at least three, and $\ev(\Gamma)$ is evaluated via \cref{FeynmanRulesA,FeynmanRulesG}.
\end{defn}
We have not yet proved that the Feynman diagrams evaluate to finite numbers --- in general there are \define{ultraviolet divergences}, and one hopes that they cancel, as we will discuss in \cref{divsection}.  

We foreshadow one remark about \cref{pathintdefn}, because it is worth emphasizing.  Our {\em ad hoc} choice for the determinant is justified, in the sense that \cref{fubinithm} holds, only when the ultraviolet divergences in the Feynman diagrams exactly cancel.  In \cref{divfreethm}, we will prove that this is the case when the Lagrangian $L$ is quadratic in velocity --- $L(\tau,v,q) = \frac12 a_{ij}(\tau,q)\,v^iv^j + b_i(\tau,q)\,v^i + c(\tau,q)$, and so by the condition that $\frac{\partial^2 L}{\partial v^i\partial v^j} = a_{ij}$ is positive-definite, $a$ is some Riemannian metric on $\RR^d$ --- and $\det a(\tau,q)$ is constant on $\RR^{d+1}$.  But we know of no other situations in which the ultraviolet divergences cancel.  We believe therefore that \cref{Udefined} is incorrect when there are ultraviolet divergences, and that there should exist a similar expression with a different choice of determinant, but we do not know what it would be.

\Section{The formal path integral depends only on the volume form} \label{coordfree}

Our goal in this paper is to define a ``formal path integral'' on a possibly curved manifold, provided only the data of a configuration space $\Nn$ with volume form, a classical Lagrangian $L$, and a distinguished path $\gamma$.  \Cref{pathintdefn} gives the formal path integral when $\Nn = \RR^d$, but this definition {\em a priori} depends on the choice of coordinates.  In this section, we show that it is invariant under volume-preserving changes of coordinates, so that \cref{pathintdefn} can be applied whenever the classical path $\gamma$ can be contained within a single coordinate patch.  In particular, we will prove:

\begin{thm} \label{coordfreethm}
  Let $\Oo \subseteq \RR^d$ be a star-shaped open neighborhood, and $f: \Oo \to \RR^d$ a locally volume-preserving smooth function.  Pick a Lagrangian $L: \RR\times \T\RR^d \to \RR$ and a path $\tilde\gamma: [t_0,t_1] \to \Oo$ such that $\gamma = f\circ \tilde\gamma$ is a classical nondegenerate path for $L$, and let $\Uu_\gamma$ be the formal path integral defined in \cref{pathintdefn} for $(\gamma,L)$.  Let $\tilde L = L\circ (\id,\d f,f): \RR\times \T\Oo \to \RR$ and write $\tilde \Uu_{\tilde \gamma}$ for the formal path integral for $(\tilde\gamma,\tilde L)$.  Then $\tilde U_{\tilde \gamma}(t_0,q_0,t_1,q_1) = U_\gamma\bigl(t_0,f(q_0),t_1,f(q_1))$ for $q_0, q_1 \in \Oo$.
\end{thm}
In fact, $f$ can depend explicitly on an external time parameter, but we prefer not to bog down the notation.
\Cref{coordfreethm} is proved in slightly more general (and less precise) form in \cite{mecoordindep}, including the case of time-dependent changes of coordinates.  We reproduce the proof here, rewriting it to apply more directly to the situation at hand.

We will use coordinates $q,q_1,\dots$ for points in $\Oo$, and $\tilde q,\dots$ for their images under $f$ in $\RR^d$.  A smooth function is \define{locally volume-preserving} if when restricted to small enough neighborhoods it is a volume-preserving diffeomorphism; equivalently, $f$ is locally volume-preserving if $\bigl| \det \frac{\partial f^i}{\partial q^j}\bigr| = 1$.  Recall that $f: \Oo \to \RR^d$ is \define{orientation-preserving} if $\det \frac{\partial f}{\partial q} > 0$.  A neighborhood $\Oo \subseteq \RR^d$ is \define{star-shaped} if $0\in \Oo$ and for each $q\in \Oo$, we have $sq \in \Oo$ for all $s\in [0,1]$.

Upon inspection of \cref{Udefined}, the following is clear: the dimension, Morse index, and classical action terms are invariant under arbitrary changes of coordinates; the determinant term is invariant under volume-preserving changes of coordinates; and the individual Feynman diagrams are invariant under affine changes of coordinates.  Therefore to prove \cref{coordfreethm} we need only to consider the sum of diagrams, and by composing $f$ with various affine maps, we can suppose that $f(0) = 0$ and $\frac{\partial f^i}{\partial q^j}(0) = \delta^i_j$.

In \cref{CF1} we will prove a lemma that allows us to reduce \cref{coordfreethm} to the case when $f$ is an infinitesimal change of coordinates.  In \cref{CF2} we perform a diagrammatic calculation to verify \cref{coordfreethm} in the infinitesimal case.

\subsection{Oriented-volume-preserving maps are homotopic to the identity}\label{CF1}

In this section we find a smooth homotopy among locally volume-preserving maps between the function $f$ considered above and the identity map $\Oo \mono \RR^d$.

\begin{lemma} \label{CFlemma}
  Let $\Oo \subseteq \RR^d$ be a star-shaped open neighborhood, and suppose that $f: \Oo \to \RR^d$ is orientation- and locally-volume-preserving.  Suppose furthermore that $f(0) = 0$ and $\frac{\partial f^i}{\partial q^j}(0) = \delta^i_j$.  Then there exists a smooth function $F: [0,1] \times \Oo \to \RR^d$ such for each $s\in [0,1]$, $F(s,-)$ is orientation- and locally-volume preserving with $F(s,0) = 0$ and $\frac{\partial F^i}{\partial q^j}(s,0) = \delta^i_j$, and such that $F(0,q) = q$ and $F(1,q) = f(q)$.
\end{lemma}

\begin{proof}
Let ${f'}^i_j = \frac{\partial f^i}{\partial q^j}$.  Then $f': \Oo \to \Mat(d)$ satisfies the following conditions:
\begin{equation}
  \label{CFlemmaeqn1}
  {f'}^i_j(0) = \delta^i_j\quad\quad\quad \det {f'} = 1 \quad\quad\quad \frac{\partial {f'}^i_j}{\partial q^k} = \frac{\partial {f'}^i_k}{\partial q^j}
\end{equation}
By the fundamental theorem of calculus, since $\Oo$ is connected, the function $f: \Oo \to \RR^d$ is completely determined by $f'$ and $f(0) = 0$.  Conversely, since $\Oo$ is simply-connected, any $f'$ satisfying the last of the above three conditions determines some function $f: \Oo \to \RR^d$ with $f(0) = 0$ and $f' = \frac{\partial f}{\partial q}$; by the middle condition, $f$ is locally-volume-preserving.

For $f$ as in the lemma, let ${F'}: [0,1] \times \Oo \to \RR^d$ be given by ${F'}(s,q) = \frac{\partial f}{\partial q}(sq)$; this is well-defined since $\Oo$ is star-shaped.  Then for each $s \in [0,1]$, $f' = {F'}(s,-)$ satisfies the conditions in \cref{CFlemmaeqn1}; the third follows by the chain rule.  Therefore, for each $s\in [0,1]$, there is a unique function $F(s,-): \Oo \to \RR^d$ with $\frac{\partial F}{\partial q} = {F'}$ and $F(s,0) = 0$, and $F$ is smooth in $s$.  When $s = 1$, $F(1,q) = f(q)$, and when $s = 0$, we have $F(0,q) = q$, as $\frac{\partial F^i}{\partial q^j}(0,q) = {F'}^i_j(0,q) = \delta^i_j$.  Therefore $F$ is the desired homotopy.
\end{proof}

\subsection{Proof of Theorem \texorpdfstring{\ref{coordfreethm}}{3.0.1}}\label{CF2}

By \cref{CFlemma} and the remarks following the statement of \cref{coordfreethm}, we can suppose that our volume-preserving function is homotopic to the identity: there is a function $F: [0,1] \times \Oo \to \RR^d$ so that for each $s\in [0,1]$, $F(s,-)$ is locally volume-preserving.  Then $E = \frac{\partial F}{\partial s}$ makes sense as a vector field on $F(s,\Oo)$, and in particular determines a family of locally-volume-preserving functions $F(s_1 , s_2,-)$ with $F(0, s,-) = F(s,-)$ and $F(s_1 , s_2,-) \circ F(s_0 , s_1,-) = F(s_0 , s_2,-)$.  Let $L^s = L \circ \bigl(dF(0,s,-),F(0,s,-)\bigr)$ be a Lagrangian on $F(0,s,\Oo)$, and let $\gamma^s = F(s, 1,-) \circ \tilde\gamma$.  Let $U^s$ be the formal path integral for $L^s$ and its classical path $\gamma^s$.  Then $U^0 = U$ and $U^1 = \tilde U$, and so to prove \cref{coordfreethm}, it suffices to show that $\frac{\partial}{\partial s}\bigl[ U^s\bigr] = 0$.

And for this, it suffices to consider \cref{coordfreethm} when $f$ is an ``infinitesimal change of coordinates''.  I.e.: $f(q) = q + \epsilon E(q)$, where $E$ is a fixed vector field on $\Oo$ and $\epsilon^2 = 0$.   We will also make the following abuse of notation: we denote the map $\gamma \mapsto f\circ \gamma$ on the space of paths in $\Oo$ by the same letter as we use for the function $f: \Oo \to \Oo$.  With all these assumptions and notation, the Feynman diagrams in the path integral $\tilde U$ are based on the action $\tilde \Aa = \Aa \circ f^{-1}$, which is the action determined by the Lagrangian $\tilde L$.  As we can ignore terms of order $\epsilon^2$, we have $f^{-1}(q) = q - \epsilon E(q)$.

Recall now the generalized chain rule of Fa\`a di Bruno \cite{Hardy2006}:
\begin{equation} \label{FdB}
  \bigl(\Aa \circ f^{-1}\bigr)^{(n)}\cdot \bigl(\xi_1 \otimes \xi_n\bigr) = \sum_{\text{partitions $S$ of }\{1,\dots,n\}} \bigl(\Aa^{(|S|)}\circ f^{-1}\bigr) \cdot \bigotimes_{s \in S} \Bigl( \bigl(f^{-1}\bigr)^{(|s|)} \cdot \bigotimes_{j \in s} \xi_j \Bigr)
\end{equation}
By a \define{partition} of $\{1,\dots,n\}$, we mean a collection $S$ of nonempty subsets of $\{1,\dots,n\}$ that are pairwise disjoint and whose union is $\{1,\dots,n\}$.
The partition determines how to contract (abbreviated ``$\cdot$'') the tensors in \cref{FdB}.  We introduce the following Feynman rules:
\begin{gather} \label{Prule1}
  \begin{tikzpicture}[baseline=(X)]
    \path node[circle,inner sep=0pt,draw] (O) {$\scriptstyle \sim$} ++(0pt,4pt) coordinate (X);
    \draw (O) -- ++(-15pt,15pt) +(0,3pt) node[anchor=base,text=black] {$\scriptstyle \xi_1$};
    \draw (O) -- ++(-6pt,15pt) +(0,3pt) node[anchor=base,text=black] {$\scriptstyle \xi_2$};
    \draw (O) -- ++(15pt,15pt) +(0,3pt) node[anchor=base,text=black] {$\scriptstyle \xi_n$};
    \path (O) ++(4pt,13pt) node[text=black] {$\scriptstyle \ldots$};
  \end{tikzpicture} = - \tilde\Aa^{(n)}(\tilde\gamma) \cdot (\xi_1 \otimes \dots \otimes \xi_n)
  \quad\quad\quad\quad
  \begin{tikzpicture}[baseline=(X)]
    \path coordinate (A) ++(0pt,4pt) coordinate (X);
    \path (A);
    \path (A) ++(20pt,0) coordinate (B);
    \draw (A) .. controls +(0,15pt) and +(0,15pt) .. node[fill=white,circle,inner sep=0pt,draw] {$\scriptstyle \sim$} (B);
  \end{tikzpicture} 
   = \bigl(\tilde\Aa^{(2)}\bigr)^{-1} \\
  \begin{tikzpicture}[baseline=(X)]
    \path node[draw,rectangle,text=black,inner sep=1pt] (O) {$\scriptstyle f^{-1}$} ++(0pt,0pt) coordinate (X);
    \draw (O) -- ++(-15pt,15pt) +(0,3pt) node[anchor=base,text=black] {$\scriptstyle \xi_1$};
    \draw (O) -- ++(-6pt,15pt) +(0,3pt) node[anchor=base,text=black] {$\scriptstyle \xi_2$};
    \draw (O) -- ++(15pt,15pt) +(0,3pt) node[anchor=base,text=black] {$\scriptstyle \xi_n$};
    \path (O) ++(4pt,13pt) node[text=black] {$\scriptstyle \ldots$};
    \draw (O) -- ++(0pt,-15pt);
  \end{tikzpicture} = \Bigl\{ x \mapsto \bigl(f^{-1}\bigr)^{(n)}(\gamma(x)) \cdot \bigl(\xi_1(x) \otimes \dots \otimes \xi_n(x)\bigr) \Bigr\} \in \T_\gamma\bigl(\Gamma(Q \to X)\bigr) \label{Prule2}
\end{gather}

Then \cref{FdB} reads:
\begin{equation}
  \begin{tikzpicture}[baseline=(X),black]
    \path node[circle,inner sep=0pt,draw] (O) {$\scriptstyle \sim$} ++(0pt,10pt) coordinate (X);
    \draw (O) -- ++(-15pt,30pt);
    \draw (O) -- ++(-6pt,30pt);
    \draw (O) -- ++(15pt,30pt);
    \path (O) ++(4pt,26pt) node[text=black] {$\scriptstyle \ldots$};
  \end{tikzpicture} =
  \sum 
  \begin{tikzpicture}[baseline=(X),black]
    \path node[dot] (O) {} ++(0pt,10pt) coordinate (X);
    \path (O)  ++(-15pt,15pt)  node[draw,rectangle,text=black,inner sep=1pt] (a1) {$\scriptstyle f^{-1}$};
    \path (O)  ++(15pt,15pt)  node[draw,rectangle,text=black,inner sep=1pt] (a2) {$\scriptstyle f^{-1}$};
    \draw (O) -- (a1);
    \draw (O) -- (a2);
    \path (O) ++(0pt,13pt) node[text=black] {$\scriptstyle \ldots$};
    \draw (a1) -- ++(0pt,15pt);
    \draw (a1) -- ++(4pt,15pt);
    \draw (a2) -- ++(0pt,15pt);
    \path (a1) ++(8pt,15pt) coordinate (b1);
    \path (a2) ++(-4pt,15pt) coordinate (b2);
    \path (a1) ++(-4pt,15pt) coordinate (b4);
    \draw (a2) -- ++(4pt,15pt);
    \draw (a1) -- (b2);
    \draw (a2) -- (b1);
    \draw (a2) -- (b4);
    \path (O) ++(2pt,29pt) node[text=black] {$\scriptstyle \ldots$};
  \end{tikzpicture} 
\end{equation}
The sum ranges over isomorphism classes of diagrams with ordered leaves but unordered \tikz[baseline=(a.base)] \node[rectangle,draw,black,text=black,anchor=base,inner sep=0pt] (a) {$f^{-1}$}; vertices.  The \tikz \node[dot,black] {}; vertex can be of arbitrary valence (non-zero, if the left-hand-side has non-zero valence), and each \tikz[baseline=(a.base)] \node[rectangle,draw,black,text=black,anchor=base,inner sep=0pt] (a) {$f^{-1}$}; vertex has one output strand and at least one input strand.  The \tikz \node[black,circle,inner sep=0pt,draw] (O) {$\scriptstyle \sim$}; vertex on the left-hand side and the  \tikz[baseline=(a.base)] \node[rectangle,draw,black,text=black,anchor=base,inner sep=0pt] (a) {$f^{-1}$}; vertices on the right hand side are evaluated at $\tilde\gamma$, and the \tikz \node[dot,black] {}; vertex on the right hand side is evaluated at $\gamma = f^{-1}\circ \tilde\gamma$.  Given that $f^{-1}(q) = q - \epsilon E(q)$, we have:
\begin{equation}
  \begin{tikzpicture}[baseline=(X),black]
    \path node[circle,inner sep=0pt,draw] (O) {$\scriptstyle \sim$} ++(0pt,10pt) coordinate (X);
    \draw (O) -- ++(-15pt,30pt);
    \draw (O) -- ++(-6pt,30pt);
    \draw (O) -- ++(15pt,30pt);
    \path (O) ++(4pt,26pt) node[text=black] {$\scriptstyle \ldots$};
  \end{tikzpicture} =
  \begin{tikzpicture}[baseline=(X),black]
    \path node[dot] (O) {} ++(0pt,10pt) coordinate (X);
    \draw (O) -- ++(-15pt,30pt);
    \draw (O) -- ++(-6pt,30pt);
    \draw (O) -- ++(15pt,30pt);
    \path (O) ++(4pt,26pt) node[text=black] {$\scriptstyle \ldots$};
  \end{tikzpicture}  -
  \epsilon \sum 
  \begin{tikzpicture}[baseline=(X),black]
    \path node[dot] (O) {} ++(0pt,10pt) coordinate (X);
    \path (O)  ++(-15pt,15pt) coordinate (a1);
    \path (O)  ++(15pt,15pt)  node[draw,rectangle,text=black,inner sep=1pt] (a2) {$E$};
    \draw (O) -- (a2);
    \draw (a1) ++(0pt,15pt) coordinate (c1);
    \draw (O) -- (c1);
    \draw (a1) ++(4pt,15pt) coordinate (c2);
    \draw (O) -- (c2);
    \draw (a2) -- ++(0pt,15pt);
    \path (a1) ++(8pt,15pt) coordinate (b1);
    \path (a2) ++(-4pt,15pt) coordinate (b2);
    \path (a1) ++(-4pt,15pt) coordinate (b4);
    \draw (a2) -- ++(4pt,15pt);
    \draw (O) -- (b2);
    \draw (a2) -- (b1);
    \draw (a2) -- (b4);
    \path (O) ++(2pt,29pt) node[text=black] {$\scriptstyle \ldots$};
  \end{tikzpicture} \;+ O(\epsilon^2) \label{proof1}
\end{equation}
Keeping with our conventions, the \tikz[baseline=(E.base)] \node[draw,rectangle,text=black,inner sep=1pt] (E) {$E$}; vertices are the obvious derivatives.  Moreover:
\begin{equation}
  \begin{tikzpicture}[baseline=(X),black]
    \path coordinate (A) ++(0pt,4pt) coordinate (X);
    \path (A);
    \path (A) ++(20pt,0) coordinate (B);
    \draw (A) .. controls +(0,15pt) and +(0,15pt) .. node[fill=white,circle,inner sep=0pt,draw] {$\scriptstyle \sim$} (B);
  \end{tikzpicture}  
  =
  \begin{tikzpicture}[baseline=(X)]
    \path coordinate (A) ++(0pt,2pt) coordinate (X);
    \path (A) ++(20pt,0) coordinate (B);
    \draw[black] (A) .. controls +(0,15pt) and +(0,15pt) .. (B);
  \end{tikzpicture}
  +\epsilon \biggl(
    \begin{tikzpicture}[baseline=(X)]
    \path coordinate (A) ++(0pt,2pt) coordinate (X);
    \path (A) ++(20pt,0) coordinate (B);
    \draw[black] (A) .. controls +(0,15pt) and +(0,15pt) .. (B);
    \path (B) node[draw,rectangle,black,text=black,inner sep=1pt,fill=white] (E) {$E$};
    \draw[black] (E) -- ++(0pt,-10pt);
  \end{tikzpicture}
  +
    \begin{tikzpicture}[baseline=(X)]
    \path coordinate (A) ++(0pt,2pt) coordinate (X);
    \path (A) ++(20pt,0) coordinate (B);
    \draw[black] (A) .. controls +(0,15pt) and +(0,15pt) .. (B);
    \path (A) node[draw,rectangle,black,text=black,inner sep=1pt,fill=white] (E) {$E$};
    \draw[black] (E) -- ++(0pt,-10pt);
  \end{tikzpicture} \biggr)
  + O(\epsilon^2) \label{proof2}
\end{equation}

Finally, we consider the sum of diagrams in $\tilde U$, and show in three steps that the extra diagrams --- those with \tikz[baseline=(E.base)] \node[draw,rectangle,text=black,inner sep=1pt] (E) {$E$};\,s --- cancel to first order in $\epsilon$.  The first step in the cancellation is essentially immediate: the extra diagrams in \cref{proof1,proof2} appear with opposite signs, and the symmetry factors $\left| \Aut\Gamma\right|$ work out, so we can cancel the diagrams from \cref{proof2} with those from \cref{proof1} in which the \tikz[baseline=(E.base)] \node[draw,rectangle,text=black,inner sep=1pt] (E) {$E$}; vertex has precisely one input string.  The second cancelation is almost as quick.  The path $\gamma$ is classical, and the Green's function is a based loop in each variable, so:
\begin{equation} \label{proof3}
  \begin{tikzpicture}[baseline=(X),black]
    \path node[dot] (O) {} ++(0pt,12pt) coordinate (X);
    \draw (O) -- ++(-10pt,18pt) node[draw,rectangle,text=black,inner sep=1pt,fill=white] (E) {$E$};
    \draw (O) .. controls +(10pt,15pt) and +(0,15pt) .. ++(15pt,0);
    \draw (E) -- ++(-10pt,15pt) ++(-2pt,2pt) coordinate (n1);
    \draw (E) -- ++(-4pt,15pt);
    \draw (E) -- ++(10pt,15pt) ++(2pt,2pt) coordinate (n2);
    \path (E) ++(2pt,13pt) node[text=black] {$\scriptstyle \ldots$};
    \draw[decorate,decoration=brace] (n1) -- coordinate(n) (n2);
    \path (n) ++(0,4pt) node[anchor=base] {$\scriptstyle n$};
  \end{tikzpicture}  
  = -
  \begin{tikzpicture}[baseline=(X),black]
    \path node[white,dot] (O) {} +(0pt,12pt) coordinate (X) +(-10pt,20pt) coordinate (t) +(15pt,0) coordinate (b);
    \draw (t) .. controls +(0pt,-8pt) and +(0pt,8pt) .. (b);
    \path (O) ++(-10pt,18pt) node[draw,rectangle,text=black,inner sep=1pt,fill=white] (E) {$E$};
    \draw (E) -- ++(-10pt,15pt) ++(-2pt,2pt) coordinate (n1);
    \draw (E) -- ++(-4pt,15pt);
    \draw (E) -- ++(10pt,15pt) ++(2pt,2pt) coordinate (n2);
    \path (E) ++(2pt,13pt) node[text=black] {$\scriptstyle \ldots$};
    \draw[decorate,decoration=brace] (n1) -- coordinate(n) (n2);
    \path (n) ++(0,4pt) node[anchor=base] {$\scriptstyle n$};
  \end{tikzpicture}
  \;,
  \quad\quad\quad\quad
  \begin{tikzpicture}[baseline=(X),black]
    \path node[dot] (O) {} ++(0pt,12pt) coordinate (X);
    \draw (O) -- ++(0pt,18pt) node[draw,rectangle,text=black,inner sep=1pt,fill=white] (E) {$E$};
    \draw (E) -- ++(-10pt,15pt) ++(-2pt,2pt) coordinate (n1);
    \draw (E) -- ++(-4pt,15pt);
    \draw (E) -- ++(10pt,15pt) ++(2pt,2pt) coordinate (n2);
    \path (E) ++(2pt,13pt) node[text=black] {$\scriptstyle \ldots$};
    \draw[decorate,decoration=brace] (n1) -- coordinate(n) (n2);
    \path (n) ++(0,4pt) node[anchor=base] {$\scriptstyle n$};
  \end{tikzpicture}  = 0
  \;,
  \quad\quad\quad\quad
  n \geq 1
\end{equation}
After the first cancellation, diagrams with either side of the first part of \cref{proof3} appear exactly when the \tikz[baseline=(E.base)] \node[draw,rectangle,text=black,inner sep=1pt] (E) {$E$}; vertex has at least two input strings, and diagrams with a component like the second part of \cref{proof3} appear whenever the \tikz[baseline=(E.base)] \node[draw,rectangle,text=black,inner sep=1pt] (E) {$E$}; vertex has at least three incoming strings.  But by \cref{proof3}, all these diagrams cancel.

Only in the final cancellation does the fact that $f$ is volume-preserving play a role.  After the cancelations in the previous paragraph, the remaining diagrams with \tikz[baseline=(E.base)] \node[draw,rectangle,text=black,inner sep=1pt] (E) {$E$};\,s in them have components of the form:
\begin{equation}
   \begin{tikzpicture}[baseline=(X)]
    \path node[dot] (O) {} ++(0pt,12pt) coordinate (X);
    \draw (O) -- ++(-10pt,18pt) node[draw,rectangle,text=black,inner sep=1pt,fill=white] (E) {$E$};
    \draw (O) .. controls +(15pt,15pt) and +(10pt,15pt) .. (E);
    \draw (E) -- ++(-10pt,15pt) ++(-2pt,2pt) coordinate (n1);
    \draw (E) -- ++(4pt,15pt) ++(2pt,2pt) coordinate (n2);
    \path (E) ++(-3pt,13pt) node[text=black] {$\scriptstyle \ldots$};
    \draw[decorate,decoration=brace] (n1) -- coordinate(n) (n2);
    \path (n) ++(0,4pt) node[anchor=base] {$\scriptstyle n$};
  \end{tikzpicture}  , \quad n \geq 1
\end{equation}
But $f(q) = q + \epsilon E(q)$ is volume-preserving up to $O(\epsilon^2)$ if and only if $\frac{\partial E^i}{\partial q^i} = 0$, and by \cref{greendefn}:
\begin{equation} \label{proofn}
  \begin{tikzpicture}[baseline=(E.base)]
    \path node[dot] (O) {} ++(0pt,15pt) coordinate (X);
    \draw (O) -- ++(-10pt,18pt) node[draw,rectangle,text=black,inner sep=1pt,fill=white] (E) {$E$};
    \draw (O) .. controls +(15pt,15pt) and +(10pt,15pt) .. (E);
    \draw (E) -- ++(-10pt,15pt) +(0,3pt) node[anchor=base,text=black] {$\scriptstyle \xi_1$};
    \draw (E) -- ++(4pt,15pt) +(0,3pt) node[anchor=base,text=black] {$\scriptstyle \xi_n$};
    \path (E) ++(-3pt,13pt) node[text=black] {$\scriptstyle \ldots$};
  \end{tikzpicture} 
  = - \int_0^t \delta(0) \,\frac{\partial^n}{\partial q^{j_1}\dots\partial q^{j_n}}\biggl[ \frac{\partial E^i}{\partial q^j}\biggr]  \biggr|_{q = \gamma(\tau)}d\tau = -\int_0^t \delta(0) \cdot 0\,d\tau = 0
\end{equation}
This completes the proof of \cref{coordfreethm}.\qedhere

\Section{A Fubini theorem for formal path integrals} \label{fubini}

In this section, we prove the following ``composition law'' for formal path integrals:

\begin{thm} \label{fubinithm}
  Fix a Lagrangian $L$ on the configuration space $\RR^d$ with $\frac{\partial^2 L}{\partial v^2}$ everywhere positive-definite.  Let $\gamma: [t_0,t_1] \to \RR^d$ be classical and nonfocal, and pick $t\in [t_0,t_1]$ such that both restrictions $\gamma_0 = \gamma|_{[t_0,t]}$ and $\gamma_1 = \gamma|_{[t,t_1]}$
  are nonfocal.  Then $\gamma(t)$ is a nondegenerate critical point for $S_{\gamma_0}(t_0,q_0,t,-) + S_{\gamma_1}(t,-,t_1,q_1)$.
  Furthermore, suppose that the formal path integrals for $L$ have no ultraviolet divergences.  Then:
  \begin{equation}  \label{complaw}
    \int_{\gamma(t)}^{\formal} U_{\gamma_0}(t_0,q_0,t,q)\,U_{\gamma_1}(t,q,t_1,q_1)\,\d q = U_\gamma(t_0,q_0,t_1,q_1)
  \end{equation}
\end{thm}
The integral in \cref{complaw} is as in \cref{formalintdefn}.

\Cref{fubinithm} provides justification for \cref{formalintdefn,pathintdefn} and so is interesting in its own right.  But it also indicates how to define formal path integrals in the absence of global coordinate systems.  Let $\Nn$ be some classical configuration space with Lagrangian $L$ and volume form $\dVol$, and let $\gamma$ be a classical nonfocal path in $\Nn$.  By \cite{Milnor1963}, sufficiently small pieces of $\gamma$ are nondegenerate, and by \cite{Moser1965} each sufficiently small piece can be included in a coordinate chart such that $\dVol$ is the pullback along the chart of the canonical volume form on $\RR^{\dim \Nn}$. (In \cite{Moser1965} this is proved even if $\dVol$ is allowed to depend on the external time parameter $\tau$, provided the coordinates also are allowed to depend on $\tau$.)  Then we can calculate the formal path integrals for each piece, and by \cref{coordfreethm} their values do not depend on the chosen charts.  To define the path integral for $\gamma$, we follow \cref{complaw} and integrate the contributions from each piece.  By interleaving different ways to cut $\gamma$ into short pieces, we see via \cref{fubinithm} that the total path integral for $\gamma$ does not depend on the choice of cuts.

In \cref{clascomplaw} we prove the first part of \cref{fubinithm} and verify \cref{complaw} up to a factor of $(1 + O(\hbar))$.  We introduce some more notation to our Feynman diagrammatic repertoire in \cref{somederiv} and use it to study the derivatives of the function $U_\gamma(t_0,q_0,t_1,q_1)$.   A diagrammatic calculation in \cref{somediagrams} verifies \cref{complaw} to all orders.

For reference, we recall \cref{formalintdefn,pathintdefn}.  We have, suppressing the $t_a$-dependence:
\begin{equation} \label{comparethisA}
  U_{\gamma}(q_0,q_1) = (2\pi i \hbar)^{-d/2} \exp\bigl( -(i\hbar)^{-1} S_{\gamma}(q_0,q_1)\bigr) (-1)^{\eta(\gamma)} \sqrt{ \left| \det \frac{\partial^2[-S_\gamma]}{\partial q_0\partial q_1}\right|} \sum_\Gamma \frac{(i\hbar)^{-\chi(\Gamma)}\ev(\Gamma)}{\left|\Aut\Gamma\right|} \hspace{-1ex}\mbox{}
\end{equation}
In \cref{comparethisA}, the Feynman diagrams are evaluated via the Feynman rules in \cref{pathintdefn}.  On the other hand:
\begin{multline} \label{comparethisB}
  \int^{\formal}_{q_{\cp}(q_0,q_1)} U_{\gamma_0}(q_0,q)\,U_{\gamma_1}(q,q_1)\,\d q = \\
   = \int^{\formal}_{q_{\cp}(q_0,q_1)} (2\pi i \hbar)^{-d/2} \exp\bigl( -(i\hbar)^{-1}S_{\gamma_0}(q_0,q)\bigr) \sqrt{ \left| \det \frac{\partial^2[-S_{\gamma_0}]}{\partial q_0\partial q}\right|} \sum_\Gamma \frac{(i\hbar)^{-\chi(\Gamma)}\ev_0(\Gamma)}{\left|\Aut\Gamma\right|} \times \\
   \times (2\pi i \hbar)^{-d/2} \exp\bigl( -(i\hbar)^{-1}S_{\gamma_1}(q,q_1)\bigr) \sqrt{ \left| \det \frac{\partial^2[-S_{\gamma_1}]}{\partial q\partial q_1}\right|} \sum_\Gamma \frac{(i\hbar)^{-\chi(\Gamma)}\ev_1(\Gamma)}{\left|\Aut\Gamma\right|} \times \d q =
   \\
   = (2\pi i \hbar)^{d/2}(2\pi i \hbar)^{-d/2}(2\pi i \hbar)^{-d/2} \times \exp\bigl( -(i\hbar)^{-1} \bigl(S_{\gamma_0}(q_0,q) + S_{\gamma}(q,q_1)\bigr) \times \\
   \times \left| \det \frac{\partial^2[-S_{\gamma_0}]}{\partial q_0\partial q}\right|^{1/2} \left| \det \frac{\partial^2[-S_{\gamma_1}]}{\partial q\partial q_1}\right|^{1/2} \left| \det \frac{\partial^2[S_{\gamma_0} + S_{\gamma_1}]}{\partial q^2}\right|^{-1/2} \times \left( 1 + \sum_\Gamma\right)
\end{multline}
In \cref{comparethisB}, the Feynman diagrams in the middle-hand side are evaluated via the Feynman rules ``$\ev_a$'' for $U_{\gamma_a}$. The right-hand side is evaluated at $q = q_{\cp}(q_0,q_1)$ a critical point of $S_{\gamma_0}(q_0,-) + S_{\gamma_1}(-,q_1)$, provided it can be chosen uniquely and is nondegenerate.  The sum of diagrams on the right-hand side follows complicated Feynman rules that we will write out in \cref{somediagrams}, and involves both the Feynman rules from \cref{pathintdefn} for $U_{\gamma_a}$ and the Feynman rules from \cref{formalintdefn} for the integral.

\subsection{The classical composition law} \label{clascomplaw}

In this section we prove the first statement of \cref{fubinithm}.  We also verify \cref{complaw} to within an accuracy of a factor of $(1 + O(\hbar))$: the product of determinants is correct, and the Morse indexes match.  We begin with the classical composition law for Lagrangian mechanics.

\begin{lemma} \label{classicalglue}
  Let $L: \RR \times \T\RR^d \to \RR$ be a Lagrangian such that $\frac{\partial^2 L}{\partial v^2}$ is everywhere positive definite.  Fix $t_0 < t < t_1$ and choose open neighborhoods $\Oo_0,\Oo,\Oo_1 \subset \RR^d$.  Suppose that we have families $\gamma_0 : \Oo_0\times \Oo \times [t_0,t] \to \RR^d$ and $\gamma_1 : \Oo\times \Oo_1 \times [t,t_1] \to \RR^d$ of classical paths --- i.e.\ for each $(q_0,q,q_1)\in \Oo_0 \times\Oo \times \Oo_1$ the paths $\gamma_0(q_0,q;-)$ and $\gamma_1(q,q_1;-)$ are classical.  Define the Hamilton principal functions $S_a = \Aa(\gamma_a)$ for $a=0,1$.  Then the critical points of $S_0(q_0,-) + S_1(-,q_1)$ are precisely those points $q\in\Oo$ such that the ``glued-together'' path $\gamma(q_0,q,q_1;-) : [t_0,t_1] \to\RR^d$ given by 
  \begin{equation} \gamma(q_0,q,q_1;\tau) = \begin{cases}
    \gamma_0(q_0,q;\tau), & \tau\leq t \\
    \gamma_1(q,q_1;\tau), & \tau \geq t
  \end{cases} 
  \label{gluedpatheqn} \end{equation}
  is smooth and classical.
\end{lemma}

\begin{proof}
The Euler-Lagrange equations are local and closed in $\tau$, so $\gamma(q_0,q,q_1;-)$ is classical if it is smooth.  Since the Euler-Lagrange equations are nondegenerate second-order, $\gamma(q_0,q,q_1;-)$ is smooth if and only if $\dot\gamma_0(q_0,q;t) = \dot\gamma_1(q,q_1;t)$.  But recall that $L$ is convex on fibers of $\T\RR^d \to \RR^d$, as $\frac{\partial^2 L}{\partial v^2}$ is positive-definite.  Therefore:
\begin{equation}
  \frac{\partial L}{\partial v}\biggr|_{(\tau,v_0,q)} = \frac{\partial L}{\partial v}\biggr|_{(\tau,v_1,q)} \text{ if and only if } v_0 = v_1.
\end{equation}
However, $\frac{\partial L}{\partial v}\bigl(t,\dot\gamma_0(t),\gamma_0(t)\bigr) = \frac{\partial S_0}{\partial q}(q_0,q)$, and $\frac{\partial L}{\partial v}\bigl(t,\dot\gamma_1(t),\gamma_1(t)\bigr) = - \frac{\partial S_1}{\partial q}(q,q_1)$.  Thus $\dot\gamma_0(q_0,q;t) = \dot\gamma_1(q,q_1;t)$ if and only if $\frac{\partial}{\partial q}[S_0(q_0,q) + S_1(q,q_1)] = 0$.  \end{proof}

\begin{lemma}
  Suppose that $\gamma,\gamma_0,\gamma_1$ are as in the statement of \cref{fubinithm}, and define the corresponding Hamilton principal functions $S_\gamma(q_0,q_1) = S_\gamma(t_0,q_0,t_1,q_1)$, $S_0(q_0,q) = S_{\gamma_0}(t_0,q_0,t,q)$, and $S_1(q,q_1) = S_{\gamma_1}(t,q,t_1,q_1)$.  Moreover, set $q_{\cp}(q_0,q_1) = \gamma(t_0,q_0,t_1,q_1;t)$.  Then $q_{\cp}$ is a nondegenerate critical point for $S_0(q_0,-) + S_1(-,q_1)$, and:
\begin{equation} \label{d2S12again}
  \frac{\partial^2 [-S_{\gamma}]}{\partial q_0^i\partial q_1^j} = \left.\frac{\partial^2 [-S_0]}{\partial q_0^i\partial q^l}\frac{\partial^2 [-S_1]}{\partial q^k\partial q_1^j} \biggl(\Bigl( \frac{\partial^2[S_0 + S_1]}{\partial q\partial q}\Bigr)^{-1}\biggr)^{kl} \right|_{q = q_{\cp}}  
\end{equation}
\end{lemma}

\begin{proof}
The additivity of the action together with \cref{classicalglue} provide:
\begin{gather}
  \label{add} S_{\gamma}(q_0,q_1)  = \bigl[ S_0(q_0,q) + S_1(q,q_1)\bigr]_{q = q_{\cp}(q_0,q_1)} \\
  \label{dif} 0  = \left[ \frac{\partial S_0}{\partial q}(q_0,q) + \frac{\partial S_1}{\partial q}(q,q_1) \right]_{q = q_{\cp}(q_0,q_1)}
\end{gather}
We evaluate the second derivative of \cref{add} with respect to $q_0,q_1$:
\begin{equation} \label{d2S12}
  \hspace{-1in} \frac{\partial^2 S_\gamma}{\partial q_0^i\partial q_1^j} = \left[ \frac{\partial^2 S_0}{\partial q_0^i\partial q^l} \frac{\partial q_{\cp}^l}{\partial q_1^j} + \frac{\partial^2 S_1}{\partial q^k\partial q_1^j} \frac{\partial q_{\cp}^k}{\partial q_0^i} + \frac{\partial^2[S_0 + S_1]}{\partial q^k\partial q^l} \frac{\partial q_{\cp}^k}{\partial q_0^i} \frac{\partial q_{\cp}^l}{\partial q_1^j} + \frac{\partial S_0}{\partial q^l}\frac{\partial^2 q_{\cp}^l}{\partial q_0^i\partial q_1^j}+ \frac{\partial S_1}{\partial q^k}\frac{\partial^2 q_{\cp}^k}{\partial q_0^i\partial q_1^j} \right]_{q = q_{\cp}} \hspace{-1in}
\end{equation}
The sum of the last two terms vanishes by \cref{dif}.  Differentiating \cref{dif} with respect to $q_0$ or $q_1$ gives:
\begin{equation} \label{intermediatestep}
  0 = \left[\frac{\partial^2 S_0}{\partial q_0^i\partial q^l}+ \frac{\partial^2[S_0 + S_1]}{\partial q^k\partial q^l} \frac{\partial q_{\cp}^k}{\partial q_0^i}\right]_{q = q_{\cp}}
  =
  \left[\frac{\partial^2 S_1}{\partial q^k\partial q_1^j}+ \frac{\partial^2[S_0 + S_1]}{\partial q^k\partial q^l} \frac{\partial q_{\cp}^l}{\partial q_1^j}\right]_{q = q_{\cp}}
\end{equation}
Since $\gamma_0,\gamma_1$ are nonfocal, the matrices $\frac{\partial^2 S_0}{\partial q_0\partial q}$ and $\frac{\partial^2 S_1}{\partial q\partial q_1}$ are invertible, and so the other terms \cref{intermediatestep} are as well.  
In particular, $q_{\cp}$ is a nondegenerate critical point of $S_0(q_0,-)+S_1(-,q_1)$.  We rearrange \cref{intermediatestep}:
\begin{align} \label{thisisadumblabel1}
  \frac{\partial q_{\cp}^k}{\partial q_0^i}  & = - \left.\frac{\partial^2 S_0}{\partial q_0^i\partial q^l} \biggl(\Bigl( \frac{\partial^2}{\partial q^2}[S_0 + S_1]\Bigr)^{-1}\biggr)^{kl} \right|_{q = q_{\cp}}
  \\
  \frac{\partial q_{\cp}^l}{\partial q_1^j} & = - \left.\frac{\partial^2 S_1}{\partial q^k\partial q_1^j} \biggl(\Bigl( \frac{\partial^2}{\partial q^2}[S_0 + S_1]\Bigr)^{-1}\biggr)^{kl} \right|_{q = q_{\cp}}
  \label{thisisadumblabel2}
\end{align}
Substituting \cref{thisisadumblabel1,thisisadumblabel2} into \cref{d2S12} gives \cref{d2S12again}.
\end{proof}

Since $q_{\cp}$ is nondegenerate as a critical point, the formal integral $\int_{q_{\cp}}^{\formal} U_{\gamma_0}U_{\gamma_1}$ is well-defined.  Moreover, the absolute value of the determinant of the left-hand side of \cref{d2S12again} is the square of the degree-zero part of $U_{\gamma}(q_0,q_1)$, whereas the absolute value of the determinant of the right-hand side is the square of the degree-zero part of $\int^{\formal} U_{\gamma_0}U_{\gamma_1}$.  Recall that the formal integral $\int^{\formal} \d q$ introduces a factor of $(2\pi i\hbar)^{d/2}$.  Then to confirm that $\int^{\formal} U_{\gamma_0}U_{\gamma_1} = U_{\gamma} \times \bigl(1 + O(\hbar)\bigr)$, we need only to check the $(-1)^{\eta}$ factors that appear in \cref{formalintdefn,Udefined}.

\begin{lemma}
  Let $\gamma: [t_0,t_1] \to \RR^d$ be classical and nonfocal with nonfocal restrictions $\gamma_0:[t_0,t] \to \RR^d$ and $\gamma_1:[t,t_1] \to \RR^d$, as in \cref{fubinithm}, and with Morse indexes $\eta(\gamma),\eta(\gamma_0),\eta(\gamma_1)$.  Let $\eta(q_{\cp})$ be the Morse index of $q_{\cp} = \gamma(t)$ with respect to the function $q \mapsto S_0(q_0,q) + S_1(q,q_1)$.  Then $\eta(\gamma) = \eta(\gamma_0) + \eta(q_{\cp}) + \eta(\gamma_1)$.
\end{lemma}
\begin{proof}
Recall that for any nonfocal classical path, by \cref{morselemma} its Morse index is the dimension of any maximal subspace of the space of based loops on which $\Aa^{(2)}$ is negative definite.  On the other hand, $\eta(q_{\cp})$ is the dimension of any maximal subspace of $\RR^d$ on which the Hessian $\frac{\partial^2[S_0+S_1]}{\partial q^2}$ is negative definite.  For $a=0,1$, there are natural embeddings that extend based loops by $0$:
\begin{align*}
  \extend_0&: \{\text{based loops with domain }[t_0,t]\} \mono \{\text{based loops with domain }[t_0,t_1]\}\\
  \extend_1&: \{\text{based loops with domain }[t,t_1]\} \mono \{\text{based loops with domain }[t_0,t_1]\}
\end{align*}
Define also the map $\extend_{\cp}: \RR^d\to \{\text{based loops with domain }[t_0,t_1]\}$ by:
\[ \extend_{\cp}(x)^i(\tau) = \begin{cases}
  x^j\,\frac{\partial \gamma_0^i}{\partial q^j}(\tau), & \tau \leq t \\
  x^j\,\frac{\partial \gamma_1^i}{\partial q^j}(\tau), & \tau \geq t \\
\end{cases} \]
The continuity of $\exp_{\cp}(x)$ follows from the equality $\frac{\partial\gamma_a^i}{\partial q^j}(t) = \delta^i_j$, $a = 0,1$.

Then for $a = 0,1$, it's clear that $\Aa^{(2)}(\gamma)\cdot \extend_a(\xi)\,\extend_a(\zeta) = \Aa^{(2)}(\gamma_a)\cdot \xi\zeta$.  Moreover, $\Aa^{(2)}(\gamma)\cdot \extend_{\cp}(x)\,\extend_{\cp}(z) = \frac{\partial^2[S_0+S_1]}{\partial q^i\partial q^j}\, x^iz^j$, by \cref{dSdQQ} or by direct calculation.  On the other hand, the images of the various extension maps are orthogonal: $\Aa^{(2)}(\gamma)\cdot \extend_0(\xi)\,\extend_1(\zeta) = 0 = \Aa^{(2)}(\gamma)\cdot \extend_a(\xi)\,\extend_{\cp}(z)$.  Let $\Vv_0$ (resp.\ $\Vv_1$) be some maximal subspace of the space of based loops with domain $[t_0,t]$ ($[t,t_1]$) on which $\Aa^{(2)}(\gamma_0)$ ($\Aa^{(2)}(\gamma_1)$) is negative definite, and let $\Vv_{\cp}$ be a maximal subspace of $\RR^d$ on which $\frac{\partial^2[S_0+S_1]}{\partial q^2}$ is negative definite.  Then $\Aa^{(2)}(\gamma)$ is negative definite on $\extend_0(\Vv_0) + \extend_{\cp}(\Vv_{\cp}) + \extend_1(\Vv_1)$.  Thus $\eta(\gamma_0) + \eta(q_{\cp}) + \eta(\gamma_1) \leq \eta(\gamma)$.

On the other hand, let 
\begin{align*}
  \res_0&: \{\text{paths with domain }[t_0,t_1]\} \epi \{\text{paths with domain }[t_0,t]\}\\
  \res_1&: \{\text{paths with domain }[t_0,t_1]\} \epi \{\text{paths with domain }[t,t_1]\}
\end{align*}
 be the natural restriction maps.  If $\xi:[t_0,t_1] \to \RR^d$, $\zeta_0: [t_0,t]\to \RR^d$, and $\zeta_1: [t,t_1] \to \RR^d$ are based loops, then $\Aa^{(2)}(\gamma) \cdot \xi\extend_a(\zeta_a) = \Aa^{(2)}(\gamma_a)\cdot \res_a(\xi)\zeta_a$, and if $z\in \RR^d$, then $\Aa^{(2)}(\gamma) \cdot \xi\, \extend_{\cp}(z) = \frac{\partial^2[S_0+S_1]}{\partial q^i\partial q^j} \, \xi^i(t)\,z^j$. Indeed:
\[\xi = \extend_0\bigl( \res_0\bigl( \xi - \extend_{\cp}\bigl(\xi(t)\bigr)\bigr)\bigr) + \extend_{\cp}\bigl(\xi(t)\bigr) +  \extend_1\bigl( \res_1\bigl( \xi - \extend_{\cp}\bigl(\xi(t)\bigr)\bigr)\bigr)\]
Suppose that $\xi:[t_0,t_1] \to \RR^d$ is a based loop such that $\Aa^{(2)}(\gamma)\cdot \xi\zeta \leq 0$ for every $\zeta \in \extend_0(\Vv_0) + \extend_{\cp}(\Vv_{\cp}) + \extend_1(\Vv_1)$.  Then $\res_a(\xi) \in \Vv_a$ and $\xi(t)\in \Vv_{\cp}$ by maximality.  Thus $\extend_0(\Vv_0) + \extend_{\cp}(\Vv_{\cp}) + \extend_1(\Vv_1)$ is a maximal negative-definite subspace of the space of based loops with domain $[t_0,t_1]$.  Therefore:
\begin{equation}\label{etaeqn}
  \eta(\gamma_0) + \eta(q_{\cp}) + \eta(\gamma_1) = \eta(\gamma)
\end{equation}
\end{proof}

\subsection{Some derivatives of the formal path integral} \label{somederiv}

Pick a classical nonfocal path $\gamma: [t_0,t_1] \to \RR^d$, extended to a family that depends smoothly on its boundary conditions $\gamma(t_0) = q_0$ and $\gamma(t_1) = q_1$.  We have already established solid vertical lines in our Feynman diagrams as referring to the vector space of all paths in $\RR^d$ with domain $[t_0,t_1]$.  (We will amend that convention in the next section.)  We now declare that a dashed line carries simply a vector in $\RR^d$, or equivalently an index $i = 1,\dots,d$.

Let $\Gamma$ be a Feynman diagram, possibly not closed.  Then its value depends on the classical path $\gamma$, and in particular on the boundary conditions $q_0,q_1$.  We represent differentiation with a dotted circle:
\begin{equation}
  \ev\Biggl(
  \begin{tikzpicture}[baseline=(G.base)]
    \path node[shape=rectangle,draw,inner sep=2pt] (G) {$\Gamma$} ++(12pt,0) coordinate (H);
    \draw (G.west) ++(0,1pt) .. controls +(-10pt,0pt) and +(5pt,-5pt) .. ++(-25pt,20pt);
    \draw (G.east) ++(0,1pt) .. controls +(10pt,0pt) and +(-5pt,-5pt) .. ++(25pt,20pt);
    \draw (G.east) ++(0,-1pt) .. controls +(10pt,0pt) and +(-5pt,5pt) .. ++(25pt,-20pt);
    \node [draw,dotted,thick, circle through=(H)] (C) at (G) {};
    \node[anchor=south west] at (C.north) {$\scriptstyle q_a$};
    \draw [dashed] (C.north) -- ++(0,15pt) +(0,3pt) node[anchor=base] {$\scriptstyle i$};
  \end{tikzpicture}
  \Biggr)
  = \frac{\partial}{\partial q_a^i} \bigl[ \ev(\Gamma)\bigr] \;,\quad\quad a = 0,1
\end{equation}
If $\Gamma$ consists of two components $\Gamma_1,\Gamma_2$, possibly connected to each other, then the product rule can be written graphically as:
\begin{align}
  \begin{tikzpicture}[baseline=(D.base)]
    \node (D) {$\scriptstyle \cdots$};
    \path (D) ++(-3pt,25pt) node[shape=rectangle,draw,inner sep=2pt] (G1) {$\Gamma_1$};
    \path (D) ++(-3pt,-25pt) node[shape=rectangle,draw,inner sep=2pt] (G2) {$\Gamma_2$};
    \draw (G1) .. controls +(-4pt,-15pt) and +(-4pt,15pt) .. (G2);
    \draw (G1) .. controls +(-10pt,-15pt) and +(-10pt,15pt) .. (G2);
    \draw (G1) .. controls +(10pt,-15pt) and +(10pt,15pt) .. (G2);
    \draw[dotted, thick] (D) ++(-3pt,0pt) ellipse (20pt and 40pt);
  \end{tikzpicture}
  =
  \begin{tikzpicture}[baseline=(D.base)]
    \node (D) {$\scriptstyle \cdots$};
    \path (D) ++(-3pt,25pt) node[shape=rectangle,draw,inner sep=2pt] (G1) {$\Gamma_1$};
    \path (D) ++(-3pt,-25pt) node[shape=rectangle,draw,inner sep=2pt] (G2) {$\Gamma_2$};
    \draw (G1) .. controls +(-4pt,-15pt) and +(-4pt,15pt) .. (G2);
    \draw (G1) .. controls +(-10pt,-15pt) and +(-10pt,15pt) .. (G2);
    \draw (G1) .. controls +(10pt,-15pt) and +(10pt,15pt) .. (G2);
    \draw[dotted, thick] (G1) circle (15pt);
  \end{tikzpicture}
  +
  \begin{tikzpicture}[baseline=(D.base)]
    \node (D) {$\scriptstyle \cdots$};
    \path (D) ++(-3pt,25pt) node[shape=rectangle,draw,inner sep=2pt] (G1) {$\Gamma_1$};
    \path (D) ++(-3pt,-25pt) node[shape=rectangle,draw,inner sep=2pt] (G2) {$\Gamma_2$};
    \draw (G1) .. controls +(-4pt,-15pt) and +(-4pt,15pt) .. (G2);
    \draw (G1) .. controls +(-10pt,-15pt) and +(-10pt,15pt) .. (G2);
    \draw (G1) .. controls +(10pt,-15pt) and +(10pt,15pt) .. (G2);
    \draw[dotted, thick] (G2) circle (15pt);
  \end{tikzpicture}  
\end{align}

Suppose then that $\Gamma_1$ is a subdiagram of $\Gamma$ whose images $\Gamma_1,\dots,\Gamma_n$ under the group of automorphisms of $\Gamma$ do not intersect, so that $\Gamma = \bar\Gamma \cup \Gamma_1\cup \dots \cup \Gamma_n$ (we do allow automorphism of $\Gamma$ to induce nontrivial automorphisms of $\Gamma_1$).  Then: 
\begin{multline}
  \tikz[baseline=(G.base)]    \node (G) [shape=ellipse,draw,dotted,thick,inner sep=3pt] {$\Gamma$};
   = 
  \tikz[baseline=(G.base)]    \node (G) [shape=ellipse,draw,dotted,thick,inner sep=3pt] {$\bar \Gamma\,\Gamma_1\,\Gamma_2\cdots\Gamma_n$};
  = 
  \tikz[baseline=(G.base)]    \node (G) [shape=ellipse,draw,dotted,thick,inner sep=2pt] {$\bar \Gamma$}; \,\Gamma_1\,\Gamma_2\cdots\Gamma_n
  +
  \bar \Gamma\, \tikz[baseline=(G.base)]    \node (G) [shape=ellipse,draw,dotted,thick,inner sep=2pt] {$\Gamma_1$};\,\Gamma_2\cdots\Gamma_n
  + \cdots +
  \bar \Gamma\,\Gamma_1\,\Gamma_2\,\cdots\,\tikz[baseline=(G.base)]    \node (G) [shape=ellipse,draw,dotted,thick,inner sep=2pt] {$\Gamma_n$};
  = \\ = 
  \tikz[baseline=(G.base)]    \node (G) [shape=ellipse,draw,dotted,thick,inner sep=2pt] {$\bar \Gamma$}; \,\Gamma_1\,\Gamma_2\cdots\Gamma_n
  +
  n\,\bar \Gamma\, \tikz[baseline=(G.base)]    \node (G) [shape=ellipse,draw,dotted,thick,inner sep=2pt] {$\Gamma_1$};\,\Gamma_2\cdots\Gamma_n
\end{multline}
It is an elementary counting lemma that $
  \Aut\left( \bar \Gamma\, \tikz[baseline=(G.base)]    \node (G) [shape=ellipse,draw,dotted,thick,inner sep=2pt] {$\Gamma_1$};\,\Gamma_2\cdots\Gamma_n\right) = \frac1n \Aut\left( \bar \Gamma\,\Gamma_1\,\Gamma_2\cdots\Gamma_n\right)
$.
  From this observation, we derive the following fundamental result:
\begin{lemma}[Product Rule] \label{productrulelemma}
  For $a = 0,1$, we have: 
  \begin{align} \frac{\partial}{\partial q_a} \sum_{\Gamma} \frac{(i\hbar)^{\chi(\Gamma)}\ev(\Gamma)}{\Aut \Gamma} = \sum_{\substack{\Gamma \textrm{ with one \, \begin{tikzpicture}[baseline=(O.base)]
    \node[dot] (O) {};
    \draw (O) -- ++(-5pt,8pt);
    \draw (O) -- ++(-2pt,9pt);
    \node at (1pt,9pt) {.};
    \node at (3pt,9pt) {.};
    \node at (5pt,9pt) {.};
    \draw (O) -- ++(6pt,7pt);
    \draw[dotted,thick] (O) circle (5pt);
  \end{tikzpicture}
   or \,
  \begin{tikzpicture}[baseline=(O.base)]
    \path coordinate (O) ++(0,-5pt) coordinate (A) ++(10pt,0) coordinate (B);
    \draw (A) .. controls +(0,10pt) and +(0,10pt) .. coordinate (C) (B);
    \draw[dotted,thick] (C) circle (4pt);    
  \end{tikzpicture}}}} \frac{(i\hbar)^{\chi(\Gamma)}\ev(\Gamma)}{\Aut \Gamma} \end{align}
  The sum on the right-hand side ranges over diagrams with precisely one differentiated basic subgraph --- either a single differentiated vertex (of valence three or more) or a single differentiated edge.
\end{lemma}

These basic derivatives are then straightforward.  Assume that $\xi_1,\dots,\xi_n$ do not depend on $q_a$.  Then: 
\begin{align}
    \begin{tikzpicture}[baseline=(X)]
    \path node[dot] (O) {} ++(0pt,4pt) coordinate (X);
    \draw (O) -- ++(-15pt,15pt) +(0,3pt) node[anchor=base] {$\scriptstyle \xi_1$};
    \draw (O) -- ++(-6pt,15pt) +(0,3pt) node[anchor=base] {$\scriptstyle \xi_2$};
    \draw (O) -- ++(15pt,15pt) +(0,3pt) node[anchor=base] {$\scriptstyle \xi_n$};
    \path (O) ++(4pt,13pt) node {$\scriptstyle \ldots$};
    \path (O) node (C) [draw,circle,dotted,thick,inner sep=5pt] {};
    \draw[dashed] (C) .. controls +(-10pt,5pt) and +(3pt,-10pt) .. ++(-30pt,25pt) +(0,3pt) node[anchor=base] {$\scriptstyle j$};
    \path (C.west) node[anchor=east] {$\scriptstyle q_a$};
  \end{tikzpicture}
  & = \frac{\partial}{\partial q_a^j} \left[ -\int_{t_0}^{t_1} \prod_{k=1}^n \left. \left( \dot\xi_k^{i_k}(\tau) \frac{\partial}{\partial v^{i_k}}  +  \xi_k^{i_k}(\tau) \frac{\partial}{\partial q^{i_k}} \right) L \right|_{\gamma(\tau)} d\tau \right] \\
    & = -\int_{t_0}^{t_1} \frac{\partial}{\partial q_a^j} \prod_{k=1}^n \left. \left( \dot\xi_k^{i_k}(\tau) \frac{\partial}{\partial v^{i_k}}  +  \xi_k^{i_k}(\tau) \frac{\partial}{\partial q^{i_k}} \right) L \right|_{\gamma(\tau)} d\tau \displaybreak[0] \\
    & = -\int_{t_0}^{t_1} \left( \frac{\partial \dot \gamma^i}{\partial q_a^j} \frac{\partial}{\partial v^i} +  \frac{\partial \gamma^i}{\partial q_a^j} \frac{\partial}{\partial q^i} \right) \prod_{k=1}^n \left. \left( \dot\xi_k^{i_k}(\tau) \frac{\partial}{\partial v^{i_k}}  +  \xi_k^{i_k}(\tau) \frac{\partial}{\partial q^{i_k}} \right) L \right|_{\gamma(\tau)} d\tau \\
    & = 
     \begin{tikzpicture}[baseline=(X)]
    \path node[dot] (O) {} ++(0pt,4pt) coordinate (X);
    \draw (O) -- ++(-15pt,15pt) +(0,3pt) node[anchor=base] {$\scriptstyle \xi_1$};
    \draw (O) -- ++(-6pt,15pt) +(0,3pt) node[anchor=base] {$\scriptstyle \xi_2$};
    \draw (O) -- ++(15pt,15pt) +(0,3pt) node[anchor=base] {$\scriptstyle \xi_n$};
    \path (O) ++(4pt,13pt) node {$\scriptstyle \ldots$};
    \path (O) ++(-30pt,10pt) node[inner sep=1pt] (gamma) {$\gamma$};
    \draw (O) -- (gamma);
    \path (gamma) node (C) [draw,circle,dotted,thick,inner sep=5pt] {};
    \draw[dashed] (C) -- ++(0,18pt) +(0,3pt) node[anchor=base] {$\scriptstyle j$};
    \path (C.north) ++(1pt,3pt) node[anchor= east] {$\scriptstyle q_a$};
  \end{tikzpicture} \label{dAdQ}
\end{align}
since the only $q_a$ dependence is in the classical path $\gamma$.   We mention that the Euler-Lagrange equations assert only that the univalent vertex vanishes on based loops; $\frac{\partial \gamma}{\partial q_a}$ does not vanish at both endpoints.

We can compute the derivative of the Green's function from the product rule.  Recall \cref{greendefn}:
\begin{equation}
  \begin{tikzpicture}[baseline=(X)]
    \path node[dot] (O) {} ++(0pt,8pt) coordinate (X);
    \draw (O) -- ++(-10pt,18pt) +(0,3pt) node[anchor=base] {$\scriptstyle \xi$};
    \draw (O) .. controls +(10pt,15pt) and +(0,15pt) .. ++(15pt,0);
  \end{tikzpicture}   
  = -
  \begin{tikzpicture}[baseline=(X)]
    \path coordinate (O) +(0pt,8pt) coordinate (X) +(-10pt,20pt) coordinate (t) +(15pt,0) coordinate (b);
    \draw (t) .. controls +(0pt,-8pt) and +(0pt,8pt) .. (b);
    \path (t) +(0,3pt) node[anchor=base] {$\scriptstyle \xi$};
  \end{tikzpicture}     
  \text{, provided $\xi$ is a based loop.}
\label{greendefngraphical} \end{equation} 
Let $\xi$ be a based loop, and consider the derivative of \cref{greendefngraphical}:
\begin{align} 0 =
  \begin{tikzpicture}[baseline=(X)]
    \path node[dot] (O) {} +(0pt,8pt) coordinate (X) +(0pt,25pt) coordinate (T);
    \draw (O) -- ++(10pt,18pt) +(0,3pt) node[anchor=base] {$\scriptstyle \xi$};
    \draw (O) .. controls +(-5pt,15pt) and +(5pt,15pt) .. coordinate (t) ++(-15pt,0);
    \node[draw,dotted,thick,inner sep=2pt,ellipse,fit = (O) (t)] (C) {};        
    \draw[dashed] (C.north) -- (C.north |- T);
  \end{tikzpicture} 
  =
  \begin{tikzpicture}[baseline=(X)]
    \path node[dot] (O) {} +(0pt,8pt) coordinate (X) +(0pt,25pt) coordinate (T);
    \draw (O) -- ++(10pt,18pt) +(0,3pt) node[anchor=base] {$\scriptstyle \xi$};
    \draw (O) .. controls +(-5pt,15pt) and +(5pt,15pt) .. coordinate (t) ++(-15pt,0);
    \node[draw,dotted,thick,inner sep=3pt,circle,fit =(t)] (C) {};        
    \draw[dashed] (C.north) -- (C.north |- T);
  \end{tikzpicture} 
  +  
  \begin{tikzpicture}[baseline=(X)]
    \path node[dot] (O) {} +(0pt,8pt) coordinate (X) +(0pt,25pt) coordinate (T);
    \draw (O) -- ++(10pt,18pt) +(0,3pt) node[anchor=base] {$\scriptstyle \xi$};
    \draw (O) .. controls +(-5pt,15pt) and +(5pt,15pt) .. coordinate (t) ++(-15pt,0);
    \node[draw,dotted,thick,inner sep=2pt,circle,fit =(O)] (C) {};        
    \draw[dashed] (C.north) -- (C.north |- T);
  \end{tikzpicture}  \label{derGreen1} \end{align}
We contract with $G$, which is a based loop in each variable, and use \cref{dAdQ,greendefngraphical}:
\begin{equation}
  \begin{tikzpicture}[baseline=(X)]
    \path coordinate (O) +(0pt,8pt) coordinate (X) +(0pt,30pt) coordinate (T);
    \draw (O) .. controls +(5pt,15pt) and +(-5pt,15pt) .. coordinate (t) ++(30pt,0);
    \node[draw,dotted,thick,inner sep=3pt,circle,fit =(t)] (C) {};        
    \draw[dashed] (C.north) -- (C.north |- T);
    \path (C.north) ++(0,2pt) node[anchor=west] {$\scriptstyle q_a$};
  \end{tikzpicture}
  = -
  \begin{tikzpicture}[baseline=(X)]
    \path node[dot] (O) {} +(0pt,8pt) coordinate (X) +(0pt,30pt) coordinate (T);
    \draw (O) .. controls +(5pt,15pt) and +(-5pt,15pt) .. ++(20pt,0);
    \draw (O) .. controls +(-5pt,15pt) and +(5pt,15pt) .. coordinate (t) ++(-20pt,0);
    \node[draw,dotted,thick,inner sep=3pt,circle,fit =(t)] (C) {};        
    \draw[dashed] (C.north) -- (C.north |- T);
    \path (C.north) ++(0,2pt) node[anchor=west] {$\scriptstyle q_a$};
  \end{tikzpicture}
  =
  \begin{tikzpicture}[baseline=(X)]
    \path node[dot] (O) {} +(0pt,8pt) coordinate (X) +(0pt,30pt) coordinate (T);
    \draw (O) .. controls +(5pt,15pt) and +(-5pt,15pt) .. ++(20pt,0);
    \draw (O) .. controls +(-5pt,15pt) and +(5pt,15pt) .. coordinate (t) ++(-20pt,0);
    \path (O) +(0pt,15pt) node[anchor=base,inner sep=1pt] (gamma) {$\gamma$};
    \draw (O) -- (gamma);
    \node[draw,dotted,thick,inner sep=0pt,circle,fit =(gamma)] (C) {};        
    \draw[dashed] (C.north) -- (C.north |- T);
    \path (C.north) ++(0,2pt) node[anchor=west] {$\scriptstyle q_a$};
  \end{tikzpicture}      \label{dGdQ}
\end{equation}
The first equality in \cref{dGdQ} requires that $\frac{\partial G}{\partial q_a}$ vanishes at both endpoints, which follows from differentiating $G(\varsigma,t_0) = 0 = G(\varsigma,t_1)$ with respect to $q_a$.

When evaluating formal integrals, we need not only the first derivative of the integrand but its full Taylor expansion.  To expand higher derivatives of diagrams, we use the product rule again:
\begin{align}
  \begin{tikzpicture}[baseline=(X)]
    \path node[dot] (O) {} ++(0pt,8pt) coordinate (X);
    \draw (O) -- ++(-10pt,25pt) ++(-2pt,2pt) coordinate (n1);
    \draw (O) -- ++(-6pt,25pt);
    \draw (O) -- ++(15pt,25pt) ++(2pt,2pt) coordinate (n2);
    \path (O) ++(4pt,23pt) node {$\scriptstyle \ldots$};
    \draw[decorate,decoration=brace] (n1) -- coordinate(n) (n2);
    \path (n) ++(0,4pt) node[anchor=base] {$\scriptstyle n$};
    \path (O) node (C1) [draw,circle,dotted,thick,inner sep=5pt] {};
    \draw[dashed,bend right] (C1) .. controls +(-10pt,5pt) and +(3pt,-10pt) .. ++(-20pt,30pt);
    \path (O) node (C2) [draw,circle,dotted,thick,inner sep=8pt] {};
    \draw[dashed,bend right] (C2) .. controls +(-15pt,5pt) and +(3pt,-10pt) .. ++(-25pt,30pt);
  \end{tikzpicture}
  & = 
  \begin{tikzpicture}[baseline=(X)]
    \path node[dot] (O) {} ++(0pt,8pt) coordinate (X);
    \draw (O) -- ++(-10pt,25pt) ++(-2pt,2pt) coordinate (n1);
    \draw (O) -- ++(-6pt,25pt);
    \draw (O) -- ++(15pt,25pt) ++(2pt,2pt) coordinate (n2);
    \path (O) ++(4pt,23pt) node {$\scriptstyle \ldots$};
    \draw[decorate,decoration=brace] (n1) -- coordinate(n) (n2);
    \path (n) ++(0,4pt) node[anchor=base] {$\scriptstyle n$};
    \path (O) ++(-20pt,10pt) node[inner sep=1pt] (gamma) {$\gamma$};
    \draw (O) -- (gamma);
    \path (gamma) node (C) [draw,circle,dotted,thick,inner sep=5pt] {};
    \draw[dashed] (C) -- ++(0,20pt);
    \node[draw,dotted,thick,inner sep=0pt,ellipse,fit = (O) (C),rotate=160] {};
  \end{tikzpicture}  
  =
  \begin{tikzpicture}[baseline=(X)]
    \path node[dot] (O) {} ++(0pt,8pt) coordinate (X);
    \draw (O) -- ++(-10pt,25pt) ++(-2pt,2pt) coordinate (n1);
    \draw (O) -- ++(-6pt,25pt);
    \draw (O) -- ++(15pt,25pt) ++(2pt,2pt) coordinate (n2);
    \path (O) ++(4pt,23pt) node {$\scriptstyle \ldots$};
    \draw[decorate,decoration=brace] (n1) -- coordinate(n) (n2);
    \path (n) ++(0,4pt) node[anchor=base] {$\scriptstyle n$};
    \path (O) ++(-20pt,15pt) node[inner sep=1pt] (gamma1) {$\gamma$};
    \draw (O) -- (gamma1);
    \path (gamma1) node (C1) [draw,circle,dotted,thick,inner sep=5pt] {};
    \draw[dashed] (C1) -- ++(0,15pt);
    \path (O) ++(-40pt,10pt) node[inner sep=1pt] (gamma2) {$\gamma$};
    \draw (O) -- (gamma2);
    \path (gamma2) node (C2) [draw,circle,dotted,thick,inner sep=5pt] {};
    \draw[dashed] (C2) -- ++(0,20pt);
  \end{tikzpicture}  
  +
  \begin{tikzpicture}[baseline=(X)]
    \path node[dot] (O) {} ++(0pt,8pt) coordinate (X);
    \draw (O) -- ++(-10pt,25pt) ++(-2pt,2pt) coordinate (n1);
    \draw (O) -- ++(-6pt,25pt);
    \draw (O) -- ++(15pt,25pt) ++(2pt,2pt) coordinate (n2);
    \path (O) ++(4pt,23pt) node {$\scriptstyle \ldots$};
    \draw[decorate,decoration=brace] (n1) -- coordinate(n) (n2);
    \path (n) ++(0,4pt) node[anchor=base] {$\scriptstyle n$};
    \path (O) ++(-20pt,10pt) node[inner sep=1pt] (gamma) {$\gamma$};
    \draw (O) -- (gamma);
    \path (gamma) node (C1) [draw,circle,dotted,thick,inner sep=5pt] {};
    \draw[dashed] (C1) -- ++(0,20pt) ++(-10pt,0) coordinate (t);
    \path (gamma) node (C2) [draw,circle,dotted,thick,inner sep=7pt] {};
    \draw[dashed,bend right] (C2.north west) .. controls +(-2pt,2pt) and +(0pt,-3pt) .. (t);
  \end{tikzpicture}  
\end{align}
In particular:
\begin{equation}
   \begin{tikzpicture}[baseline=(X)]
    \path node[dot] (O) {} ++(0pt,8pt) coordinate (X);
    \draw (O) -- ++(20pt,30pt) +(0,3pt) node[anchor=base] {$\scriptstyle \xi$};
    \path (O) ++(-1pt,15pt) node[anchor=base,inner sep=1pt] (gamma1) {$\gamma$};
    \draw (O) -- (gamma1);
    \path (gamma1) node (C1) [draw,circle,dotted,thick,inner sep=5pt] {};
    \draw[dashed] (C1) -- ++(0,23pt);
    \path (O) ++(-18pt,15pt) node[anchor=base,inner sep=1pt] (gamma2) {$\gamma$};
    \draw (O) -- (gamma2);
    \path (gamma2) node (C2) [draw,circle,dotted,thick,inner sep=5pt] {};
    \draw[dashed] (C2) -- ++(0,23pt);
  \end{tikzpicture} 
  +
  \begin{tikzpicture}[baseline=(X)]
    \path node[dot] (O) {} ++(0pt,8pt) coordinate (X);
    \path (O) ++(-10pt,15pt) node[anchor=base,inner sep=1pt] (gamma) {$\gamma$};
    \draw (O) -- (gamma);
    \draw (O) -- ++(20pt,30pt) +(0,3pt) node[anchor=base] {$\scriptstyle \xi$};
    \path (gamma) node (C) [draw,circle,dotted,thick,inner sep=5pt] {};
    \draw[dashed] (C) -- ++(0,23pt) coordinate (t);
    \path (gamma) node (C2) [draw,circle,dotted,thick,inner sep=7pt] {};
    \draw[dashed] (C2.north west) -- (C2.north west |- t);
  \end{tikzpicture} 
  =
  \frac{\partial^2}{\partial q^2} \Aa^{(1)}[\gamma]\cdot \xi
\end{equation}
This vanishes if $\xi$ is a based loop.  Thus:
\begin{equation}  0  = 
  \begin{tikzpicture}[baseline=(X)]
    \path node[dot] (O) {} ++(0pt,8pt) coordinate (X);
    \path (O) ++(-10pt,15pt) node[anchor=base,inner sep=1pt] (gamma) {$\gamma$};
    \draw (O) -- (gamma);
    \path (gamma) node (C) [draw,circle,dotted,thick,inner sep=5pt] {};
    \draw[dashed] (C) -- ++(0,23pt) coordinate (t);
    \path (gamma) node (C2) [draw,circle,dotted,thick,inner sep=7pt] {};
    \draw[dashed] (C2.north west) -- (C2.north west |- t);
    \draw (O) .. controls +(10pt,15pt) and +(0,15pt) .. ++(20pt,0);
  \end{tikzpicture} 
  +
   \begin{tikzpicture}[baseline=(X)]
    \path node[dot] (O) {} ++(0pt,8pt) coordinate (X);
    \path (O) ++(-1pt,15pt) node[anchor=base,inner sep=1pt] (gamma1) {$\gamma$};
    \draw (O) -- (gamma1);
    \path (gamma1) node (C1) [draw,circle,dotted,thick,inner sep=5pt] {};
    \draw[dashed] (C1) -- ++(0,23pt);
    \path (O) ++(-18pt,15pt) node[anchor=base,inner sep=1pt] (gamma2) {$\gamma$};
    \draw (O) -- (gamma2);
    \path (gamma2) node (C2) [draw,circle,dotted,thick,inner sep=5pt] {};
    \draw[dashed] (C2) -- ++(0,23pt);
    \draw (O) .. controls +(10pt,15pt) and +(0,15pt) .. ++(20pt,0);
  \end{tikzpicture} \label{contractedeqn}
\end{equation}
And $\frac{\partial^2\gamma}{\partial q_a\partial q_b}$ vanishes at both endpoints, so \cref{greendefngraphical} applies.  Therefore:
\begin{equation}
  \frac{\partial^2 \gamma^i}{\partial q_a^{j}\partial q_b^{k}}(\tau) = 
  \begin{tikzpicture}[baseline=(gamma.base)]
    \path node[inner sep=1pt] (gamma) {$\gamma$};
    \draw (gamma) -- ++(0,-15pt) +(0,-8pt) node[anchor=base] {$\scriptstyle \tau,i$};
    \path (gamma) node (C) [draw,circle,dotted,thick,inner sep=5pt] {};
    \path (gamma) node (C2) [draw,circle,dotted,thick,inner sep=6pt] {};
    \draw[dashed] (C2.north west) -- ++(0,18pt) +(0,3pt) node[anchor=base] {$\scriptstyle j$};    
    \draw[dashed] (C2.north east) -- ++(0,18pt) +(0,3pt) node[anchor=base] {$\scriptstyle k$};    
    \path (C2.north east) ++(-1pt,3pt) node[anchor= west] {$\scriptstyle q_b$};
    \path (C2.north west) ++(1pt,3pt) node[anchor= east] {$\scriptstyle q_a$};
  \end{tikzpicture} 
  =
   \begin{tikzpicture}[baseline=(X.base)]
    \path node[dot] (O) {} ++(0pt,8pt) coordinate (X);
    \path (O) ++(-1pt,15pt) node[anchor=base,inner sep=1pt] (gamma1) {$\gamma$};
    \draw (O) -- (gamma1);
    \path (gamma1) node (C1) [draw,circle,dotted,thick,inner sep=5pt] {};
    \draw[dashed] (C1) -- ++(0,17pt) +(0,3pt) node[anchor=base] {$\scriptstyle k$};
    \path (O) ++(-18pt,15pt) node[anchor=base,inner sep=1pt] (gamma2) {$\gamma$};
    \draw (O) -- (gamma2);
    \path (gamma2) node (C2) [draw,circle,dotted,thick,inner sep=5pt] {};
    \draw[dashed] (C2) -- ++(0,17pt) +(0,3pt) node[anchor=base] {$\scriptstyle j$};
    \draw (O) .. controls +(10pt,15pt) and +(0,15pt) .. ++(20pt,-5pt) +(0,-8pt) node[anchor=base] {$\scriptstyle \tau,i$};
    \path (C1.north) ++(-1pt,3pt) node[anchor= west] {$\scriptstyle q_b$};
    \path (C2.north) ++(1pt,3pt) node[anchor= east] {$\scriptstyle q_a$};
  \end{tikzpicture}   
\label{dphithree} \end{equation}

Thus in general to take the second derivative of a vertex one either adds two $\frac{\partial \gamma}{\partial q_a}$s or an edge connecting to a trivalent vertex with two $\frac{\partial\gamma}{\partial q_a}$s.  The lowest-valence example is:
\begin{align}\label{dSdQQ}
  \begin{tikzpicture}[baseline=(T.base)]
    \path node[Sdot] (S) {} ++(0,5pt) coordinate (T);
    \path (S) node[draw,dotted,thick,inner sep=5pt,circle] (C1) {};        
    \path (S) node[draw,dotted,thick,inner sep=6pt,circle] (C2) {};        
    \path (S) +(-15pt,25pt) coordinate (t1) +(15pt,25pt) coordinate (t2);
    \draw[dashed] (C2.north west) .. controls +(-5pt,5pt) and +(0pt,-10pt) .. (t1);
    \draw[dashed] (C2.north east) .. controls +(5pt,5pt) and +(0pt,-10pt) .. (t2);
    \path (C2.north west) node[anchor=east] {$\scriptstyle q_a$};
    \path (C2.north east) node[anchor=west] {$\scriptstyle q_b$};
  \end{tikzpicture}
  \, = \,
  \begin{tikzpicture}[baseline=(T.base)]
    \path node[dot] (O) {}  ++(0,13pt) coordinate (T);
    \path (O) ++(-15pt,15pt)node[inner sep=0pt] (gamma1) {$\gamma$};
    \draw (O) -- (gamma1);
    \node[draw,dotted,thick,inner sep=2pt,circle,fit =(gamma1)] (C1) {};        
    \path (O) ++(15pt,15pt)node[inner sep=0pt] (gamma2) {$\gamma$};
    \draw (O) -- (gamma2);
    \node[draw,dotted,thick,inner sep=2pt,circle,fit =(gamma2)] (C2) {};      
    \path (O) +(-15pt,33pt) coordinate (t1) +(15pt,33pt) coordinate (t2);
    \draw[dashed] (C1) -- (t1);
    \draw[dashed] (C2) -- (t2);
    \path (C1.north) ++(1pt,3pt) node[anchor= east] {$\scriptstyle q_a$};
    \path (C2.north) ++(-1pt,3pt) node[anchor= west] {$\scriptstyle q_b$};
  \end{tikzpicture}    
  \, + \,
  \begin{tikzpicture}[baseline=(X.base)]
    \path node[dot] (O) {} ++(0pt,13pt) coordinate (X);
    \path (O) ++(-1pt,15pt) node[anchor=base,inner sep=1pt] (gamma1) {$\gamma$};
    \draw (O) -- (gamma1);
    \path (gamma1) node (C1) [draw,circle,dotted,thick,inner sep=5pt] {};
    \draw[dashed] (C1) -- ++(0,17pt) ;
    \path (O) ++(-18pt,15pt) node[anchor=base,inner sep=1pt] (gamma2) {$\gamma$};
    \draw (O) -- (gamma2);
    \path (gamma2) node (C2) [draw,circle,dotted,thick,inner sep=5pt] {};
    \draw[dashed] (C2) -- ++(0,17pt) ;
    \draw (O) .. controls +(10pt,15pt) and +(0,15pt) .. ++(20pt,0pt) node[dot] {};
    \path (C1.north) ++(-1pt,3pt) node[anchor= west] {$\scriptstyle q_b$};
    \path (C2.north) ++(1pt,3pt) node[anchor= east] {$\scriptstyle q_a$};
  \end{tikzpicture}   
  \, = \,
  \begin{tikzpicture}[baseline=(T.base)]
    \path node[dot] (O) {}  ++(0,13pt) coordinate (T);
    \path (O) ++(-15pt,15pt)node[inner sep=0pt] (gamma1) {$\gamma$};
    \draw (O) -- (gamma1);
    \node[draw,dotted,thick,inner sep=2pt,circle,fit =(gamma1)] (C1) {};        
    \path (O) ++(15pt,15pt)node[inner sep=0pt] (gamma2) {$\gamma$};
    \draw (O) -- (gamma2);
    \node[draw,dotted,thick,inner sep=2pt,circle,fit =(gamma2)] (C2) {};      
    \path (O) +(-15pt,33pt) coordinate (t1) +(15pt,33pt) coordinate (t2);
    \draw[dashed] (C1) -- (t1);
    \draw[dashed] (C2) -- (t2);
    \path (C1.north) ++(1pt,3pt) node[anchor= east] {$\scriptstyle q_a$};
    \path (C2.north) ++(-1pt,3pt) node[anchor= west] {$\scriptstyle q_b$};
  \end{tikzpicture}    
  \, + \, 0
\end{align}
The second summand vanishes because $G$ is a based loop in each variable.  We record one further derivative here, as it indicates how to proceed to higher derivatives:
\begin{multline}
  \begin{tikzpicture}[baseline=(T.base)]
    \path node[dot] (O) {}  ++(0,13pt) coordinate (T);
    \path (O) ++(-15pt,15pt)node[inner sep=0pt] (gamma1) {$\gamma$};
    \draw (O) -- (gamma1);
    \node[draw,dotted,thick,inner sep=2pt,circle,fit =(gamma1)] (C1) {};        
    \path (O) ++(15pt,15pt)node[inner sep=0pt] (gamma2) {$\gamma$};
    \draw (O) -- (gamma2);
    \node[draw,dotted,thick,inner sep=2pt,circle,fit =(gamma2)] (C2) {};      
    \path (O) +(-15pt,40pt) coordinate (t1) +(15pt,40pt) coordinate (t2) +(0pt,40pt) coordinate (t3);
    \draw[dashed] (C1) -- (t1);
    \draw[dashed] (C2) -- (t2);
    \node[draw,dotted,thick,inner sep=0pt,ellipse,fit =(O) (C1) (C2)] (C3) {};          
    \draw[dashed] (C3) -- (t3);
    \path (C1.north) ++(1pt,3pt) node[anchor= east] {$\scriptstyle q_a$};
    \path (C2.north) ++(-1pt,3pt) node[anchor= west] {$\scriptstyle q_b$};
    \path (C3.north) ++(-1pt,3pt) node[anchor= west] {$\scriptstyle q_{c}$};
  \end{tikzpicture}
  \,  = \,  
  \begin{tikzpicture}[baseline=(T.base)]
    \path node[dot] (O) {}  ++(0,13pt) coordinate (T);
    \path (O) ++(-15pt,15pt)node[inner sep=0pt] (gamma1) {$\gamma$};
    \draw (O) -- (gamma1);
    \node[draw,dotted,thick,inner sep=2pt,circle,fit =(gamma1)] (C1) {};        
    \path (O) ++(15pt,15pt)node[inner sep=0pt] (gamma2) {$\gamma$};
    \draw (O) -- (gamma2);
    \node[draw,dotted,thick,inner sep=2pt,circle,fit =(gamma2)] (C2) {};      
    \path (O) +(-15pt,40pt) coordinate (t1) +(15pt,40pt) coordinate (t2) +(0pt,40pt) coordinate (t3);
    \draw[dashed] (C1.north west) -- (C1.north west |- t1);
    \draw[dashed] (C2) -- (t2);
    \node[draw,dotted,thick,inner sep=3pt,ellipse,fit =(gamma1)] (C3) {};          
    \draw[dashed] (C3.north east) -- (C3.north east |- t1);
    \path (C3.north west) ++(1pt,3pt) node[anchor= east] {$\scriptstyle q_a$};
    \path (C2.north) ++(-1pt,3pt) node[anchor= west] {$\scriptstyle q_b$};
    \path (C3.north east) ++(-1pt,3pt) node[anchor= west] {$\scriptstyle q_{c}$};
  \end{tikzpicture}  
  \, + \,
  \begin{tikzpicture}[baseline=(T.base)]
    \path node[dot] (O) {}  ++(0,13pt) coordinate (T);
    \path (O) ++(-15pt,15pt)node[inner sep=0pt] (gamma1) {$\gamma$};
    \draw (O) -- (gamma1);
    \node[draw,dotted,thick,inner sep=2pt,circle,fit =(gamma1)] (C1) {};        
    \path (O) ++(15pt,15pt)node[inner sep=0pt] (gamma2) {$\gamma$};
    \draw (O) -- (gamma2);
    \node[draw,dotted,thick,inner sep=2pt,circle,fit =(gamma2)] (C2) {};      
    \node[draw,dotted,thick,inner sep=3pt,circle,fit =(O)] (C3) {};      
    \path (O)  +(0pt,40pt) coordinate (t);
    \draw[dashed] (C1) -- (C1 |- t);
    \draw[dashed] (C2) -- (C2 |- t);
    \draw[dashed] (C3) -- (C3 |- t);
    \path (C1.north) ++(1pt,3pt) node[anchor= east] {$\scriptstyle q_a$};
    \path (C2.north) ++(-1pt,3pt) node[anchor= west] {$\scriptstyle q_b$};
    \path (C3.south east) ++(-1pt,3pt) node[anchor= west] {$\scriptstyle q_{c}$};
  \end{tikzpicture}
  \, + \,  
  \begin{tikzpicture}[baseline=(T.base)]
    \path node[dot] (O) {}  ++(0,13pt) coordinate (T);
    \path (O) ++(-15pt,15pt)node[inner sep=0pt] (gamma1) {$\gamma$};
    \draw (O) -- (gamma1);
    \node[draw,dotted,thick,inner sep=2pt,circle,fit =(gamma1)] (C2) {};        
    \path (O) ++(15pt,15pt)node[inner sep=0pt] (gamma2) {$\gamma$};
    \draw (O) -- (gamma2);
    \node[draw,dotted,thick,inner sep=2pt,circle,fit =(gamma2)] (C1) {};      
    \path (O) +(-15pt,40pt) coordinate (t1) +(15pt,40pt) coordinate (t2) +(0pt,40pt) coordinate (t3);
    \draw[dashed] (C1.north west) -- (C1.north west |- t1);
    \draw[dashed] (C2) -- (t1);
    \node[draw,dotted,thick,inner sep=3pt,ellipse,fit =(gamma2)] (C3) {};          
    \draw[dashed] (C3.north east) -- (C3.north east |- t1);
    \path (C2.north) ++(1pt,3pt) node[anchor= east] {$\scriptstyle q_a$};
    \path (C3.north east) ++(-1pt,3pt) node[anchor= west] {$\scriptstyle q_b$};
    \path (C3.north west) ++(1pt,3pt) node[anchor= east] {$\scriptstyle q_{c}$};
  \end{tikzpicture}  
  \, = \\  = \,
  \begin{tikzpicture}[baseline=(T.base)]
    \path node[dot] (O) {}  ++(0,13pt) coordinate (T);
    \path (O) ++(25pt,0) node[dot] (O1) {};
    \draw (O) .. controls +(10pt,15pt) and +(-10pt,15pt) .. (O1);
    \path (O) ++(-20pt,15pt)node[inner sep=0pt] (gamma1) {$\gamma$};
    \draw (O) -- (gamma1);
    \node[draw,dotted,thick,inner sep=2pt,circle,fit =(gamma1)] (C1) {};        
    \path (O1) ++(10pt,15pt)node[inner sep=0pt] (gamma2) {$\gamma$};
    \draw (O1) -- (gamma2);
    \node[draw,dotted,thick,inner sep=2pt,circle,fit =(gamma2)] (C2) {};      
    \path (O) ++(0pt,18pt)node[inner sep=0pt] (gamma3) {$\gamma$};
    \draw (O) -- (gamma3);
    \node[draw,dotted,thick,inner sep=2pt,circle,fit =(gamma3)] (C3) {};      
    \path (O)  +(0pt,40pt) coordinate (t);
    \draw[dashed] (C1) -- (C1 |- t);
    \draw[dashed] (C2) -- (C2 |- t);
    \draw[dashed] (C3) -- (C3 |- t);
    \path (C1.north) ++(1pt,3pt) node[anchor= east] {$\scriptstyle q_a$};
    \path (C2.north) ++(-1pt,3pt) node[anchor= west] {$\scriptstyle q_b$};
    \path (C3.north) ++(-1pt,3pt) node[anchor= west] {$\scriptstyle q_{c}$};
  \end{tikzpicture}  
  \, + \,
  \begin{tikzpicture}[baseline=(T.base)]
    \path node[dot] (O) {}  ++(0,13pt) coordinate (T);
    \path (O) ++(-20pt,15pt)node[inner sep=0pt] (gamma1) {$\gamma$};
    \draw (O) -- (gamma1);
    \node[draw,dotted,thick,inner sep=2pt,circle,fit =(gamma1)] (C1) {};        
    \path (O) ++(20pt,15pt)node[inner sep=0pt] (gamma2) {$\gamma$};
    \draw (O) -- (gamma2);
    \node[draw,dotted,thick,inner sep=2pt,circle,fit =(gamma2)] (C2) {};      
    \path (O) ++(0pt,18pt)node[inner sep=0pt] (gamma3) {$\gamma$};
    \draw (O) -- (gamma3);
    \node[draw,dotted,thick,inner sep=2pt,circle,fit =(gamma3)] (C3) {};      
    \path (O)  +(0pt,40pt) coordinate (t);
    \draw[dashed] (C1) -- (C1 |- t);
    \draw[dashed] (C2) -- (C2 |- t);
    \draw[dashed] (C3) -- (C3 |- t);
    \path (C1.north) ++(1pt,3pt) node[anchor= east] {$\scriptstyle q_a$};
    \path (C2.north) ++(-1pt,3pt) node[anchor= west] {$\scriptstyle q_b$};
    \path (C3.north) ++(-1pt,3pt) node[anchor= west] {$\scriptstyle q_{c}$};
  \end{tikzpicture}  
  \, + \,
  \begin{tikzpicture}[baseline=(T.base)]
    \path node[dot] (O) {}  ++(0,13pt) coordinate (T);
    \path (O) ++(-25pt,0) node[dot] (O1) {};
    \draw (O) .. controls +(-10pt,15pt) and +(10pt,15pt) .. (O1);
    \path (O1) ++(-10pt,15pt)node[inner sep=0pt] (gamma1) {$\gamma$};
    \draw (O1) -- (gamma1);
    \node[draw,dotted,thick,inner sep=2pt,circle,fit =(gamma1)] (C1) {};        
    \path (O) ++(20pt,15pt)node[inner sep=0pt] (gamma2) {$\gamma$};
    \draw (O) -- (gamma2);
    \node[draw,dotted,thick,inner sep=2pt,circle,fit =(gamma2)] (C2) {};      
    \path (O) ++(0pt,18pt)node[inner sep=0pt] (gamma3) {$\gamma$};
    \draw (O) -- (gamma3);
    \node[draw,dotted,thick,inner sep=2pt,circle,fit =(gamma3)] (C3) {};      
    \path (O)  +(0pt,40pt) coordinate (t);
    \draw[dashed] (C1) -- (C1 |- t);
    \draw[dashed] (C2) -- (C2 |- t);
    \draw[dashed] (C3) -- (C3 |- t);
    \path (C1.north) ++(1pt,3pt) node[anchor= east] {$\scriptstyle q_a$};
    \path (C2.north) ++(-1pt,3pt) node[anchor= west] {$\scriptstyle q_b$};
    \path (C3.north) ++(-1pt,3pt) node[anchor= west] {$\scriptstyle q_{c}$};
  \end{tikzpicture}  
 \, = \, 0 \, + \,
  \begin{tikzpicture}[baseline=(T.base)]
    \path node[dot] (O) {}  ++(0,13pt) coordinate (T);
    \path (O) ++(-20pt,15pt)node[inner sep=0pt] (gamma1) {$\gamma$};
    \draw (O) -- (gamma1);
    \node[draw,dotted,thick,inner sep=2pt,circle,fit =(gamma1)] (C1) {};        
    \path (O) ++(20pt,15pt)node[inner sep=0pt] (gamma2) {$\gamma$};
    \draw (O) -- (gamma2);
    \node[draw,dotted,thick,inner sep=2pt,circle,fit =(gamma2)] (C2) {};      
    \path (O) ++(0pt,18pt)node[inner sep=0pt] (gamma3) {$\gamma$};
    \draw (O) -- (gamma3);
    \node[draw,dotted,thick,inner sep=2pt,circle,fit =(gamma3)] (C3) {};      
    \path (O)  +(0pt,40pt) coordinate (t);
    \draw[dashed] (C1) -- (C1 |- t);
    \draw[dashed] (C2) -- (C2 |- t);
    \draw[dashed] (C3) -- (C3 |- t);
    \path (C1.north) ++(1pt,3pt) node[anchor= east] {$\scriptstyle q_a$};
    \path (C2.north) ++(-1pt,3pt) node[anchor= west] {$\scriptstyle q_b$};
    \path (C3.north) ++(-1pt,3pt) node[anchor= west] {$\scriptstyle q_{c}$};
  \end{tikzpicture} 
  \, + 0 \label{dSdQQQ}
\end{multline}
The outer terms vanish because $\Aa^{(2)} \cdot \frac{\partial \gamma}{\partial q_a}\xi = 0$ when $\xi$ is a based loop.  One more derivative gives:
\begin{multline}
  \frac{\partial^4[-S_\gamma]}{\partial q_a^{i_1}\partial q_b^{i_2}\partial q_c^{i_3}\partial q_d^{i_4}}\, = \,
  \begin{tikzpicture}[baseline=(T.base)]
    \path node[dot] (O) {}  ++(0,13pt) coordinate (T);
    \path (O) ++(-20pt,15pt)node[inner sep=0pt] (gamma1) {$\gamma$};
    \draw (O) -- (gamma1);
    \node[draw,dotted,thick,inner sep=2pt,circle,fit =(gamma1)] (C1) {};        
    \path (O) ++(20pt,15pt)node[inner sep=0pt] (gamma2) {$\gamma$};
    \draw (O) -- (gamma2);
    \node[draw,dotted,thick,inner sep=2pt,circle,fit =(gamma2)] (C2) {};      
    \path (O) ++(0pt,18pt)node[inner sep=0pt] (gamma3) {$\gamma$};
    \draw (O) -- (gamma3);
    \node[draw,dotted,thick,inner sep=2pt,circle,fit =(gamma3)] (C3) {};      
    \path (O)  +(0pt,45pt) coordinate (t);
    \draw[dashed] (C1) -- (C1 |- t) +(0,3pt) node[anchor=base] {$\scriptstyle i_2$};
    \draw[dashed] (C2) -- (C2 |- t) +(0,3pt) node[anchor=base] {$\scriptstyle i_4$};
    \draw[dashed] (C3) -- (C3 |- t) +(0,3pt) node[anchor=base] {$\scriptstyle i_3$};
    \node[draw,dotted,thick,inner sep=0pt,ellipse,fit =(O) (C1) (C2) (C3)] (C4) {};          
    \draw[dashed] (C4.north west) -- (C4.north west |- t) +(0,3pt) node[anchor=base] {$\scriptstyle i_1$};
    \path (C1.north) ++(1pt,3pt) node[anchor=east] {$\scriptstyle q_b$};
    \path (C2.north) ++(1pt,3pt) node[anchor=east] {$\scriptstyle q_c$};
    \path (C3.north) ++(1pt,3pt) node[anchor=east] {$\scriptstyle q_d$};
    \path (C4.north west) ++(1pt,3pt) node[anchor=east] {$\scriptstyle q_a$};
  \end{tikzpicture}  
  \, = \\ = \,
  \begin{tikzpicture}[baseline=(T.base)]
    \path node[dot] (O) {}  ++(0,13pt) coordinate (T);
    \path (O) ++(-31pt,20pt)node[inner sep=0pt] (gamma1) {$\gamma$};
    \draw (O) -- (gamma1);
    \node[draw,dotted,thick,inner sep=2pt,circle,fit =(gamma1)] (C1) {};        
    \path (O) ++(-11pt,25pt)node[inner sep=0pt] (gamma2) {$\gamma$};
    \draw (O) -- (gamma2);
    \node[draw,dotted,thick,inner sep=2pt,circle,fit =(gamma2)] (C2) {};      
    \path (O) ++(11pt,25pt)node[inner sep=0pt] (gamma3) {$\gamma$};
    \draw (O) -- (gamma3);
    \node[draw,dotted,thick,inner sep=2pt,circle,fit =(gamma3)] (C3) {};      
    \path (O) ++(31pt,20pt)node[inner sep=0pt] (gamma4) {$\gamma$};
    \draw (O) -- (gamma4);
    \node[draw,dotted,thick,inner sep=2pt,circle,fit =(gamma4)] (C4) {};        
    \path (O)  +(0pt,45pt) coordinate (t);
    \draw[dashed] (C1) -- (C1 |- t) +(0,3pt) node[anchor=base] {$\scriptstyle i_1$};
    \draw[dashed] (C2) -- (C2 |- t) +(0,3pt) node[anchor=base] {$\scriptstyle i_2$};
    \draw[dashed] (C3) -- (C3 |- t) +(0,3pt) node[anchor=base] {$\scriptstyle i_3$};
    \draw[dashed] (C4) -- (C4 |- t) +(0,3pt) node[anchor=base] {$\scriptstyle i_4$};
    \path (C1.north) ++(1pt,3pt) node[anchor=east] {$\scriptstyle q_a$};
    \path (C2.north) ++(1pt,3pt) node[anchor=east] {$\scriptstyle q_b$};
    \path (C3.north) ++(1pt,3pt) node[anchor=east] {$\scriptstyle q_c$};
    \path (C4.north) ++(1pt,3pt) node[anchor=east] {$\scriptstyle q_d$};
  \end{tikzpicture}  
  \, + \,  
  \begin{tikzpicture}[baseline=(T.base)]
    \path coordinate (O)  ++(0,13pt) coordinate (T);
    \path (O) ++(-31pt,20pt)node[inner sep=0pt] (gamma1) {$\gamma$};
    \node[draw,dotted,thick,inner sep=2pt,circle,fit =(gamma1)] (C1) {};        
    \path (O) ++(-11pt,25pt)node[inner sep=0pt] (gamma2) {$\gamma$};
    \node[draw,dotted,thick,inner sep=2pt,circle,fit =(gamma2)] (C2) {};      
    \path (O) ++(11pt,25pt)node[inner sep=0pt] (gamma3) {$\gamma$};
    \node[draw,dotted,thick,inner sep=2pt,circle,fit =(gamma3)] (C3) {};      
    \path (O) ++(31pt,20pt)node[inner sep=0pt] (gamma4) {$\gamma$};
    \node[draw,dotted,thick,inner sep=2pt,circle,fit =(gamma4)] (C4) {};        
    \path (O)  +(0pt,45pt) coordinate (t);
    \draw[dashed] (C1) -- (C1 |- t) +(0,3pt) node[anchor=base] {$\scriptstyle i_1$};
    \draw[dashed] (C2) -- (C2 |- t) +(0,3pt) node[anchor=base] {$\scriptstyle i_2$};
    \draw[dashed] (C3) -- (C3 |- t) +(0,3pt) node[anchor=base] {$\scriptstyle i_3$};
    \draw[dashed] (C4) -- (C4 |- t) +(0,3pt) node[anchor=base] {$\scriptstyle i_4$};
    \path (O) +(-15pt,0) node[dot] (O1) {} +(15pt,0) node[dot] (O2) {};
    \draw (O1) .. controls +(15pt,10pt) and +(-15pt,10pt) .. (O2);
    \draw (O1) -- (gamma1);
    \draw (O1) -- (gamma2);
    \draw (O2) -- (gamma3);
    \draw (O2) -- (gamma4);
    \path (C1.north) ++(1pt,3pt) node[anchor=east] {$\scriptstyle q_a$};
    \path (C2.north) ++(1pt,3pt) node[anchor=east] {$\scriptstyle q_b$};
    \path (C3.north) ++(1pt,3pt) node[anchor=east] {$\scriptstyle q_c$};
    \path (C4.north) ++(1pt,3pt) node[anchor=east] {$\scriptstyle q_d$};
  \end{tikzpicture}  
  \, + \,  
  \begin{tikzpicture}[baseline=(T.base)]
    \path coordinate (O)  ++(0,13pt) coordinate (T);
    \path (O) ++(-31pt,20pt)node[inner sep=0pt] (gamma1) {$\gamma$};
    \node[draw,dotted,thick,inner sep=2pt,circle,fit =(gamma1)] (C1) {};        
    \path (O) ++(-11pt,25pt)node[inner sep=0pt] (gamma2) {$\gamma$};
    \node[draw,dotted,thick,inner sep=2pt,circle,fit =(gamma2)] (C2) {};      
    \path (O) ++(11pt,25pt)node[inner sep=0pt] (gamma3) {$\gamma$};
    \node[draw,dotted,thick,inner sep=2pt,circle,fit =(gamma3)] (C3) {};      
    \path (O) ++(31pt,20pt)node[inner sep=0pt] (gamma4) {$\gamma$};
    \node[draw,dotted,thick,inner sep=2pt,circle,fit =(gamma4)] (C4) {};        
    \path (O)  +(0pt,45pt) coordinate (t);
    \draw[dashed] (C1) -- (C1 |- t) +(0,3pt) node[anchor=base] {$\scriptstyle i_1$};
    \draw[dashed] (C2) -- (C2 |- t) +(0,3pt) node[anchor=base] {$\scriptstyle i_2$};
    \draw[dashed] (C3) -- (C3 |- t) +(0,3pt) node[anchor=base] {$\scriptstyle i_3$};
    \draw[dashed] (C4) -- (C4 |- t) +(0,3pt) node[anchor=base] {$\scriptstyle i_4$};
    \path (O) +(-15pt,0) node[dot] (O1) {} +(15pt,0) node[dot] (O2) {};
    \draw (O1) .. controls +(15pt,10pt) and +(-15pt,10pt) .. (O2);
    \draw (O1) -- (gamma1);
    \draw (O2) -- (gamma2);
    \draw (O1) -- (gamma3);
    \draw (O2) -- (gamma4);
    \path (C1.north) ++(1pt,3pt) node[anchor=east] {$\scriptstyle q_a$};
    \path (C2.north) ++(1pt,3pt) node[anchor=east] {$\scriptstyle q_b$};
    \path (C3.north) ++(1pt,3pt) node[anchor=east] {$\scriptstyle q_c$};
    \path (C4.north) ++(1pt,3pt) node[anchor=east] {$\scriptstyle q_d$};
  \end{tikzpicture}  
  \, + \,  
  \begin{tikzpicture}[baseline=(T.base)]
    \path coordinate (O)  ++(0,13pt) coordinate (T);
    \path (O) ++(-31pt,20pt)node[inner sep=0pt] (gamma1) {$\gamma$};
    \node[draw,dotted,thick,inner sep=2pt,circle,fit =(gamma1)] (C1) {};        
    \path (O) ++(-11pt,25pt)node[inner sep=0pt] (gamma2) {$\gamma$};
    \node[draw,dotted,thick,inner sep=2pt,circle,fit =(gamma2)] (C2) {};      
    \path (O) ++(11pt,25pt)node[inner sep=0pt] (gamma3) {$\gamma$};
    \node[draw,dotted,thick,inner sep=2pt,circle,fit =(gamma3)] (C3) {};      
    \path (O) ++(31pt,20pt)node[inner sep=0pt] (gamma4) {$\gamma$};
    \node[draw,dotted,thick,inner sep=2pt,circle,fit =(gamma4)] (C4) {};        
    \path (O)  +(0pt,45pt) coordinate (t);
    \draw[dashed] (C1) -- (C1 |- t) +(0,3pt) node[anchor=base] {$\scriptstyle i_1$};
    \draw[dashed] (C2) -- (C2 |- t) +(0,3pt) node[anchor=base] {$\scriptstyle i_2$};
    \draw[dashed] (C3) -- (C3 |- t) +(0,3pt) node[anchor=base] {$\scriptstyle i_3$};
    \draw[dashed] (C4) -- (C4 |- t) +(0,3pt) node[anchor=base] {$\scriptstyle i_4$};
    \path (O) +(-15pt,0) node[dot] (O1) {} +(15pt,0) node[dot] (O2) {};
    \draw (O1) .. controls +(15pt,10pt) and +(-15pt,10pt) .. (O2);
    \draw (O1) -- (gamma1);
    \draw (O2) -- (gamma2);
    \draw (O2) -- (gamma3);
    \draw (O1) .. controls +(5pt,10pt) and +(-15pt,-5pt) .. (gamma4);
    \path (C1.north) ++(1pt,3pt) node[anchor=east] {$\scriptstyle q_a$};
    \path (C2.north) ++(1pt,3pt) node[anchor=east] {$\scriptstyle q_b$};
    \path (C3.north) ++(1pt,3pt) node[anchor=east] {$\scriptstyle q_c$};
    \path (C4.north) ++(1pt,3pt) node[anchor=east] {$\scriptstyle q_d$};
  \end{tikzpicture}  
\end{multline}
In general, the $n$th derivative will be a sum of trees.

The final component of \cref{Udefined} that is not locally constant in $q_0,q_1$ is the determinant $\det \frac{\partial^2[-S]}{\partial q_0\partial q_1}$.  Recall the derivative of a determinant of a matrix-valued function: $\frac{\partial}{\partial z}\bigl[ \det M(z)\bigr] = \bigl( \det M(z)\bigr) \frac{\partial}{\partial z} \bigl[ \log \det M(z)\bigr] =  \bigl( \det M(z)\bigr) \frac{\partial}{\partial z} \bigl[ \tr \log M(z)\bigr] =  \bigl( \det M(z)\bigr) \tr \bigl( M(z)^{-1}\frac{\partial}{\partial z} \bigl[  M(z)\bigr] \bigr)$.  To denote the derivatives of  $\det \frac{\partial^2[-S]}{\partial q_0\partial q_1}$ graphically, we introduce the Feynman rule $_0\tikz[baseline=(S.base)] \node[draw,ellipse,inner sep=1pt] (S) {$(\partial^2[-S])^{-1}$};{}_1$ for the inverse matrix $\bigl( \frac{\partial^2[-S]}{\partial q_0\partial q_1}\bigr)^{-1}$:
\begin{equation}
  \begin{tikzpicture}[baseline=(c)]
    \path coordinate (b) ++(0pt,40pt) node[inner sep=1pt,ellipse,draw] (S1) {$(\partial^2[-S])^{-1}$} ++(0,20pt) coordinate (t) ++(0,-30pt) coordinate (c);
    \draw[dashed] (S1.south west) -- (S1.south west |- b);
    \path (S1.south east) ++(20pt,-20pt) node[Sdot] (S) {};
    \node (C1) [draw,circle,dotted,thick,inner sep=2pt, fit=(S)] {};    
    \node (C2) [draw,circle,dotted,thick,inner sep=3pt, fit=(S)] {};    
    \draw[dashed] (S1.south east) -- (C2.north west);
    \draw[dashed] (C2.north east) -- (C2.north east |- t);
     \path (S1.south east) ++(1pt,-2pt) node[anchor= west] {$\scriptstyle 1$};
    \path (S1.south west) ++(0pt,-3pt) node[anchor= east] {$\scriptstyle 0$};
    \path (C2.north west) node[anchor=east] {$\scriptstyle q_1$};
    \path (C2.north east) node[anchor=west] {$\scriptstyle q_0$};
  \end{tikzpicture}
  = \quad
  \begin{tikzpicture}[baseline=(c)]
    \path coordinate (b) ++(0pt,60pt) coordinate (t) ++(0,-30pt) coordinate (c);
    \draw[dashed] (b) -- (t);
  \end{tikzpicture}
  \quad =
  \begin{tikzpicture}[baseline=(c)]
    \path coordinate (b) ++(0pt,40pt) node[inner sep=1pt,ellipse,draw] (S1) {$(\partial^2[-S])^{-1}$} ++(0,20pt) coordinate (t) ++(0,-30pt) coordinate (c);
    \draw[dashed] (S1.south east) -- (S1.south east |- b);
    \path (S1.south west) ++(-20pt,-20pt) node[Sdot] (S) {};
    \node (C1) [draw,circle,dotted,thick,inner sep=2pt, fit=(S)] {};    
    \node (C2) [draw,circle,dotted,thick,inner sep=3pt, fit=(S)] {};    
    \draw[dashed] (S1.south west) -- (C2.north east);
    \draw[dashed] (C2.north west) -- (C2.north west |- t);
     \path (S1.south east) ++(0pt,-3pt) node[anchor= west] {$\scriptstyle 1$};
    \path (S1.south west) ++(-1pt,-2pt) node[anchor= east] {$\scriptstyle 0$};
    \path (C2.north west) node[anchor=east] {$\scriptstyle q_1$};
    \path (C2.north east) node[anchor=west] {$\scriptstyle q_0$};
  \end{tikzpicture}
\end{equation}
Then \cref{dSdQQQ} gives:
\begin{align} \label{dDETdQ}
  \frac{\partial}{\partial q_a} \left[ \log \left| \det \frac{\partial^2[-S]}{\partial q_0\partial q_1}\right|\right] & = 
  \begin{tikzpicture}[baseline=(X)]
    \path node[inner sep=1pt,ellipse,draw] (S) {$(\partial^2[-S])^{-1}$} +(0pt,-15pt) coordinate (X) ++(0pt,-30pt) node[Sdot] (O) {};
    \path (S.south west) ++(2pt,2pt) node[anchor=north east] {$\scriptstyle 0$};
    \path (S.south east) ++(-2pt,2pt) node[anchor=north west] {$\scriptstyle 1$};
    \path (O) node[draw,dotted,thick,inner sep=5pt,circle] (C1) {};        
    \path (O) node[draw,dotted,thick,inner sep=6pt,circle] (C2) {};        
    \path (O) node[draw,dotted,thick,inner sep=7pt,circle] (C) {};        
    \draw[dashed] (C2.north west) .. controls +(-5pt,5pt) and +(0pt,-10pt) .. (S.south west);
    \draw[dashed] (C2.north east) .. controls +(5pt,5pt) and +(0pt,-10pt) .. (S.south east);
    \path (C2.north west) node[anchor=east] {$\scriptstyle q_0$};
    \path (C2.north east) node[anchor=west] {$\scriptstyle q_1$};
    \path (C.west) ++(2pt,0pt) node[anchor=north east] {$\scriptstyle q_a$};
    \draw[dashed] (C.west) .. controls +(-20pt,0pt) and +(0pt,-40pt) .. ++(-50pt,50pt);    
  \end{tikzpicture}
  =
  \begin{tikzpicture}[baseline=(X)]
    \path node[inner sep=1pt,ellipse,draw] (S) {$(\partial^2[-S])^{-1}$} +(0pt,-15pt) coordinate (X) ++(0pt,-30pt) node[dot] (O) {};
    \path (S.south west) ++(2pt,2pt) node[anchor=north east] {$\scriptstyle 0$};
    \path (S.south east) ++(-2pt,2pt) node[anchor=north west] {$\scriptstyle 1$};
    \path (O) ++(-15pt,10pt) node[inner sep=0pt] (gamma1) {$\gamma$}; 
    \path (O) ++(15pt,10pt) node[inner sep=0pt] (gamma2) {$\gamma$}; 
    \path (O) ++(-50pt,20pt) node[inner sep=0pt] (gamma3) {$\gamma$}; 
    \path (gamma1) node (C1) [draw,circle,dotted,thick,inner sep=5pt] {};
    \path (gamma2) node (C2) [draw,circle,dotted,thick,inner sep=5pt] {};
    \path (gamma3) node (C3) [draw,circle,dotted,thick,inner sep=5pt] {};
    \draw[dashed] (C1.west) .. controls +(-5pt,5pt) and +(0pt,-10pt) .. (S.south west);
    \draw[dashed] (C2.east) .. controls +(5pt,5pt) and +(0pt,-10pt) .. (S.south east);
    \path (C1.west) ++(2pt,-2pt) node[anchor=east] {$\scriptstyle q_0$};
    \path (C2.east) ++(-2pt,-2pt) node[anchor=west] {$\scriptstyle q_1$};
    \path (C3.north) ++(1pt,3pt) node[anchor= east] {$\scriptstyle q_a$};
    \draw (O) -- (gamma1); \draw (O) -- (gamma2);
    \draw[draw] (O) .. controls +(-30pt,0pt) and +(5pt,-15pt) .. (gamma3);
    \draw[dashed] (C3) -- ++(0pt,30pt);    
  \end{tikzpicture}
\end{align}

An important comparison is in order.  In the graphical notation, \cref{greeneqn} reads:
\begin{equation} \label{greeneqn2}
  \ev\left(  \begin{tikzpicture}[baseline=(X)]
    \path coordinate (A) ++(0pt,2pt) coordinate (X);
    \path (A) +(0,-8pt) node[anchor=base] {$\scriptstyle \varsigma$};
    \path (A) ++(20pt,0) coordinate (B) +(0,-8pt) node[anchor=base] {$\scriptstyle \tau$};
    \draw (A) .. controls +(0,15pt) and +(0,15pt) .. (B);
  \end{tikzpicture}  \right)
  =
  \ev\left(
  \begin{tikzpicture}[baseline=(X)]
    \path node[inner sep=1pt,ellipse,draw] (S) {$(\partial^2[-S])^{-1}$} +(0,-20pt) coordinate (X);
    \path (S.south west) ++(0pt,-20pt) node (gamma1) {$\gamma$};
    \path (S.south east) ++(0pt,-20pt) node (gamma2) {$\gamma$};
    \path (S.south east) ++(0pt,-3pt) node[anchor= west] {$\scriptstyle 1$};
    \path (S.south west) ++(0pt,-3pt) node[anchor= east] {$\scriptstyle 0$};
    \draw (gamma1) -- ++(0,-15pt) +(0,-6pt) node[anchor=base] {$\scriptstyle \varsigma$};
    \path (gamma1) node (C1) [draw,circle,dotted,thick,inner sep=5pt] {};
    \path (C1.north) ++(1pt,3pt) node[anchor= west] {$\scriptstyle q_0$};
    \draw (gamma2) -- ++(0,-15pt) +(0,-6pt) node[anchor=base] {$\scriptstyle \tau$};
    \path (gamma2) node (C2) [draw,circle,dotted,thick,inner sep=5pt] {};
    \path (C2.north) ++(1pt,3pt) node[anchor= east] {$\scriptstyle q_1$};
    \draw[dashed] (C1.north) -- (S.south west);
    \draw[dashed] (C2.north) -- (S.south east);
  \end{tikzpicture}
  \right)\Theta(\varsigma - \tau)
  +
  \ev\left(
  \begin{tikzpicture}[baseline=(X)]
    \path node[inner sep=1pt,ellipse,draw] (S) {$(\partial^2[-S])^{-1}$} +(0,-20pt) coordinate (X);
    \path (S.south west) ++(0pt,-20pt) node (gamma1) {$\gamma$};
    \path (S.south east) ++(0pt,-20pt) node (gamma2) {$\gamma$};
    \path (S.south east) ++(0pt,-3pt) node[anchor= west] {$\scriptstyle 0$};
    \path (S.south west) ++(0pt,-3pt) node[anchor= east] {$\scriptstyle 1$};
    \draw (gamma1) -- ++(0,-15pt) +(0,-6pt) node[anchor=base] {$\scriptstyle \varsigma$};
    \path (gamma1) node (C1) [draw,circle,dotted,thick,inner sep=5pt] {};
    \path (C1.north) ++(1pt,3pt) node[anchor= west] {$\scriptstyle q_1$};
    \draw (gamma2) -- ++(0,-15pt) +(0,-6pt) node[anchor=base] {$\scriptstyle \tau$};
    \path (gamma2) node (C2) [draw,circle,dotted,thick,inner sep=5pt] {};
    \path (C2.north) ++(1pt,3pt) node[anchor= east] {$\scriptstyle q_0$};
    \draw[dashed] (C1.north) -- (S.south west);
    \draw[dashed] (C2.north) -- (S.south east);
  \end{tikzpicture}
  \right)\Theta(\tau - \varsigma)
\end{equation}

Recall that implicit in vertices are integrals, which might not converge: the Green's function is not smooth at $\varsigma=\tau$.  Suppose, however, that the diagram   \begin{tikzpicture}[baseline=(X)]
    \path node[dot] (O) {} ++(0pt,4pt) coordinate (X);
    \draw (O) .. controls +(15pt,15pt) and +(0pt,17pt) .. (O);
    \draw (O) -- ++(-7pt,10pt);
\end{tikzpicture}\!\!\!\! does converge.  Then we claim that:
\begin{equation} \label{claimoneloop}
  \begin{tikzpicture}[baseline=(X)]
    \path node[inner sep=1pt,ellipse,draw] (S) {$(\partial^2[-S])^{-1}$} +(0pt,-15pt) coordinate (X) ++(0pt,-30pt) node[dot] (O) {};
    \path (S.south west) ++(2pt,2pt) node[anchor=north east] {$\scriptstyle 0$};
    \path (S.south east) ++(-2pt,2pt) node[anchor=north west] {$\scriptstyle 1$};
    \path (O) ++(-15pt,10pt) node[inner sep=0pt] (gamma1) {$\gamma$}; 
    \path (O) ++(15pt,10pt) node[inner sep=0pt] (gamma2) {$\gamma$}; 
    \path (O) ++(-50pt,20pt) node[inner sep=0pt] (gamma3) {}; 
    \path (gamma1) node (C1) [draw,circle,dotted,thick,inner sep=5pt] {};
    \path (gamma2) node (C2) [draw,circle,dotted,thick,inner sep=5pt] {};
    \draw[dashed] (C1.west) .. controls +(-5pt,5pt) and +(0pt,-10pt) .. (S.south west);
    \draw[dashed] (C2.east) .. controls +(5pt,5pt) and +(0pt,-10pt) .. (S.south east);
    \path (C1.west) ++(2pt,-2pt) node[anchor=east] {$\scriptstyle q_1$};
    \path (C2.east) ++(-2pt,-2pt) node[anchor=west] {$\scriptstyle q_2$};
    \draw (O) -- (gamma1); \draw (O) -- (gamma2);
    \draw[draw] (O) .. controls +(-30pt,0pt) and +(5pt,-15pt) .. (gamma3);
  \end{tikzpicture}
  =
  \begin{tikzpicture}[baseline=(X)]
    \path node[dot] (O) {} ++(0pt,12pt) coordinate (X);
    \draw (O) .. controls +(20pt,20pt) and +(0pt,25pt) .. (O);
    \path (O) ++(-15pt,20pt)node[inner sep=0pt] (gamma) {};
    \draw (O) -- (gamma);
  \end{tikzpicture}  
\end{equation}
Indeed, the difference comes only from derivatives of the Heaviside step functions in \cref{greeneqn2}, and of these only $\delta(\varsigma - \tau)$ can contribute, but if it contributes at all to \begin{tikzpicture}[baseline=(X)]
    \path node[dot] (O) {} ++(0pt,4pt) coordinate (X);
    \draw (O) .. controls +(15pt,15pt) and +(0pt,17pt) .. (O);
    \draw (O) -- ++(-7pt,10pt);
\end{tikzpicture}\!\!\!\! then it contributes a divergent term proportional to $\delta(0)$.  This proves \cref{claimoneloop} when the formal path integral has no ultraviolet divergences.

We remark that \cref{claimoneloop} says nothing more or less than that our {\em ad hoc} determinant has the same derivatives as would be had by the undefined determinant $\bigl|\det \Aa^{(2)}\bigr|^{-1}$, which is what \cref{formalintdefn} says should be in \cref{Udefined}.  When there are ultraviolet divergences, we believe that some sort of ``divergent'' derivative should replace our {\em ad hoc} definition.

In \cref{Udefined}, the determinant appears with exponent $\frac12$.  This fraction appears with new meaning: $\partial \biggl[ \sqrt{ \bigl| \det \frac{\partial^2[-S]}{\partial q_0\partial q_1}\bigr|} \biggr] = \sqrt{ \bigl| \det \frac{\partial^2[-S]}{\partial q_0\partial q_1}\bigr|} \times \frac12 \begin{tikzpicture}[baseline=(X)]
    \path node[dot] (O) {} ++(0pt,2pt) coordinate (X);
    \draw (O) .. controls +(15pt,15pt) and +(0pt,17pt) .. (O);
    \draw (O) -- ++(-7pt,10pt);
\end{tikzpicture}\!\!\!\!$, and on the right-hand-side the $\frac12$ can be understood as the symmetry factor of the diagram.  All together, \cref{productrulelemma,dAdQ,dGdQ,dphithree,claimoneloop} imply:
\begin{prop} \label{derivativesofU}
  Let $L$ be a Lagrangian on $\RR^d$ and $\gamma$ a classical nonfocal path depending smoothly on its boundary conditions such that the formal path integral $U_\gamma(q_0,q_1)$ has no ultraviolet divergences.  Then:
  \begin{gather} \label{dUeqn1}
    -S_\gamma = \ev( \tikz \node[Sdot] {};)
    \quad\quad\quad
    \frac{\partial[-S_\gamma]}{\partial q_a} =  \ev \Biggl( \begin{tikzpicture}[baseline=(T.base)]
      \path node[dot] (O) {}  ++(0,13pt) coordinate (T);
      \path (O) ++(-10pt,15pt)node[inner sep=0pt] (gamma1) {$\gamma$};
      \draw (O) -- (gamma1);
      \node[draw,dotted,thick,inner sep=2pt,circle,fit =(gamma1)] (C1) {};        
      \path (O) +(-10pt,33pt) coordinate (t1);
      \draw[dashed] (C1) -- (t1);
      \path (C1.north) ++(1pt,3pt) node[anchor= east] {$\scriptstyle q_a$};
    \end{tikzpicture}    \;
    \Biggr)
    \quad\quad\quad
    \frac{\partial^2[-S_\gamma]}{\partial q_a\partial q_b} =  \ev \Biggl( \begin{tikzpicture}[baseline=(T.base)]
    \path node[dot] (O) {}  ++(0,13pt) coordinate (T);
    \path (O) ++(-15pt,15pt)node[inner sep=0pt] (gamma1) {$\gamma$};
    \draw (O) -- (gamma1);
    \node[draw,dotted,thick,inner sep=2pt,circle,fit =(gamma1)] (C1) {};        
    \path (O) ++(15pt,15pt)node[inner sep=0pt] (gamma2) {$\gamma$};
    \draw (O) -- (gamma2);
    \node[draw,dotted,thick,inner sep=2pt,circle,fit =(gamma2)] (C2) {};      
    \path (O) +(-15pt,33pt) coordinate (t1) +(15pt,33pt) coordinate (t2);
    \draw[dashed] (C1) -- (t1);
    \draw[dashed] (C2) -- (t2);
    \path (C1.north) ++(1pt,3pt) node[anchor= east] {$\scriptstyle q_a$};
    \path (C2.north) ++(-1pt,3pt) node[anchor= west] {$\scriptstyle q_b$};
  \end{tikzpicture}    
  \Biggr)
     \\ \label{dUeqn2}
    \frac{\partial^n[-S_\gamma]}{(\partial q)^n} = \sum_{\substack{\text{trees $\Gamma$ with trivalent} \\ \text{and higher vertices} \\ \text{and $n$ leaves}}} \ev(\Gamma),\quad n\geq 3
    \end{gather} \\[-3pc] \begin{multline} \label{dUeqn3}
    \frac{\partial^n}{(\partial q)^n} \left[ \sqrt{ \left| \det \frac{\partial^2[-S_\gamma]}{\partial q_0\partial q_1}\right|} \sum_{\substack{\text{diagrams $\Gamma$ with} \\ \text{no leaves and trivalent} \\ \text{and higher vertices}}} \frac{(i\hbar)^{-\chi(\Gamma)}\ev(\Gamma)}{\left| \Aut\Gamma\right|} \right] = \hfill \\
   \hfill = \sqrt{ \left| \det \frac{\partial^2[-S_\gamma]}{\partial q_0\partial q_1}\right|} \sum_{\substack{\text{diagrams $\Gamma$ with} \\ \text{$n$ leaves and trivalent} \\ \text{and higher vertices} \\ \text{and no trees}}} \frac{(i\hbar)^{-\chi(\Gamma)}\ev(\Gamma)}{\left| \Aut\Gamma\right|}
  \end{multline}
  In \cref{dUeqn2,dUeqn3}, the leaves are ordered and are attached to $\frac{\partial \gamma}{\partial q}$s.  In \cref{dUeqn2}, we recognize that a tree with ordered leaves has no automorphisms.  In each case, you could just as well have divided the left-hand side by $n!$, working with $\frac1{n!} \frac{\partial^n}{\partial q^n}$, in which case we would not order the leaves on the right-hand sides and would have to divide by $\left|\Aut\Gamma\right|$ in \cref{dUeqn2}.
  \qedhere
\end{prop}

There is a cute way of rewriting \cref{derivativesofU}.  By recalling the formal integral of \cref{formalintdefn}, and making the same {\em ad hoc} choices as in \cref{pathintdefn}, \cref{dUeqn1,dUeqn2,dUeqn3} can be packaged together as saying that we can ``differentiate under the formal path integral'':
\begin{equation} \label{dUeqncute}
  \frac{\partial^n}{(\partial q_a)^n}\bigl[U_\gamma(q_0,q_1)\bigr] = \int^{\formal}_\gamma \frac{\partial^n}{(\partial q_a)^n} \left[ \exp\bigl( -(i\hbar)^{-1}\Aa(\varphi)\bigr)\right] \d\varphi
\end{equation}
By $\frac{\partial}{\partial q_0}[\Bb(\varphi)]$, say, we mean the following.  The paths $\varphi$ range among paths with boundary values $\varphi(t_a) = q_a$, $a = 0,1$.  Arbitrarily pick a collection of path $\xi_j: [t_0,t_1] \to \RR^d$ with $\xi(t_1) = 0$ and $\xi^i_j(t_0) =  \delta^i_j$.  Then $\frac{\partial}{\partial q_0^j}[\Bb(\varphi)]$ is the functional derivative $\lim_{\epsilon \to 0} \epsilon^{-1}\bigl( \Bb(\varphi+\epsilon\xi_j) - \Bb(\varphi)\bigr) = \Bb^{(1)}(\varphi) \cdot \xi_j$.  By the Euler-Lagrange equations, the choice of $\xi$ does not effect the value of the formal integral in \cref{dUeqncute}.  If the integral made sense analytically, the choice of $\xi$ would be ``integrated out'' by the integral $\int \d\varphi$.

\subsection{The Feynman-diagrammatic part of equation \texorpdfstring{\ref{complaw}}{4.0.2}} \label{somediagrams}

We now return to the situation in \cref{fubinithm}.  In this section, we compare the higher-order terms in $U_{\gamma}$ and $\int^{\formal} U_{\gamma_0}U_{\gamma_1}$.  We have:
\begin{equation} \label{comparethis}
  U_{\gamma}(q_0,q_1) = (2\pi i \hbar)^{-d/2} e^{\frac i \hbar S_{\gamma}} (-1)^{\eta(\gamma)} \sqrt{ \left| \det \frac{\partial^2[-S_\gamma]}{\partial q_0\partial q_1}\right|} \sum_\Gamma \frac{(i\hbar)^{-\chi(\Gamma)}\ev(\Gamma)}{\left|\Aut\Gamma\right|}
\end{equation}
The sum ranges over unmarked diagrams, which we henceforth draw with doubled edges to distinguish from the diagrams in the integral $\int^{\formal} U_{\gamma_0}U_{\gamma_1}$, and $\ev\biggl(
  \begin{tikzpicture}[baseline=(X)]
    \path node[dot] (O) {} ++(0pt,4pt) coordinate (X);
    \draw[twice] (O) -- ++(-10pt,15pt) ++(-2pt,2pt) coordinate (n1);
    \draw[twice] (O) -- ++(-6pt,15pt);
    \draw[twice] (O) -- ++(10pt,15pt) ++(2pt,2pt) coordinate (n2);
    \path (O) ++(2pt,13pt) node {$\scriptstyle \ldots$};
    \draw[decorate,decoration=brace] (n1) -- coordinate(n) (n2);
    \path (n) ++(0,4pt) node[anchor=base] {$\scriptstyle n$};
  \end{tikzpicture}
  \biggr) 
  = -\Aa^{(n)}(\gamma)$ and $
  \ev\biggl(
  \begin{tikzpicture}[baseline=(X)]
    \path coordinate (A) ++(0pt,4pt) coordinate (X);
    \path (A) +(0,-8pt);
    \path (A) ++(20pt,0) coordinate (B) +(0,-8pt);
    \draw[twice] (A) .. controls +(0,15pt) and +(0,15pt) .. (B);
  \end{tikzpicture} 
  \biggr) 
  = G_{\gamma}
$ are the corresponding Feynman rules.

On the other hand, \cref{add,d2S12,etaeqn,formalintdefn} give:
\begin{equation} \label{comparethat}
  \int^{\formal}_{q_{\cp}(q_0,q_1)} U_{\gamma_0}(q_0,q)\,U_{\gamma_1}(q,q_1)\,\d q = (2\pi i \hbar)^{-d/2} e^{\frac i \hbar S_{\gamma}} (-1)^{\eta(\gamma)} \sqrt{\left| \det \frac{\partial^2[-S_{\gamma}]}{\partial q_0\partial q_1}\right|} \sum_{\Gamma\text{ marked}} \frac{(i\hbar)^{-\chi(\Gamma)}\ev(\Gamma)}{\left|\Aut\Gamma\right|}
\end{equation}
where the sum ranges over diagrams with precisely two marked vertices $\star_0,\star_1$ of arbitrary valence.  We continue to use dashed edges to denote the finite-dimensional space $\RR^d$.  Then the Feynman rules are:
\begin{gather*}
  \ev\biggl(
  \begin{tikzpicture}[baseline=(X)]
    \path node[dot] (O) {} ++(0pt,4pt) coordinate (X);
    \draw[dashed] (O) -- ++(-10pt,15pt) ++(-2pt,2pt) coordinate (n1);
    \draw[dashed] (O) -- ++(-6pt,15pt);
    \draw[dashed] (O) -- ++(10pt,15pt) ++(2pt,2pt) coordinate (n2);
    \path (O) ++(2pt,13pt) node {$\scriptstyle \ldots$};
    \draw[decorate,decoration=brace] (n1) -- coordinate(n) (n2);
    \path (n) ++(0,4pt) node[anchor=base] {$\scriptstyle n$};
  \end{tikzpicture}
  \biggr) 
   = - \frac{\partial^n}{\partial q^n}\bigl[ S_0(q_0,q) + S_1(q,q_1)\bigr] 
   \hspace{1in}
  \ev\biggl(
  \begin{tikzpicture}[baseline=(X)]
    \path coordinate (A) ++(0pt,4pt) coordinate (X);
    \path (A) +(0,-8pt);
    \path (A) ++(20pt,0) coordinate (B) +(0,-8pt);
    \draw[dashed] (A) .. controls +(0,15pt) and +(0,15pt) .. (B);
  \end{tikzpicture} 
  \biggr) 
   = \left( \frac{\partial^2[S_0 + S_1]}{\partial q^2}\right)^{-1} \\
  \ev\biggl(
  \begin{tikzpicture}[baseline=(X)]
    \path node[inner sep=1pt] (O) {$\star$} +(4pt,-2pt) node {$\scriptstyle a$} ++(0pt,4pt) coordinate (X);
    \draw[dashed] (O) -- ++(-10pt,15pt) ++(-2pt,2pt) coordinate (n1);
    \draw[dashed] (O) -- ++(-6pt,15pt);
    \draw[dashed] (O) -- ++(10pt,15pt) ++(2pt,2pt) coordinate (n2);
    \path (O) ++(2pt,13pt) node {$\scriptstyle \ldots$};
    \draw[decorate,decoration=brace] (n1) -- coordinate(n) (n2);
    \path (n) ++(0,4pt) node[anchor=base] {$\scriptstyle n$};
  \end{tikzpicture}
  \biggr)    =   \frac1{\sqrt{\left|\det \frac{\partial^2[-S_a]}{\partial q_a\partial q}\right|}} \, \frac{\partial^n}{\partial q^n}\left[ \sqrt{\left|\det \frac{\partial^2[-S_a]}{\partial q_a\partial q}\right|}  \sum_{\Gamma} \frac{(i\hbar)^{-\chi(\Gamma)}\ev_{a}(\Gamma)}{\left|\Aut\Gamma\right|} \right]
\end{gather*}
Here $a = 0,1$, $\ev_a$ denotes the Feynman rules used to define $U_{\gamma_a}$, and every expression is evaluated at $q = q_{\cp}(q_0,q_1)$.

\Cref{derivativesofU} gives:
\begin{gather}  \label{formulaforstar}
  \ev\biggl(
  \begin{tikzpicture}[baseline=(X)]
    \path node[inner sep=1pt] (O) {$\star$} +(4pt,-2pt) node {$\scriptstyle a$} ++(0pt,4pt) coordinate (X);
    \draw[dashed] (O) -- ++(-10pt,15pt) ++(-2pt,2pt) coordinate (n1);
    \draw[dashed] (O) -- ++(-6pt,15pt);
    \draw[dashed] (O) -- ++(10pt,15pt) ++(2pt,2pt) coordinate (n2);
    \path (O) ++(2pt,13pt) node {$\scriptstyle \ldots$};
    \draw[decorate,decoration=brace] (n1) -- coordinate(n) (n2);
    \path (n) ++(0,4pt) node[anchor=base] {$\scriptstyle n$};
  \end{tikzpicture}
  \biggr)
  =
  \sum_{\substack{\Gamma \text{ with no trees and} \\ n \text{ exterior \tikz[baseline=(gamma.base)]{ \path (0,0) node[inner sep=0pt,draw,dotted,thick,circle] (gamma) {$\scriptstyle \gamma_a$}; \path (gamma.east) [draw,dashed] -- ++(10pt,0pt) (gamma.north east) ++(2pt,0pt) node {$\scriptscriptstyle q$}; \path (0,0) [draw,solid] ++(-2pt,-2pt) -- ++(-4pt,-4pt);} s}}} \frac{(i\hbar)^{\chi(\Gamma)}\,\ev_a(\Gamma)}{\left|\Aut \Gamma\right|}  
  \\
  \label{sumoftrees}
  \ev\biggl(
  \begin{tikzpicture}[baseline=(X)]
    \path node[dot] (O) {} ++(0pt,4pt) coordinate (X);
    \draw[dashed] (O) -- ++(-10pt,15pt) ++(-2pt,2pt) coordinate (n1);
    \draw[dashed] (O) -- ++(-6pt,15pt);
    \draw[dashed] (O) -- ++(10pt,15pt) ++(2pt,2pt) coordinate (n2);
    \path (O) ++(2pt,13pt) node {$\scriptstyle \ldots$};
    \draw[decorate,decoration=brace] (n1) -- coordinate(n) (n2);
    \path (n) ++(0,4pt) node[anchor=base] {$\scriptstyle n$};
  \end{tikzpicture}
  \biggr) 
  =
  \sum_{\substack{\text{trees $\Gamma$ with} \\
  n \text{ exterior \tikz[baseline=(gamma.base)]{ \path (0,0) node[inner sep=0pt,draw,dotted,thick,circle] (gamma) {$\scriptstyle \gamma_0$}; \path (gamma.east) [draw,dashed] -- ++(10pt,0pt) (gamma.north east) ++(2pt,0pt) node {$\scriptscriptstyle q$}; \path (0,0) [draw,solid] ++(-2pt,-2pt) -- ++(-4pt,-4pt);} s}}} \frac{(i\hbar)^{\chi(\Gamma)}\,\ev_0(\Gamma)}{\left|\Aut \Gamma\right|}  
  +
  \sum_{\substack{\text{trees $\Gamma$ with} \\
  n \text{ exterior \tikz[baseline=(gamma.base)]{ \path (0,0) node[inner sep=0pt,draw,dotted,thick,circle] (gamma) {$\scriptstyle \gamma_1$}; \path (gamma.east) [draw,dashed] -- ++(10pt,0pt) (gamma.north east) ++(2pt,0pt) node {$\scriptscriptstyle q$}; \path (0,0) [draw,solid] ++(-2pt,-2pt) -- ++(-4pt,-4pt);} s}}} \frac{(i\hbar)^{\chi(\Gamma)}\,\ev_1(\Gamma)}{\left|\Aut \Gamma\right|}  
\end{gather}
In \cref{formulaforstar}, the diagrams in the sum may be disconnected, but no connected component can be a tree.  In total, each diagram must have $n$ occurrences of $\frac{\partial \gamma_a}{\partial q}$.  The two sums in \cref{sumoftrees} are the same, but in one we evaluate each diagram with respect to the Feynman rules for $U_{\gamma_0}$ and contract each leaf with $\frac{\partial \gamma_0}{\partial q}$, and in the other we use $U_{\gamma_1}$ and $\frac{\partial \gamma_1}{\partial q}$.

Finally, we modify the notation slightly so that we can drop the ``$\ev_a$'' notation but still consider diagrams with both $\star_0,\star_1$ expanded out:
\begin{equation*}
  \ev\biggl(
  \begin{tikzpicture}[baseline=(X)]
    \path node[dot] (O) {} +(4pt,-2pt) node {$\scriptstyle a$} ++(0pt,4pt) coordinate (X);
    \draw (O) -- ++(-10pt,15pt) ++(-2pt,2pt) coordinate (n1);
    \draw (O) -- ++(-6pt,15pt);
    \draw (O) -- ++(10pt,15pt) ++(2pt,2pt) coordinate (n2);
    \path (O) ++(2pt,13pt) node {$\scriptstyle \ldots$};
    \draw[decorate,decoration=brace] (n1) -- coordinate(n) (n2);
    \path (n) ++(0,4pt) node[anchor=base] {$\scriptstyle n$};
  \end{tikzpicture}
  \biggr)
   =
  \ev_a\biggl(
  \begin{tikzpicture}[baseline=(X)]
    \path node[dot] (O) {} ++(0pt,4pt) coordinate (X);
    \draw (O) -- ++(-10pt,15pt) ++(-2pt,2pt) coordinate (n1);
    \draw (O) -- ++(-6pt,15pt);
    \draw (O) -- ++(10pt,15pt) ++(2pt,2pt) coordinate (n2);
    \path (O) ++(2pt,13pt) node {$\scriptstyle \ldots$};
    \draw[decorate,decoration=brace] (n1) -- coordinate(n) (n2);
    \path (n) ++(0,4pt) node[anchor=base] {$\scriptstyle n$};
  \end{tikzpicture}
  \biggr)  
  \hspace{1in}
  \ev\biggl(
  \begin{tikzpicture}[baseline=(X)]
    \path coordinate (A) ++(0pt,4pt) coordinate (X);
    \path (A) +(0,-8pt);
    \path (A) ++(20pt,0) coordinate (B) +(0,-8pt);
    \draw (A) .. controls +(0,15pt) and +(0,15pt) .. coordinate(T) (B);
    \path (T) ++(2pt,3pt) node {$\scriptstyle a$};
  \end{tikzpicture} 
  \biggr) 
  =
  \ev_{a}\biggl(
  \begin{tikzpicture}[baseline=(X)]
    \path coordinate (A) ++(0pt,4pt) coordinate (X);
    \path (A) +(0,-8pt);
    \path (A) ++(20pt,0) coordinate (B) +(0,-8pt);
    \draw (A) .. controls +(0,15pt) and +(0,15pt) .. (B);
  \end{tikzpicture} 
  \biggr) 
\end{equation*}
Then \cref{formulaforstar,sumoftrees} give:
\begin{equation}
  \sum_{\Gamma\text{ with }\star_0\text{ and }\star_1} \frac{(i\hbar)^{-\chi(\Gamma)}\ev(\Gamma)}{\left|\Aut\Gamma\right|} = \sum_{\Gamma\text{ decorated}} \frac{(i\hbar)^{-\chi(\Gamma)}\ev(\Gamma)}{\left|\Aut\Gamma\right|}
\end{equation}
By ``$\Gamma$ decorated'' we mean that the sum ranges over all diagrams made from the following ingredients (in the vertices, we require $n\geq 3$):
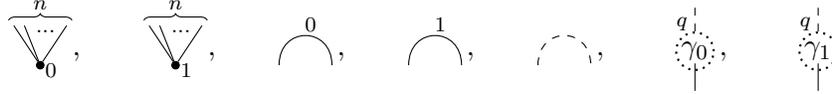
\[
  \begin{tikzpicture}[baseline=(X)]
    \path node[dot] (O) {} +(4pt,-2pt) node {$\scriptstyle 0$} ++(0pt,4pt) coordinate (X);
    \draw (O) -- ++(-10pt,15pt) ++(-2pt,2pt) coordinate (n1);
    \draw (O) -- ++(-6pt,15pt);
    \draw (O) -- ++(10pt,15pt) ++(2pt,2pt) coordinate (n2);
    \path (O) ++(2pt,13pt) node {$\scriptstyle \ldots$};
    \draw[decorate,decoration=brace] (n1) -- coordinate(n) (n2);
    \path (n) ++(0,4pt) node[anchor=base] {$\scriptstyle n$};
  \end{tikzpicture}
  ,\quad\quad
  \begin{tikzpicture}[baseline=(X)]
    \path node[dot] (O) {} +(4pt,-2pt) node {$\scriptstyle 1$} ++(0pt,4pt) coordinate (X);
    \draw (O) -- ++(-10pt,15pt) ++(-2pt,2pt) coordinate (n1);
    \draw (O) -- ++(-6pt,15pt);
    \draw (O) -- ++(10pt,15pt) ++(2pt,2pt) coordinate (n2);
    \path (O) ++(2pt,13pt) node {$\scriptstyle \ldots$};
    \draw[decorate,decoration=brace] (n1) -- coordinate(n) (n2);
    \path (n) ++(0,4pt) node[anchor=base] {$\scriptstyle n$};
  \end{tikzpicture}
  ,\quad\quad
  \begin{tikzpicture}[baseline=(X)]
    \path coordinate (A) ++(0pt,4pt) coordinate (X);
    \path (A) +(0,-8pt);
    \path (A) ++(20pt,0) coordinate (B) +(0,-8pt);
    \draw (A) .. controls +(0,15pt) and +(0,15pt) .. coordinate(T) (B);
    \path (T) ++(2pt,4pt) node {$\scriptstyle 0$};
  \end{tikzpicture} 
  \,,\quad\quad
  \begin{tikzpicture}[baseline=(X)]
    \path coordinate (A) ++(0pt,4pt) coordinate (X);
    \path (A) +(0,-8pt);
    \path (A) ++(20pt,0) coordinate (B) +(0,-8pt);
    \draw (A) .. controls +(0,15pt) and +(0,15pt) .. coordinate(T) (B);
    \path (T) ++(2pt,4pt) node {$\scriptstyle 1$};
  \end{tikzpicture} 
  \,,\quad\quad
  \begin{tikzpicture}[baseline=(X)]
    \path coordinate (A) ++(0pt,4pt) coordinate (X);
    \path (A) +(0,-8pt);
    \path (A) ++(20pt,0) coordinate (B) +(0,-8pt);
    \draw[dashed] (A) .. controls +(0,15pt) and +(0,15pt) .. coordinate(T) (B);
  \end{tikzpicture} 
  \,,\quad\quad
  \begin{tikzpicture}[baseline=(gamma.base)]
    \path node[inner sep=1pt] (gamma) {$\gamma_0$};
    \draw (gamma) -- ++(0,-15pt);
    \path (gamma) node (C) [draw,circle,dotted,thick,inner sep=5pt] {};
    \draw[dashed] (C) -- ++(0,18pt);    
    \path (C.north) ++(1pt,3pt) node[anchor= east] {$\scriptstyle q$};
  \end{tikzpicture}  
  \,,\quad\quad
  \begin{tikzpicture}[baseline=(gamma.base)]
    \path node[inner sep=1pt] (gamma) {$\gamma_1$};
    \draw (gamma) -- ++(0,-15pt);
    \path (gamma) node (C) [draw,circle,dotted,thick,inner sep=5pt] {};
    \draw[dashed] (C) -- ++(0,18pt);    
    \path (C.north) ++(1pt,3pt) node[anchor= east] {$\scriptstyle q$};
  \end{tikzpicture}  
\]

Since $\int_{t_0}^{t_1} = \int_{t_0}^{t} + \int_{t}^{t_1}$, the vertices in $U_{\gamma}$ decompose as:   $
  \begin{tikzpicture}[baseline=(X)]
    \path node[dot] (O) {}  ++(0pt,4pt) coordinate (X);
    \draw[twice] (O) -- ++(-10pt,15pt) ++(-2pt,2pt) coordinate (n1);
    \draw[twice] (O) -- ++(-6pt,15pt);
    \draw[twice] (O) -- ++(10pt,15pt) ++(2pt,2pt) coordinate (n2);
    \path (O) ++(2pt,13pt) node {$\scriptstyle \ldots$};
    \draw[decorate,decoration=brace] (n1) -- coordinate(n) (n2);
    \path (n) ++(0,4pt) node[anchor=base] {$\scriptstyle n$};
  \end{tikzpicture}
  =
  \begin{tikzpicture}[baseline=(X)]
    \path node[dot] (O) {} +(4pt,-2pt) node {$\scriptstyle 0$} ++(0pt,4pt) coordinate (X);
    \draw (O) -- ++(-10pt,15pt) ++(-2pt,2pt) coordinate (n1);
    \draw (O) -- ++(-6pt,15pt);
    \draw (O) -- ++(10pt,15pt) ++(2pt,2pt) coordinate (n2);
    \path (O) ++(2pt,13pt) node {$\scriptstyle \ldots$};
    \draw[decorate,decoration=brace] (n1) -- coordinate(n) (n2);
    \path (n) ++(0,4pt) node[anchor=base] {$\scriptstyle n$};
  \end{tikzpicture}
  +
  \begin{tikzpicture}[baseline=(X)]
    \path node[dot] (O) {} +(4pt,-2pt) node {$\scriptstyle 1$} ++(0pt,4pt) coordinate (X);
    \draw (O) -- ++(-10pt,15pt) ++(-2pt,2pt) coordinate (n1);
    \draw (O) -- ++(-6pt,15pt);
    \draw (O) -- ++(10pt,15pt) ++(2pt,2pt) coordinate (n2);
    \path (O) ++(2pt,13pt) node {$\scriptstyle \ldots$};
    \draw[decorate,decoration=brace] (n1) -- coordinate(n) (n2);
    \path (n) ++(0,4pt) node[anchor=base] {$\scriptstyle n$};
  \end{tikzpicture}
$.  Similarly, we will decompose the Green's function $  \begin{tikzpicture}[baseline=(X)]
    \path coordinate (A) ++(0pt,2pt) coordinate (X);
    \path (A) +(0,-8pt);
    \path (A) ++(20pt,0) coordinate (B) +(0,-8pt);
    \draw[twice] (A) .. controls +(0,15pt) and +(0,15pt) .. coordinate(T) (B);
  \end{tikzpicture} 
= G_{\gamma}$ by showing that a certain sum of six terms satisfies its defining relation.

We claim that if $\xi: [t_0,t_1] \to \RR^d$ has $\xi(t_0) = 0$, then:
\begin{equation}  \label{aclaimtocheck}
  \begin{tikzpicture}[baseline=(X)]
    \path node[dot] (O) {} ++(0pt,10pt) coordinate (X);
    \draw (O) -- ++(-15pt,22pt)  +(0,4pt) node[anchor=base] {$\scriptstyle \res_0(\xi)$};
    \draw (O) .. controls +(8pt,25pt) and +(-8pt,25pt) .. coordinate (t) ++(30pt,0) +(0,-8pt) node[anchor=base] {$\scriptstyle \varsigma,i$};
    \path (O) node[anchor=east] {$\scriptstyle 0$};
    \path (t) node[anchor=south west] {$\scriptstyle 0$};
  \end{tikzpicture}
  +
  \begin{tikzpicture}[baseline=(X)]
    \path node[dot] (O) {} ++(0pt,10pt) coordinate (X);
    \draw (O) -- ++(-15pt,22pt)  +(0,4pt) node[anchor=base] {$\scriptstyle \res_0(\xi)$};
    \path (O) ++(10pt,15pt) node[inner sep=1pt] (gamma1) {$\gamma_0$};
    \draw (O) -- (gamma1);
    \path (gamma1) node (C1) [draw,circle,dotted,thick,inner sep=5pt] {};
    \path (gamma1) ++(30pt,0) node[inner sep=1pt] (gamma2) {$\gamma_0$};
    \draw (gamma2) -- ++(0,-15pt) +(0,-8pt) node[anchor=base] {$\scriptstyle \varsigma,i$};
    \path (gamma2) node (C2) [draw,circle,dotted,thick,inner sep=5pt] {};
    \draw[dashed] (C1) .. controls +(10pt,15pt) and +(-10pt,15pt) .. (C2);  
    \path (O) node[anchor=east] {$\scriptstyle 0$};
    \path (C1.north east) node[anchor=south] {$\scriptstyle q$};
    \path (C2.north west) node[anchor=south] {$\scriptstyle q$};
  \end{tikzpicture}\,
  +
  \begin{tikzpicture}[baseline=(X)]
    \path node[dot] (O) {} ++(0pt,10pt) coordinate (X);
    \draw (O) -- ++(-15pt,22pt)  +(0,4pt) node[anchor=base] {$\scriptstyle \res_1(\xi)$};
    \path (O) ++(10pt,15pt) node[inner sep=1pt] (gamma1) {$\gamma_1$};
    \draw (O) -- (gamma1);
    \path (gamma1) node (C1) [draw,circle,dotted,thick,inner sep=5pt] {};
    \path (gamma1) ++(30pt,0) node[inner sep=1pt] (gamma2) {$\gamma_0$};
    \draw (gamma2) -- ++(0,-15pt) +(0,-8pt) node[anchor=base] {$\scriptstyle \varsigma,i$};
    \path (gamma2) node (C2) [draw,circle,dotted,thick,inner sep=5pt] {};
    \draw[dashed] (C1) .. controls +(10pt,15pt) and +(-10pt,15pt) .. (C2);  
    \path (O) node[anchor=east] {$\scriptstyle 1$};
    \path (C1.north east) node[anchor=south] {$\scriptstyle q$};
    \path (C2.north west) node[anchor=south] {$\scriptstyle q$};
  \end{tikzpicture}
  = - \res_0(\xi)^i(\varsigma)
\end{equation}
Indeed, integrating by parts gives:
\begin{multline}
  \begin{tikzpicture}[baseline=(X)]
    \path node[dot] (O) {} ++(0pt,15pt) coordinate (X);
    \draw (O) -- ++(-15pt,22pt)  +(0,4pt) node[anchor=base] {$\scriptstyle \res_0(\xi)$};
    \path (O) ++(10pt,15pt) node[inner sep=1pt] (gamma1) {$\gamma_0$};
    \draw (O) -- (gamma1);
    \path (gamma1) node (C1) [draw,circle,dotted,thick,inner sep=5pt] {};
    \draw[dashed] (C1) -- ++(0,15pt) +(0,3pt) node[anchor=base] {$\scriptstyle k$};
    \path (O) node[anchor= east] {$\scriptstyle 0$};
    \path (C1.north east) node[anchor=south] {$\scriptstyle q$};
  \end{tikzpicture}    
  = 0 - \left.\left( \xi^i \frac{\partial^2 L}{\partial v^i\partial v^j} \frac{\partial \dot\gamma_0^j}{\partial q^k} + \xi^i  \frac{\partial^2 L}{\partial v^i\partial q^j} \frac{\partial \gamma_0^j}{\partial q^k} \right)\right|_{t} =  \\
    = -\xi^i(t)\frac{\partial}{\partial q^k} \left[ \left. \frac{\partial L}{\partial v^i}\right|_{\gamma_0(t_0,q_0,t,q;t)}\right]  
    = \xi^i(t)\frac{\partial^2[-S_0]}{\partial q^k\partial q^i} 
\end{multline}
Similarly $
  \begin{tikzpicture}[baseline=(X)]
    \path node[dot] (O) {} ++(0pt,15pt) coordinate (X);
    \draw (O) -- ++(-15pt,22pt)  +(0,4pt) node[anchor=base] {$\scriptstyle \res_1(\xi)$};
    \path (O) ++(10pt,15pt) node[inner sep=1pt] (gamma1) {$\gamma_1$};
    \draw (O) -- (gamma1);
    \path (gamma1) node (C1) [draw,circle,dotted,thick,inner sep=5pt] {};
    \draw[dashed] (C1) -- ++(0,15pt) +(0,3pt) node[anchor=base] {$\scriptstyle k$};
    \path (O) node[anchor= east] {$\scriptstyle 1$};
    \path (C1.north east) node[anchor=south] {$\scriptstyle q$};
  \end{tikzpicture}    
  =   \xi^i(t) \frac{\partial^2[-S_1]}{\partial q^k\partial q^i} 
$.
We contract with \!\!\!\begin{tikzpicture}[baseline=(gamma2.base)]
    \path coordinate (gamma1);
    \path (gamma1) ++(30pt,0) node[inner sep=1pt] (gamma2) {$\gamma_0$};
    \draw (gamma2) -- ++(0,-10pt);
    \path (gamma2) node (C2) [draw,circle,dotted,thick,inner sep=5pt] {};
    \path (gamma1) node (C1) [circle,inner sep=5pt] {};
    \draw[dashed] (C1) .. controls +(10pt,15pt) and +(-10pt,15pt) .. (C2);  
    \path (C2.north west) node[anchor=south] {$\scriptstyle q$};
  \end{tikzpicture}:
\begin{multline}
  \begin{tikzpicture}[baseline=(X)]
    \path node[dot] (O) {} ++(0pt,10pt) coordinate (X);
    \draw (O) -- ++(-15pt,22pt)  +(0,4pt) node[anchor=base] {$\scriptstyle \res_0(\xi)$};
    \path (O) ++(10pt,15pt) node[inner sep=1pt] (gamma1) {$\gamma_0$};
    \draw (O) -- (gamma1);
    \path (gamma1) node (C1) [draw,circle,dotted,thick,inner sep=5pt] {};
    \path (gamma1) ++(30pt,0) node[inner sep=1pt] (gamma2) {$\gamma_0$};
    \draw (gamma2) -- ++(0,-15pt) +(0,-8pt) node[anchor=base] {$\scriptstyle \varsigma,j$};
    \path (gamma2) node (C2) [draw,circle,dotted,thick,inner sep=5pt] {};
    \draw[dashed] (C1) .. controls +(10pt,15pt) and +(-10pt,15pt) .. (C2);  
    \path (O) node[anchor=east] {$\scriptstyle 0$};
    \path (C1.north east) node[anchor=south] {$\scriptstyle q$};
    \path (C2.north west) node[anchor=south] {$\scriptstyle q$};
  \end{tikzpicture}\,
  +
  \begin{tikzpicture}[baseline=(X)]
    \path node[dot] (O) {} ++(0pt,10pt) coordinate (X);
    \draw (O) -- ++(-15pt,22pt)  +(0,4pt) node[anchor=base] {$\scriptstyle \res_1(\xi)$};
    \path (O) ++(10pt,15pt) node[inner sep=1pt] (gamma1) {$\gamma_1$};
    \draw (O) -- (gamma1);
    \path (gamma1) node (C1) [draw,circle,dotted,thick,inner sep=5pt] {};
    \path (gamma1) ++(30pt,0) node[inner sep=1pt] (gamma2) {$\gamma_0$};
    \draw (gamma2) -- ++(0,-15pt) +(0,-8pt) node[anchor=base] {$\scriptstyle \varsigma,j$};
    \path (gamma2) node (C2) [draw,circle,dotted,thick,inner sep=5pt] {};
    \draw[dashed] (C1) .. controls +(10pt,15pt) and +(-10pt,15pt) .. (C2);  
    \path (O) node[anchor=east] {$\scriptstyle 1$};
    \path (C1.north east) node[anchor=south] {$\scriptstyle q$};
    \path (C2.north west) node[anchor=south] {$\scriptstyle q$};
  \end{tikzpicture}  \, = \\
  = \left(  \xi^i(t )\frac{\partial^2[-S_0]}{\partial q^k\partial q^i}  + \xi^i(t )\frac{\partial^2[-S_1]}{\partial q^k\partial q^i}  \right) \left( \Bigl( \frac{\partial^2 [S_0 + S_1]}{\partial q\partial q} \bigr)^{-1} \right)^{kl} \frac{\partial \gamma_0^j}{\partial q^l}(\varsigma)= \\
   = - \xi^i(t ) \frac{\partial \gamma_0^j}{\partial q^i}(\varsigma)
\end{multline}
On the other hand, by another integration by parts and recalling \cref{Greeneqntheother,greeneqn}:
\begin{multline}
  \begin{tikzpicture}[baseline=(X)]
    \path node[dot] (O) {} ++(0pt,10pt) coordinate (X);
    \draw (O) -- ++(-15pt,22pt)  +(0,4pt) node[anchor=base] {$\scriptstyle \res_0(\xi)$};
    \draw (O) .. controls +(8pt,25pt) and +(-8pt,25pt) .. coordinate (t) ++(30pt,0) +(0,-8pt) node[anchor=base] {$\scriptstyle \varsigma,i$};
    \path (O) node[anchor=east] {$\scriptstyle 0$};
    \path (t) node[anchor=south west] {$\scriptstyle 0$};
  \end{tikzpicture}
  = \int_{t_0}^{t} \xi^k(\tau)\,\Dd_{1,jk}[G_0^{ij}(\varsigma,\tau)]\,d\tau - \left[ \xi^k(\tau)\,\frac{\partial^2 L}{\partial v^j \partial v^k} \,\frac{\partial G_0^{ij}}{\partial \tau}(\varsigma,\tau) + \xi^k(\tau)\, \frac{\partial^2 L}{\partial q^j \partial v^k} \,\frac{\partial G_0^{ij}}{\partial \tau}(\varsigma,\tau) \right]_{t = t} \\
  = -\int_{t_0}^{t} \xi^k(\tau)\, \delta_k^i \delta(\tau - \varsigma)\,d\tau - \xi^k(t)\, \frac{\partial^2 L}{\partial v^j \partial v^k}\,\frac{\partial G_0^{ij}}{\partial \tau}(\varsigma,t) = \\
  = -\xi^i(\varsigma) + \xi^k(t)\, \frac{\partial \gamma_0^i}{\partial q^k}(\varsigma)
\end{multline}
This proves \cref{aclaimtocheck}.  Along with a similar formula for $\res_1(\xi)$ when $\xi(t_1) = 0$, we have proven that:
\begin{equation}
    \begin{tikzpicture}[baseline=(X)]
    \path coordinate (O) +(0pt,8pt) coordinate (X);
    \draw[twice] (O) .. controls +(0pt,25pt) and +(-0pt,25pt) .. coordinate (t) ++(30pt,0);
  \end{tikzpicture}
 = 
    \begin{tikzpicture}[baseline=(X)]
    \path coordinate (O) +(0pt,8pt) coordinate (X);
    \draw (O) .. controls +(0pt,25pt) and +(-0pt,25pt) .. coordinate (t) ++(30pt,0);
    \path (t) node[anchor=south west] {$\scriptstyle 0$};
  \end{tikzpicture}
+
  \begin{tikzpicture}[baseline=(X)]
    \path coordinate (O) +(0pt,8pt) coordinate (X);
    \draw (O) .. controls +(0pt,25pt) and +(-0pt,25pt) .. coordinate (t) ++(30pt,0);
    \path (t) node[anchor=south west] {$\scriptstyle 1$};
  \end{tikzpicture}
+
  \begin{tikzpicture}[baseline=(X)]
    \path node[inner sep=1pt] (gamma1) {$\gamma_0$};
    \draw (gamma1) -- ++(0,-15pt) coordinate (O);
    \path (O) +(0pt,8pt) coordinate (X);
    \path (gamma1) node (C1) [draw,circle,dotted,thick,inner sep=5pt] {};
    \path (gamma1) ++(30pt,0) node[inner sep=1pt] (gamma2) {$\gamma_0$};
    \draw (gamma2) -- ++(0,-15pt);
    \path (gamma2) node (C2) [draw,circle,dotted,thick,inner sep=5pt] {};
    \draw[dashed] (C1) .. controls +(10pt,15pt) and +(-10pt,15pt) .. (C2);  
    \path (C1.north) ++(1pt,3pt) node[anchor= east] {$\scriptstyle q$};
    \path (C2.north) ++(-1pt,3pt) node[anchor= west] {$\scriptstyle q$};
  \end{tikzpicture}  
+
  \begin{tikzpicture}[baseline=(X)]
    \path node[inner sep=1pt] (gamma1) {$\gamma_0$};
    \draw (gamma1) -- ++(0,-15pt) coordinate (O);
    \path (O) +(0pt,8pt) coordinate (X);
    \path (gamma1) node (C1) [draw,circle,dotted,thick,inner sep=5pt] {};
    \path (gamma1) ++(30pt,0) node[inner sep=1pt] (gamma2) {$\gamma_1$};
    \draw (gamma2) -- ++(0,-15pt);
    \path (gamma2) node (C2) [draw,circle,dotted,thick,inner sep=5pt] {};
    \draw[dashed] (C1) .. controls +(10pt,15pt) and +(-10pt,15pt) .. (C2);  
    \path (C1.north) ++(1pt,3pt) node[anchor= east] {$\scriptstyle q$};
    \path (C2.north) ++(-1pt,3pt) node[anchor= west] {$\scriptstyle q$};
  \end{tikzpicture}  
+
  \begin{tikzpicture}[baseline=(X)]
    \path node[inner sep=1pt] (gamma1) {$\gamma_1$};
    \draw (gamma1) -- ++(0,-15pt) coordinate (O);
    \path (O) +(0pt,8pt) coordinate (X);
    \path (gamma1) node (C1) [draw,circle,dotted,thick,inner sep=5pt] {};
    \path (gamma1) ++(30pt,0) node[inner sep=1pt] (gamma2) {$\gamma_0$};
    \draw (gamma2) -- ++(0,-15pt);
    \path (gamma2) node (C2) [draw,circle,dotted,thick,inner sep=5pt] {};
    \draw[dashed] (C1) .. controls +(10pt,15pt) and +(-10pt,15pt) .. (C2);  
    \path (C1.north) ++(1pt,3pt) node[anchor= east] {$\scriptstyle q$};
    \path (C2.north) ++(-1pt,3pt) node[anchor= west] {$\scriptstyle q$};
  \end{tikzpicture}  
+
  \begin{tikzpicture}[baseline=(X)]
    \path node[inner sep=1pt] (gamma1) {$\gamma_1$};
    \draw (gamma1) -- ++(0,-15pt) coordinate (O);
    \path (O) +(0pt,8pt) coordinate (X);
    \path (gamma1) node (C1) [draw,circle,dotted,thick,inner sep=5pt] {};
    \path (gamma1) ++(30pt,0) node[inner sep=1pt] (gamma2) {$\gamma_1$};
    \draw (gamma2) -- ++(0,-15pt);
    \path (gamma2) node (C2) [draw,circle,dotted,thick,inner sep=5pt] {};
    \draw[dashed] (C1) .. controls +(10pt,15pt) and +(-10pt,15pt) .. (C2);  
    \path (C1.north) ++(1pt,3pt) node[anchor= east] {$\scriptstyle q$};
    \path (C2.north) ++(-1pt,3pt) node[anchor= west] {$\scriptstyle q$};
  \end{tikzpicture}  
\end{equation}
as the right-hand side satisfies the defining relation of the left-hand side (the boundary conditions are clear).  Because this sum comprises precisely the possible ways to connect vertices in the sum of diagrams in $\int^{\formal} U_{\gamma_0}U_{\gamma_1}$, the sums of diagrams in $U_{\gamma}$ and in $\int^{\formal} U_{\gamma_0}U_{\gamma_1}$ match identically.  This completes the proof of \cref{fubinithm}. \qedhere

\Section{Handling ultraviolet divergences} \label{divsection}

Although \cref{coordfreethm} is formally true even when the formal path integral of \cref{pathintdefn} has ultraviolet divergences, in \cref{fubinithm} we required that all ultraviolet divergences cancel.  In this section, we discuss the ultraviolet divergences that can occur in the formal-path-integral approach to quantum mechanics, thereby completing the proof of the Main Theorem.  

Ultraviolet divergences in quantum-mechanical path integrals have been addressed before.  Notably, Manuel and Tarrach \cite{MT1994} consider the case of motion on flat space with a mildly divergent potential (too divergent and the problem is hopeless: the corresponding Heisenberg operator is not self-adjoint).  They handle the corresponding path-integral divergences through a system of regularization and renormalization with counterterms, just as is standardly done in quantum field theory, and achieve finite physical results.  Closer to our approach, Kleinert and Chervyakov \cite{KC1,KC2,KC3,KC4,KC5,KC6} discuss the divergences that arise from using ``the wrong coordinates'' --- as we will see in this section, divergences do not arise when using the ``correct'' volume form --- within the framework of dimensional renormalization.  In both approaches, Planck's constant $\hbar$ is set to unity from the beginning, forcing the authors to incorporate their perturbation parameters into the potential energy functions.  Our semiclassical ($\hbar \approx 0$) approach allows us more freedom to consider quantizations of very nonlinear classical theories, and in particular Riemannian manifolds in which the metric is not flat.

We begin this section with a natural example in \cref{divergenteg} to emphasize that, even in the case of totally smooth classical (flat!) physics, ultraviolet divergences really can be a problem --- this is in contradiction with the many texts that suggest that, absent singular potentials, ultraviolet divergences are a feature of quantum field theories in dimensions $2$ and higher.  In \cref{DVTsub}, however, we prove that in the most important case of ``nonrelativistic'' quantum mechanics on a manifold, if the metric and measure are compatible then all ultraviolet divergences cancel.

We remark that even in ``divergence-free'' path integrals, individual Feynman diagrams may represent divergent integrals.  We say that a path integral \define{has no ultraviolet divergences} if for each $n$, the divergent parts of the integrals that contribute to the coefficient of $\hbar^n$ in the path integral cancel.  One can express this in several equivalent ways.  The divergent part of any of our Feynman diagrams is always of the form ``$\int \bigl(\delta(0)\bigr)^m[\dots]$'', where ``$\int$'' represents some finite-dimensional integral and ``$[\dots]$'' some piecewise-continuous bounded function.  
Our informal approach will be to write $\int \bigl(\delta(0)\bigr)^m[\dots] =  \bigl(\delta(0)\bigr)^m \int[\dots]$, and assert that at the end of the day the polynomial in $\delta(0)$ is degree-zero.  
Although somewhat obscured, our proof finds coordinates in which to write the sum of all diagrams with the same Euler characteristic as a single integral, and show that for this integral, the integrand is bounded.  Alternately, one can introduce a small parameter $\epsilon$, replace the Green's function by a smoothing that differs from the true Green's function only by $\epsilon$, and prove that as $\epsilon \to 0$, the limit of the corresponding ``regularized'' formal path integral is a well-defined power series in $\hbar$.

\subsection{A divergent example: geodesic motion on \texorpdfstring{$\RR$}{R} in the wrong coordinates}
  \label{divergenteg}
We consider free motion on the line in exponential coordinates.  Under the map $x \mapsto q = \exp x$, the usual Lagrangian $L = \frac12 \dot x^2$ transforms to:
\[ L(v,q) = \frac12 \frac{v^2}{q^2} \]
Here $q$ is the usual coordinate on $\Nn = \RR_{>0}$ and $v$ is the corresponding fiber coordinate --- we can trivialize the tangent bundle as $\T\Nn = \RR \times \RR_{> 0}$.  The Euler-Lagrange equations of motion in these coordinates are:
\[ \frac{\ddot \gamma}{\gamma^2} - 2\frac{\dot \gamma^2}{\gamma^3} = -\frac{\dot \gamma^2}{\gamma^3} \]
and so $\gamma(\tau) = q_0^{(t_1 - \tau)/(t_1 - t_0)}q_1^{(\tau - t_0)/(t_1 - t_0)}$ is the unique solution with $\gamma(t_0) = q_0$, $\gamma(t_1) = q_1$.  For notational convenience, we write $\ell = \frac1{t_1 - t_0} \log \frac{q_1}{q_0}$.  Then $\dot \gamma / \gamma = \ell$, and the Green function $G(\varsigma,\tau)$ must satisfy:
\[ \left( \frac{\d}{\d\tau} - \ell\right)^2 G = -\delta(\tau-\varsigma)\,\gamma(\tau)^2 \]
The solution with $G(\varsigma,t_0) = 0 = G(\varsigma,t_1)$ is:
\[ G(\varsigma,\tau) = \gamma(\varsigma)\,\gamma(\tau)\, \left( \frac12(\varsigma+\tau) - \frac{\varsigma\tau}{t_1 - t_0} - \frac12|\tau-\varsigma|\right) \]
Then $G, \frac{\partial G}{\partial \varsigma}, \frac{\partial G}{\partial \tau}$ are bounded, but:
\[ \frac{\partial^2 G}{\partial \varsigma\partial \tau} = \ell^2 \,G - \ell\, \gamma(\varsigma)\,\gamma(\tau) \left(\frac{\varsigma+\tau}{t_1 - t_0}\right) + \gamma(\varsigma)\,\gamma(\tau) \left( -\frac1{t_1 - t_0} + \delta(\tau-\varsigma)\right) \]

The derivatives of the Lagrangian are:
\[ \left. \frac{\partial^nL}{\partial v^k\partial q^{n-k}}\right|_{(v,q) = (\dot\gamma(\tau),\gamma(\tau))} = (-1)^{n-k} \frac{(n+1-k)!}{(2-k)!} \frac{\ell^k}{\gamma(\tau)^n} \]
where by convention $(-m)! = \pm\infty$ for $m$ a positive integer, so $1/(2-k)!$ vanishes for $k \geq 3$.

We can now (try to) evaluate any diagram we wish.  There are no diagrams with one loop, and three diagrams with two loops:
\[ 
  \hspace{-1.5ex}\begin{tikzpicture}[baseline=(X)]
    \path node[dot] (O) {} ++ (100pt,0pt) node[dot] (P) {} ++ (90pt,0pt) node[dot] (Q) {};
    \draw (O) .. controls +(20pt,20pt) and +(0pt,25pt) .. (O);
    \path (O) ++(-20pt,0pt) node[dot] (O1) {};
    \draw (O1) .. controls +(-20pt,20pt) and +(0pt,25pt) .. (O1);
    \draw (O) .. controls +(-7pt,20pt) and +(7pt,20pt) .. (O1);
    \path (P) ++(-20pt,0pt) node[dot] (P1) {};
    \draw (P) .. controls +(-7pt,10pt) and +(7pt,10pt) .. (P1);
    \draw (P) .. controls +(0pt,17pt) and +(0pt,17pt) .. (P1);
    \draw (P) .. controls +(15pt,25pt) and +(-15pt,25pt) .. (P1);
    \draw (Q) .. controls +(30pt,20pt) and +(10pt,25pt) .. (Q);
    \draw (Q) .. controls +(-30pt,20pt) and +(-10pt,25pt) .. (Q);
    \path (O) ++(-10pt,-20pt) node[anchor=base,text centered] {``barbell''};
    \path (P) ++(-10pt,-20pt) node[anchor=base,text centered] {``theta''};
    \path (Q) ++(0pt,-20pt) node[anchor=base,text centered] {``infinity sign''};
  \end{tikzpicture}\hspace{-2ex}
\]
Whenever the Green's function is twice-differentiated, it contributes a $\delta$ function to the integrand.  These become problems for loops in diagrams, as then there can be as many $\delta$ functions as integration variables.  In particular,
the values of the infinity and theta graphs are of the form $[\text{finite}]\delta(0) + [\text{finite}]$.  In the barbell, the two loops do not overlap, and each one can diverge.  Thus, in addition to terms of the form  $[\text{finite}]\delta(0) + [\text{finite}]$, the barbell has a divergence equal to $\frac{(t_1-t_0)^3}{24}(\delta(0))^2$, after performing all integrals.  This term will not be canceled by divergences from other diagrams.

\subsection{Nonrelativistic quantum mechanics is divergence-free} \label{DVTsub}

As \cref{divergenteg} shows, there can be ultraviolet divergences even for geodesic flow on flat space.  However, in this section, we show that the problem in \cref{divergenteg} really is that the coordinates are wrong, in the sense that they are not compatible with the volume form induced by the metric:

\begin{thm} \label{divfreethm}
  Let $L$ be a Lagrangian on $\RR^d$ of the form $L(\tau,v,q) = \frac12 a_{ij}(\tau,q)\,v^iv^j + b_i(\tau,q)v^i + c(\tau,q)$, where $a_{ij}(\tau,q) = a_{ji}(\tau,q)$ and $\det a(\tau,q) = 1$ for all $(\tau,q)\in \RR^{d+1}$.  Then the formal path integral for $L$ has no ultraviolet divergences.
\end{thm}

In order for the Morse index in \cref{pathintdefn} to be defined, we in fact need $a_{ij}(q)$ to be positive-definite for each $q$, so it provides some metric on $\RR^d$, and $\frac12 a_{ij}(q)\,v^iv^j$ is a ``kinetic energy'' term.  The one-form $b$ is a ``magnetic potential'' on $\RR^d$, and the function $c$ is an ``electric potential,'' so the Lagrangian in \cref{divfreethm} describes the classical nonrelativistic motion of a charged particle moving through an external electromagnetic field on a curved background.

Consider now the case when $\Nn$ is a Riemannian manifold with metric $a$, a one-form $b$, and a function $c$.  The metric determines a volume form $\sqrt{\det a}$ on $\Nn$.  If we find local coordinates compatible with this volume form, then in local coordinates the Lagrangian $L(v,q) = \frac12 a(q)\cdot v^2 + b(q)\cdot v + c(q)$ is of the form in \cref{divfreethm}, and the compatibility condition guarantees that in these coordinates $\det a_{ij} = 1$.  Along with \cref{coordfreethm,fubinithm}, we can compute formal path integrals for the system $(\Nn,\sqrt{\det a},L)$ by cutting and pasting and finding local volume-compatible coordinates.  Even if $a,b,c$ depend on an external time variable, \cite{Moser1965} guarantees the existence of (time-dependent) volume-compatible local coordinates.  Therefore \cref{divfreethm} completes the proof of our Main Theorem.

\begin{proof}[of \cref{divfreethm}]
\Cref{greendefn} implies that:
\begin{equation}\label{DF1}
\frac{\partial^2}{\partial \varsigma \partial \tau}  \biggl[  \begin{tikzpicture}[baseline=(X)]
    \path coordinate (A) ++(0pt,2pt) coordinate (X);
    \path (A) +(0,-8pt) node[anchor=base] {$\scriptstyle \varsigma,i$};
    \path (A) ++(20pt,0) coordinate (B) +(0,-8pt) node[anchor=base] {$\scriptstyle \tau,j$};
    \draw (A) .. controls +(0,15pt) and +(0,15pt) .. (B);
  \end{tikzpicture} \biggr]
  = \delta(\varsigma-\tau)\,\bigl(a(\tau,\gamma(\tau))^{-1}\bigr)^{ij} + \text{finite}
\end{equation}
This is the only source of divergences in Feynman diagrams: regardless of the type of Lagrangian, provided it depends on position and velocity alone no vertex differentiates an incoming edge more than once, so no Green's function is differentiated more than twice.  Note that by and large, Dirac-delta functions are not a problem in integrals: they simply identify integration variables.  So the $\delta(\varsigma-\tau)$ in \cref{DF1} is a problem only when $\varsigma$ and $\tau$ are already identified.  This can happen only when the Feynman diagram has a loop of Green's functions, all of which are differentiated twice: ultraviolet divergences live on loops in Feynman diagrams.

However, since the Lagrangian $L$ is quadratic in velocity, no vertex differentiates more than two of its incoming edges.  Thus, divergent loops in the same Feynman diagram cannot intersect.  Our strategy, then, is as follows.  For each Euler characteristic, we record all possible Feynman diagrams, expand the summations implicit in each vertex (\cref{Anformula}), and keep only the divergent diagrams, labeling individually the divergent loops.  We can then grade each diagram by the multiset that records the number of external edges attached to each divergent loop.  By ``pulling the loops far away from each other,'' one can express the sum of divergent Feynman diagrams as essentially the exponential of a sum of individual divergent loops, contracted with some convergent parts.  In particular, to prove \cref{divfreethm}, it suffices to prove the following lemma:
\begin{lemma}  \label{DFlemma}
  For each $n$, we have:
  \begin{equation} \label{DFlemma1}
    \sum_{\substack{\text{loops $\Gamma$ with} \\ \text{$n$ exterior edges}}} \frac{\ev(\Gamma)}{\left|\Aut\Gamma\right|}  = \text{finite}
  \end{equation}
\end{lemma}

In \cref{DFlemma1}, the $n$ external edges are ordered and contracted with based loops $\xi_1,\dots,\xi_n$.  For example, the left-hand side of \cref{DFlemma1} for $n=1,2,3$ are:
\begin{gather*}
  n=1: \quad
  \frac12
  \begin{tikzpicture}[baseline=(X)]
    \path node[dot] (O) {} ++(0pt,6pt) coordinate (X);
    \draw (O) .. controls +(20pt,20pt) and +(0pt,25pt) .. (O);
    \draw (O) -- ++(-10pt,17pt) +(0,3pt) node[anchor=base,black] {$\scriptstyle \xi$};
  \end{tikzpicture} \quad\quad\quad\quad
  n=2: \quad
  \frac12
  \begin{tikzpicture}[baseline=(X)]
    \path node[dot] (O) {} ++(0pt,4pt) coordinate (X);
    \draw (O) .. controls +(20pt,20pt) and +(0pt,25pt) .. (O);
    \draw (O) -- ++(-15pt,17pt) +(0,3pt) node[anchor=base,black] {$\scriptstyle \xi_1$};
    \draw (O) -- ++(-7pt,17pt) +(0,3pt) node[anchor=base,black] {$\scriptstyle \xi_2$};
  \end{tikzpicture}
  \hspace{-1.5ex}
  +\;
  \frac12
  \begin{tikzpicture}[baseline=(X)]
    \path node[dot] (O) {} ++(0pt,4pt) coordinate (X);
    \path (O) ++(-20pt,0pt) node[dot] (O1) {};
    \draw (O) .. controls +(-7pt,10pt) and +(7pt,10pt) .. (O1);
    \draw (O) .. controls +(0pt,17pt) and +(0pt,17pt) .. (O1);
    \draw (O) -- ++(7pt,17pt) +(0,3pt) node[anchor=base,black] {$\scriptstyle \xi_2$};
    \draw (O1) -- ++(-7pt,17pt) +(0,3pt) node[anchor=base,black] {$\scriptstyle \xi_1$};
  \end{tikzpicture}
  \\
  n=3 : \quad 
  \frac12
  \begin{tikzpicture}[baseline=(X)]
    \path node[dot] (O) {} ++(0pt,4pt) coordinate (X);
    \draw (O) .. controls +(20pt,20pt) and +(0pt,25pt) .. (O);
    \draw (O) -- ++(-23pt,17pt) +(0,3pt) node[anchor=base,black] {$\scriptstyle \xi_1$};
    \draw (O) -- ++(-15pt,17pt) +(0,3pt) node[anchor=base,black] {$\scriptstyle \xi_2$};
    \draw (O) -- ++(-7pt,17pt) +(0,3pt) node[anchor=base,black] {$\scriptstyle \xi_3$};
  \end{tikzpicture}
  \hspace{-1.5ex}
  +\;
  \frac12
  \begin{tikzpicture}[baseline=(X)]
    \path node[dot] (O) {} ++(0pt,4pt) coordinate (X);
    \path (O) ++(-20pt,0pt) node[dot] (O1) {};
    \draw (O) .. controls +(-7pt,10pt) and +(7pt,10pt) .. (O1);
    \draw (O) .. controls +(0pt,17pt) and +(0pt,17pt) .. (O1);
    \draw (O) -- ++(7pt,17pt) +(0,3pt) node[anchor=base,black] {$\scriptstyle \xi_3$};
    \draw (O1) -- ++(-7pt,17pt) +(0,3pt) node[anchor=base,black] {$\scriptstyle \xi_2$};
    \draw (O1) -- ++(-15pt,17pt) +(0,3pt) node[anchor=base,black] {$\scriptstyle \xi_1$};
  \end{tikzpicture}
  +
  \begin{tikzpicture}[baseline=(X)]
    \path node[dot] (O) {} ++(0pt,4pt) coordinate (X);
    \path (O) ++(-20pt,0pt) node[dot] (O1) {} ++(-20pt,0pt) node[dot] (O2) {};
    \draw (O) .. controls +(-7pt,10pt) and +(7pt,10pt) .. (O1);
    \draw (O1) .. controls +(-7pt,10pt) and +(7pt,10pt) .. (O2);
    \draw (O) .. controls +(0pt,17pt) and +(0pt,17pt) .. (O2);
    \draw (O) -- ++(7pt,17pt) +(0,3pt) node[anchor=base,black] {$\scriptstyle \xi_3$};
    \draw (O1) -- ++(0pt,17pt) +(0,3pt) node[anchor=base,black] {$\scriptstyle \xi_2$};
    \draw (O2) -- ++(-7pt,17pt) +(0,3pt) node[anchor=base,black] {$\scriptstyle \xi_1$};
  \end{tikzpicture}
\end{gather*}

Consider again \cref{DF1,Anformula}.  Then for $n=1$ the left-hand side of \cref{DFlemma1} is:
\begin{equation} \label{DF2}
  \frac12 \ev \biggl( \begin{tikzpicture}[baseline=(X)]
    \path node[dot] (O) {} ++(0pt,6pt) coordinate (X);
    \draw (O) .. controls +(20pt,20pt) and +(0pt,25pt) .. (O);
    \draw (O) -- ++(-10pt,15pt) +(0,3pt) node[anchor=base,black] {$\scriptstyle \xi$};
  \end{tikzpicture} \hspace{-1.5ex}  \biggr)
     = \frac12  \int_0^t \delta(\tau - \tau)\times \frac{\partial a_{jk}}{\partial q^i} \, \bigl(a(\tau,q)^{-1}\bigr)^{jk}\biggr|_{q = \gamma(\tau)}\xi(\tau)^i\,d\tau    +  \text{finite}
\end{equation}
But by assumption, $\det a = 1$, and so $0 = \frac{1}{{\det a}} \frac{\partial}{\partial q}\bigl[{\det a(\tau,q)}\bigr] = \frac{\partial a_{jk}}{\partial q^i} \, \bigl(a(\tau,q)^{-1}\bigr)^{jk}$.  Therefore the divergent part of \cref{DF2} vanishes.  Differentiating again shows that the $n=2$ term is finite, and in general:
\begin{multline} \label{DFend}
    \sum_{\substack{\text{loops $\Gamma$ with} \\ \text{$n$ exterior edges} \\ \text{connected to $\xi_1,\dots,\xi_n$}}} \frac{\ev(\Gamma)}{\left|\Aut\Gamma\right|}  = \\
    =   \int_{n=0}^t \delta(\tau - \tau)\, \frac1{\sqrt{\det a(\tau,q)}}\frac{\partial^n\bigl[ \sqrt{\det a(\tau,q)}\bigr]}{\partial q^{i_1}\cdots\partial q^{i_n}} \biggr|_{q = \gamma(\tau)} \xi_1^{i_1}(\tau)\cdots \xi_n^{i_n}(\tau)\,d\tau 
    + \text{finite} = \\
    = \text{finite}
  \end{multline}
This completes the proof of \cref{DFlemma} and therefore of \cref{divfreethm}.
\end{proof}

%\bibliography{Edited}{}
%\bibliographystyle{plain}

%

\end{document}